\documentclass[twocolumn,floatfix,showpacs,prd,aps,tightenlines]{revtex4}
\usepackage{amsmath}
\usepackage{graphicx}
\usepackage{psfrag}
\usepackage{color}
\usepackage{dcolumn}
\usepackage{bm}
\usepackage{longtable}

\newcommand{\bea}{\begin{eqnarray}}
\newcommand{\eea}{\end{eqnarray}}
\newcommand{\beq}{\begin{equation}}
\newcommand{\eeq}{\end{equation}}
\newcommand{\KMS}{\rm km\ s^{-1}}

\begin{document}

\def\fun#1#2{\lower3.6pt\vbox{\baselineskip0pt\lineskip.9pt
  \ialign{$\mathsurround=0pt#1\hfil##\hfil$\crcr#2\crcr\sim\crcr}}}
\def\lap{\mathrel{\mathpalette\fun <}}
\def\gap{\mathrel{\mathpalette\fun >}}
\def\kms{{\rm km\ s}^{-1}}
\def\vk{V_{\rm recoil}}

\title{Intermediate-mass-ratio black hole binaries:\\
intertwining numerical and perturbative techniques}

\author{
Carlos O. Lousto,
Hiroyuki Nakano,
Yosef Zlochower,
Manuela Campanelli
}

\affiliation{Center for Computational Relativity and Gravitation,\\
and School of Mathematical Sciences, Rochester Institute of
Technology, 85 Lomb Memorial Drive, Rochester, New York 14623}

\begin{abstract}
We describe in detail full numerical and perturbative techniques
to compute the gravitational radiation from intermediate-mass-ratio
black-hole-binary  inspirals and mergers.
We perform a series of full numerical simulations of nonspinning
black holes with mass ratios $q=1/10$ and $q=1/15$ from different
initial separations and for different finite-difference resolutions.
In order to perform those full numerical runs, we adapt the gauge
of the moving punctures approach with a variable damping term
for the shift. We also derive an extrapolation (to infinite radius)
formula for the waveform extracted at finite radius.
For the perturbative evolutions we use the full numerical tracks,
transformed into the Schwarzschild gauge, in the source terms of the
Regge-Wheller-Zerilli Schwarzschild perturbations formalism.
We then extend this perturbative formalism to take into account
small intrinsic spins of the large black hole, and validate it by
computing the quasinormal mode frequencies, where we find good agreement
for spins $|a/M|<0.3$. Including the final spins improves the
overlap functions when comparing full numerical and perturbative 
waveforms, reaching 99.5\% for the leading $(\ell,m)=(2,2)$ and (3,3) modes,
and 98.3\% for the nonleading (2,1) mode in the $q=1/10$ case,
which includes 8 orbits before merger. For
the $q=1/15$ case, we obtain overlaps near 99.7\% for all three
modes.
We discuss the modeling of the full inspiral and merger based on a
combined
matching of post-Newtonian, full numerical, and geodesic trajectories.
\end{abstract}

\pacs{04.25.dg, 04.30.Db, 04.25.Nx, 04.70.Bw} \maketitle

\section{Introduction}\label{sec:Introduction}

There is strong indirect evidence for the existence
of black holes (BHs) of a few solar masses ($M_\odot$)
residing in  galaxies and for supermassive BHs
(SMBHs), with masses $10^5M_\odot$-$10^{10}M_\odot$ in the central
cores of active galaxies. These BHs can form binaries and the
 mergers of black-hole binaries (BHBs)
are expected to be the strongest sources of gravitational radiation and
the most energetic event in the Universe.  The current generation of
ground-based interferometric gravitational wave detectors, such as
LIGO, VIRGO, and GEO, are most sensitive to BHB mergers with total
masses of a few tens to hundreds of solar masses, while the
space-based LISA detector will be  sensitive to mergers of BHBs with
a few million solar masses.

The existence of intermediate-mass BHs (IMBH), 
from a few hundred to tens of thousand of solar masses, is still uncertain.
If they exist, then these IMBH can form binaries with solar-mass-sized
objects, leading to compact-object mergers with  
mass ratios in the range $0.001<q=m_1/m_2<0.1$, which could be detected
by advanced LIGO.
 The detection of
gravitational waves from these encounters, as well as the correct modeling
of the waveform as a function of the BHBs physical parameters,
would allow us to estimate the population of such objects in the Universe.
Likewise, encounters of IMBH with SMBHs in the
centers of a galaxies would lead to mergers with mass
ratios in the range $0.001<q<0.1$, detectable by LISA.
On the other hand, theoretical N-body simulations \cite{Volonteri:2008gj},
assuming direct cosmological collisions of galaxies with 
central SMBHs,
set the most likely SMBH binary mass ratios in the range $0.01<q<0.1$. 

In Refs.~\cite{Brown:2006pj,Mandel:2007hi}
the prospects of  detecting IMBH binary (IMBHB)
 inspirals with advanced LIGO  was
discussed, and in Ref.~\cite{Mandel:2008bc} it was shown that 
intermediate-mass-ratio
(IMR) inspirals of IMBHs plunging into supermassive BHs would be
relevant to LISA, while IMR mergers of IMBHs with stellar
objects can be detected by LIGO/VIRGO. In both cases
the accuracy of the post-Newtonian (PN) approach (which was used to model the
gravitational radiation) was questioned
and the need for more accurate waveforms was stressed.

After the 2005 breakthroughs in numerical relativity
\cite{Pretorius:2005gq, Campanelli:2005dd, Baker:2005vv},
simulations of BHBs became routine.  The exploration of generic
binaries~\cite{Campanelli:2007ew} led to the discovery of large
recoils 
acquired by the remnant BH. While long term generic BHB evolutions
are possible, including
the last few tens of orbits \cite{Campanelli:2008nk, Szilagyi:2009qz},
two very interesting corners of the intrinsic parameter space
of the BHBs remain largely unexplored: maximally spinning binaries
and the small mass ratio limit.

In a previous letter \cite{Lousto:2010tb} we introduced
a new technique that makes use of nonlinear numerical 
trajectories and efficient perturbative evolutions to 
compute waveforms at large radii for the leading and nonleading modes. 
As a proof-of-concept, we computed waveforms for a relatively
close binary with $q=1/10$. In this paper we will describe
these techniques in detail, extend them to slowly spinning black holes,
and reach smaller mass ratios, to
the $q=1/15$ case, with full numerical simulations.

The paper is organized as follows. In Sec.\ \ref{Sec:Numerical} we describe
the full numerical techniques employed in the evolution of BHBs. Those
are based in the moving puncture approach \cite{Campanelli:2005dd,
Baker:2005vv} 
with a gauge choice
that allows a spatial and time variation of the gamma-driver parameter
$\eta(x^a,t)$. We describe the results of the simulations for two different
mass ratios $q=1/10,1/15$ and two different initial separations leading
to evolutions with BHs performing between 4 and 8 orbits prior
to merger, the latter representing the longest waveform published so far
in the small $q$ regime. The gauge has also been shown to work for
evolutions of a nonspinning $q=1/100$ BHB~\cite{Lousto:2010ut}.
In Sec.\ \ref{Sec:Perturbations} we describe the perturbative techniques
used to evolve a particle around a massive black hole. We extend the
Regge-Wheeler-Zerilli (RWZ) techniques to include, perturbatively,
a term linear 
in the spin of the larger black hole. This takes the form of second-order
perturbations and adds a source term to the usual Schwarzschild perturbations
(SRWZ).
We also study the asymptotic behavior of the perturbative solutions for
large $r$ and come up with a practical
way of correcting finite observer location effects
perturbatively on the numerical waveforms. 
In Sec. \ref{sec:NvsP} we describe the results of comparing full numerical
waveforms with perturbative ones that use the full numerical tracks for the
particle motion. We compute matching overlaps for the leading modes 
$(\ell,m)\ =\ (2,2);\ (2,1);\ (3,3)$. We verify the scaling  
of the waveform amplitudes with the reduced
mass $\mu$ for the mass ratios $q=1/10,1/15$.
We also quantify the effects of adding the spin of the final black hole into
the perturbative integrations.
In Sec. \ref{sec:Discussion} we discuss the properties of the full numerical
trajectories in the two cases studied $q=1/10,1/15$ that can be 
generalized to smaller
mass ratios and hence can help in providing a modeling for the tracks used
in the perturbative integration, in particular, the final ``universal
plunge''
and the use of resummed PN trajectories for the stages prior to the 
full numerical simulation.
Finally in the Appendix \ref{app:WEdetail} we give further evidence of
the accuracy and validity of the SRWZ formalism here developed by
computing the quasinormal modes
(QNM) and comparing them with the exact Kerr black-hole 
modes for different values of the spin parameter.

\section{Numerical Relativity Techniques}\label{Sec:Numerical}

To compute the numerical initial data, we use the puncture
approach~\cite{Brandt97b} along with the {\sc
TwoPunctures}~\cite{Ansorg:2004ds} thorn.  In this approach the
3-metric on the initial slice has the form $\gamma_{a b} = (\psi_{BL}
+ u)^4 \delta_{a b}$, where $\psi_{BL}$ is the Brill-Lindquist
conformal factor, $\delta_{ab}$ is the Euclidean metric, and $u$ is
(at least) $C^2$ on the punctures.  The Brill-Lindquist conformal
factor is given by $ \psi_{BL} = 1 + \sum_{i=1}^n m_{i}^p / (2 |\vec r
- \vec r_i|), $ where $n$ is the total number of `punctures',
$m_{i}^p$ is the mass parameter of puncture $i$ ($m_{i}^p$ is {\em
not} the horizon mass associated with puncture $i$), and $\vec r_i$ is
the coordinate location of puncture $i$.  We evolve these
black-hole-binary data-sets using the {\sc
LazEv}~\cite{Zlochower:2005bj} implementation of the moving puncture
approach~\cite{Campanelli:2005dd,Baker:2005vv} with the conformal
factor $W=\sqrt{\chi}=\exp(-2\phi)$ suggested by~\cite{Marronetti:2007wz}
For the runs presented here
we use centered, eighth-order finite differencing in
space~\cite{Lousto:2007rj} and an RK4 time integrator. (Note that we do
not upwind the advection terms.)

We use the Carpet~\cite{Schnetter-etal-03b} mesh refinement driver to
provide a ``moving boxes'' style of mesh refinement. In this approach
refined grids of fixed size are arranged about the coordinate centers
of both holes.  The Carpet code then moves these fine grids about the
computational domain by following the trajectories of the two black
holes.

We use {\sc AHFinderDirect}~\cite{Thornburg2003:AH-finding} to locate
apparent horizons.  We measure the magnitude of the horizon spin using
the Isolated Horizon algorithm detailed in~\cite{Dreyer02a}. This
algorithm is based on finding an approximate rotational Killing vector
(i.e.\ an approximate rotational symmetry) on the horizon $\varphi^a$. Given
this approximate Killing vector $\varphi^a$, the spin magnitude is
\begin{equation}
 \label{isolatedspin} S_{[\varphi]} =
 \frac{1}{8\pi}\int_{AH}(\varphi^aR^bK_{ab})d^2V,
\end{equation}
where $K_{ab}$ is the extrinsic curvature of the 3D-slice, $d^2V$ is
the natural volume element intrinsic to the horizon, and $R^a$ is the
outward pointing unit vector normal to the horizon on the 3D-slice.
We measure the direction of the spin by finding the coordinate line
joining the poles of this Killing vector field using the technique
introduced in~\cite{Campanelli:2006fy}.  Our algorithm for finding the
poles of the Killing vector field has an accuracy of $\sim 2^\circ$
(see~\cite{Campanelli:2006fy} for details). Note that once we have the
horizon spin, we can calculate the horizon mass via the Christodoulou
formula
\begin{equation}
{m^H} = \sqrt{m_{\rm irr}^2 +
 S^2/(4 m_{\rm irr}^2)},
\end{equation}
where $m_{\rm irr} = \sqrt{A/(16 \pi)}$ and $A$ is the surface area of
the horizon.
We measure radiated energy, linear momentum, and angular momentum, in
terms of $\psi_4$, using the formulae provided in
Refs.~\cite{Campanelli99,Lousto:2007mh}. However, rather than using
the full $\psi_4$, we decompose it into $\ell$ and $m$ modes and solve
for the radiated linear momentum, dropping terms with $\ell \geq 5$.
The formulae in Refs.~\cite{Campanelli99,Lousto:2007mh} are valid at
$r=\infty$. 
Typically, we would extract the radiated energy-momentum at finite
radius and extrapolate to $r=\infty$. However, for the smaller mass
ratios examined here, noise in the waveform introduces spurious
effects that make these extrapolations inaccurate. We therefore use
the average of these quantities extracted at radii $r=70$, $80$,
$90$, $100$ and use the difference between these quantities at
different radii as a measure of the error.
 We found that extrapolating
the waveform itself to $r=\infty$ introduced  phase errors due to
uncertainties in the areal radius of the observers, as well as
numerical noise. Thus when comparing perturbative to numerical waveforms, we use
the waveform extracted at $r=100M$. 
 In Sec.~\ref{sss:OLE} we provide an alternative
method of extrapolation of waveforms based on a perturbative propagation
of the asymptotic form of $\psi_4$ at large distances from the sources
leading to the following simple expression

\begin{eqnarray}
&&\lim_{r\to\infty}[r \,\psi_{4}^{\ell m}(r,t)]  \nonumber \\
&&=\left[r \,\psi_{4}^{\ell m}(r,t) 
- \frac{(\ell -1)(\ell +2)}{2} \int_0^t dt \, \psi_{4}^{\ell m}(r,t)
\right]_{r=r_{\rm Obs}}\nonumber \\
&& 
 + O(R_{\rm Obs}^{-2}) \,,
\label{eq:asymtpsi4ext}
\end{eqnarray}
where $r_{\rm Obs}$ is the approximate areal radius of the
sphere $R_{\rm Obs}=const$ [Add a factor $(1/2 - M/r)$ multiplying
the square bracket to correct for a difference in normalization 
between the Psikadelia and Kinnersley tetrads at large
distances.]
We have found that this formula gives reliable extrapolations 
for $R_{\rm Obs}\gtrsim100M$.

\subsection{Gauge}

We obtain accurate, convergent waveforms and horizon parameters by
evolving this system in conjunction with a modified 1+log lapse and a
modified Gamma-driver shift
condition~\cite{Alcubierre02a,Campanelli:2005dd}, and an initial lapse
$\alpha(t=0) = 2/(1+\psi_{BL}^{4})$.  The lapse and shift are evolved
with
\begin{subequations}
\label{eq:gauge}
  \begin{eqnarray}
(\partial_t - \beta^i \partial_i) \alpha &=& - 2 \alpha K,\\
 \partial_t \beta^a &=& (3/4) \tilde \Gamma^a - \eta(x^a,t) \beta^a,
 \label{eq:Bdot}
 \end{eqnarray}
 \end{subequations}
where different functional dependences for $\eta(x^a,t)$ have been
proposed in 
\cite{Alcubierre:2004bm, Zlochower:2005bj, Mueller:2009jx, Muller:2010zze, Schnetter:2010cz,Alic:2010wu}. 
Here we use a modification of the form proposed
in~\cite{Mueller:2009jx},
\begin{equation}
  \eta(x^a,t) =  R_0 \frac{\sqrt{\partial_i W \partial_j W \tilde
\gamma^{ij}}}{ \left(1 - W^a\right)^b},
\end{equation}
where we chose $R_0=1.31$.
The above gauge condition is inspired by, but differs from Ref.~\cite{Mueller:2009jx}
between the BHs and in the outer zones when $a\neq1$ and $b\neq2$.
Once the conformal factor settles down to its asymptotic
$\psi=C/\sqrt{r} + O(1)$ form near the puncture, $\eta$ will have the
form  $\eta = (R_0/C^2) ( 1+ b (r/C^2)^a)$ near the puncture and
$\eta= R_0 r^{b-2} M/(a M)^b$ as $r\to \infty$. In practice we used
$a=2$ and $b=2$, which reduces $\eta$ by a factor of $4$ at infinity
when compared to the original version of this gauge proposed
by~\cite{Mueller:2009jx}.
 We note that if we set $b=1$ then $\eta$
will have a $1/r$ falloff at $r=\infty$ as suggested
by~\cite{Schnetter:2010cz}. Our tests indicate that the choices
$(a=2$, $b=1)$ and $(a=1, b=1)$ lead to more noise in the waveform
than $(a=2,b=2)$.

\subsection{Simulations and results}\label{SubSec:Runs}

In order to obtain low-eccentricity initial data parameters, we
started with quasicircular post-Newtonian initial data parameters for
 the momenta and particle positions. We then evolved for 1-2 orbits,
and used the procedure detailed in~\cite{Pfeiffer:2007yz} to obtain lower
eccentricity parameters. In practice we performed between 3 and 4
iterations of the above procedure. In Table~\ref{table:ID} we show the
initial data parameters, horizon masses and mass ratio, and initial
orbital eccentricities  for the three configurations considered here.

\begin{widetext}

\begin{table}[!h]
\caption{Initial data parameters. The punctures are located on the
$x$-axis at positions $x_1$ and $x_2$, with puncture mass parameters
(not horizon masses) $m_1$ and $m_2$, and momentum $\pm\vec p$. In
all cases, the punctures have zero spin. Configuration $q10r7.3PN$
is based on the original PN parameters, prior to any eccentricity
removal iteration. The lower part of the table shows the horizon masses $m_{H_1}$
and $m_{H_2}$, the mass ratio $q$, the ADM mass, and the eccentricity
$e$.}
\label{table:ID}
\begin{ruledtabular}
\begin{tabular}{l|llllll}
Config & $x_1$ & $x_2$ & $p_x$ & $p_y$ & $m_1$ & $m_2$   \\
\hline
$q10r8.4$ & 7.633129 & -0.7531758 & -0.000168519 & 0.0366988 &
0.08523727 & 0.90739686 \\
$q10r7.3$   & 6.604383 & -0.6715184  & -0.000219713 & 0.0410386 &
0.08438951 & 0.90703855 \\
$q10r7.3PN$ & 6.604383 & -0.6715184  & -0.000326708 & 0.0404057 &
0.08438951 & 0.90703855 \\
$q15r7.3$ & 6.806173 & -0.4438775 & -0.000160518 & 0.0290721 &
0.05756623 & 0.93622418 \\
\hline
\hline
Config & $m_{H_1}$ & $m_{H_2}$ & $q$ & $M_{\rm ADM}$ & $e$ \\
\hline
$q10r8.4$   & 0.091289 & 0.912545 & 0.10004 & 1.0000428  & $0.0004$ \\
$q10r7.3$   & 0.091378 & 0.913010 & 0.10008 & 1.00025882 & $0.0017$ \\
$q10r7.3PN$ & 0.091329 & 0.912990 & 0.10003 & 1.00000000 & $0.008 $ \\
$q15r7.3$   & 0.062536 & 0.940421 & 0.06650 & 1.00005083 & $<0.0015$\\

\end{tabular}
\end{ruledtabular}
\end{table}

\begin{table}[!h]
\caption{Remnant horizon parameters and radiated energy-momentum}
\label{table:Remnant}
\begin{ruledtabular}
\begin{tabular}{l|llllll}
Config & $E_{rad}$ & $ J_{rad}$ & $M_{H} - M_{ADM}$ & 
$S_{ADM}-S_{H}$ & $\alpha$ & Kick $\KMS$\\
\hline
$q10r8.4$ & $0.00446\pm0.0001$ & $0.0517\pm0.001$ &
$0.00046\pm0.00003$ & $0.05028 \pm0.00001$ & $0.25986\pm0.00001$ &
$59.4\pm3.0$\\
$q10r7.3$  & $0.00400\pm0.00001$ & $0.0386\pm0.003$ &
$0.00415\pm0.00001$ & $0.04028\pm0.00001$ & $0.26034\pm0.00001$ & $65.8\pm2.0$ \\
$q15r7.3$ & $0.00216\pm0.00001$ & $0.0235\pm0.0004$ &
$0.00225\pm0.00001$ & $0.02289\pm 0.0004$ & $0.18872\pm0.00001$ & $
33.5\pm2.1$\\

\end{tabular}
\end{ruledtabular}
\end{table}

\end{widetext}

In all the simulations presented here, the outer boundaries were
placed at $400M$. We performed runs with three resolutions, with a
global refinement factor of $1.2$ between resolutions. For the
$q=1/10$ runs, we used 11 levels of refinement around the smaller BH,
with a central resolution of $h=M/307.2$ for the coarsest runs, while
for the $q=1/15$ run we used 12 levels of refinement, with a central
resolution of $M/614.4$. In Table~\ref{table:Remnant} we show the
radiated energy-momentum and remnant BH parameters for these
configurations. In the figures and tables below we refer to the
different resolution runs using the gridspacing on the coarsest grid
relative to  $h_0 = 10/3 M$.

\begin{figure}[!h]
  \caption{The puncture separation as a function of time for three
  $q=1/10$ simulations. The solid curve shows a high-eccentricity
  simulation obtained from PN quasicircular parameters (with particle
  limit corrections); the dotted curve shows results from a
 simulations with similar initial separation after a few 
   iterations to reduce eccentricity; the dot-dashed curve
  shows an even further separated binary with still smaller
  eccentricity. Note that the initial jump in the orbit does not
  appear to be a strong function of the eccentricity or initial separation.} 
  \includegraphics[width=3.2in]{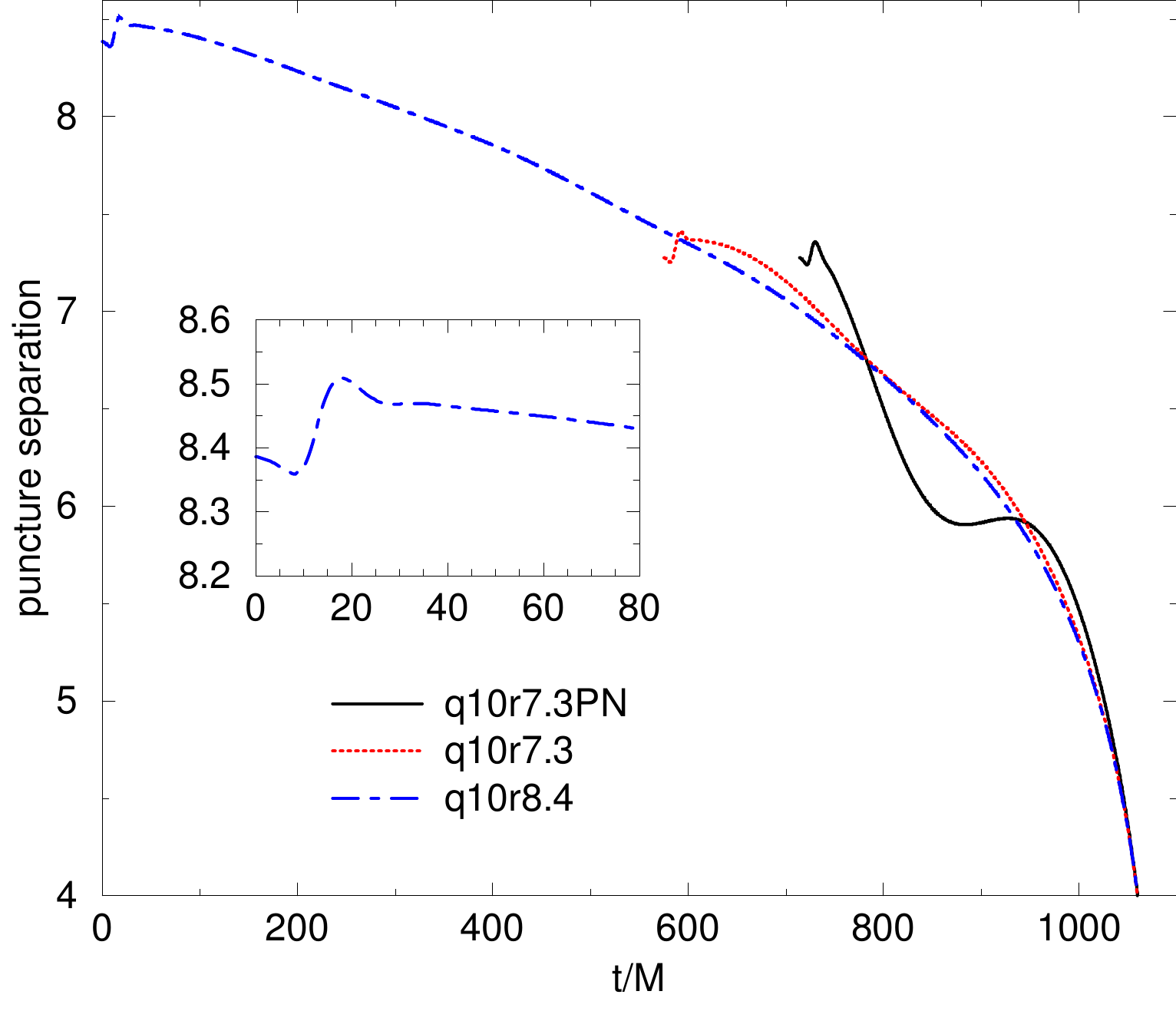}
  \label{fig:q10_r}
\end{figure}

\begin{figure}[!h]
  \caption{The magnitude of the puncture separation ($|\vec x_1 - \vec x_2|$) as a function of time for 
  a $q=1/10$ and $q=1/15$ binary at similar initial separations. Note
  that the initial jump in the orbit appears to be independent of $q$.
  Also note that the $q=1/15$ run inspirals more slowly.}
  \includegraphics[width=3.2in]{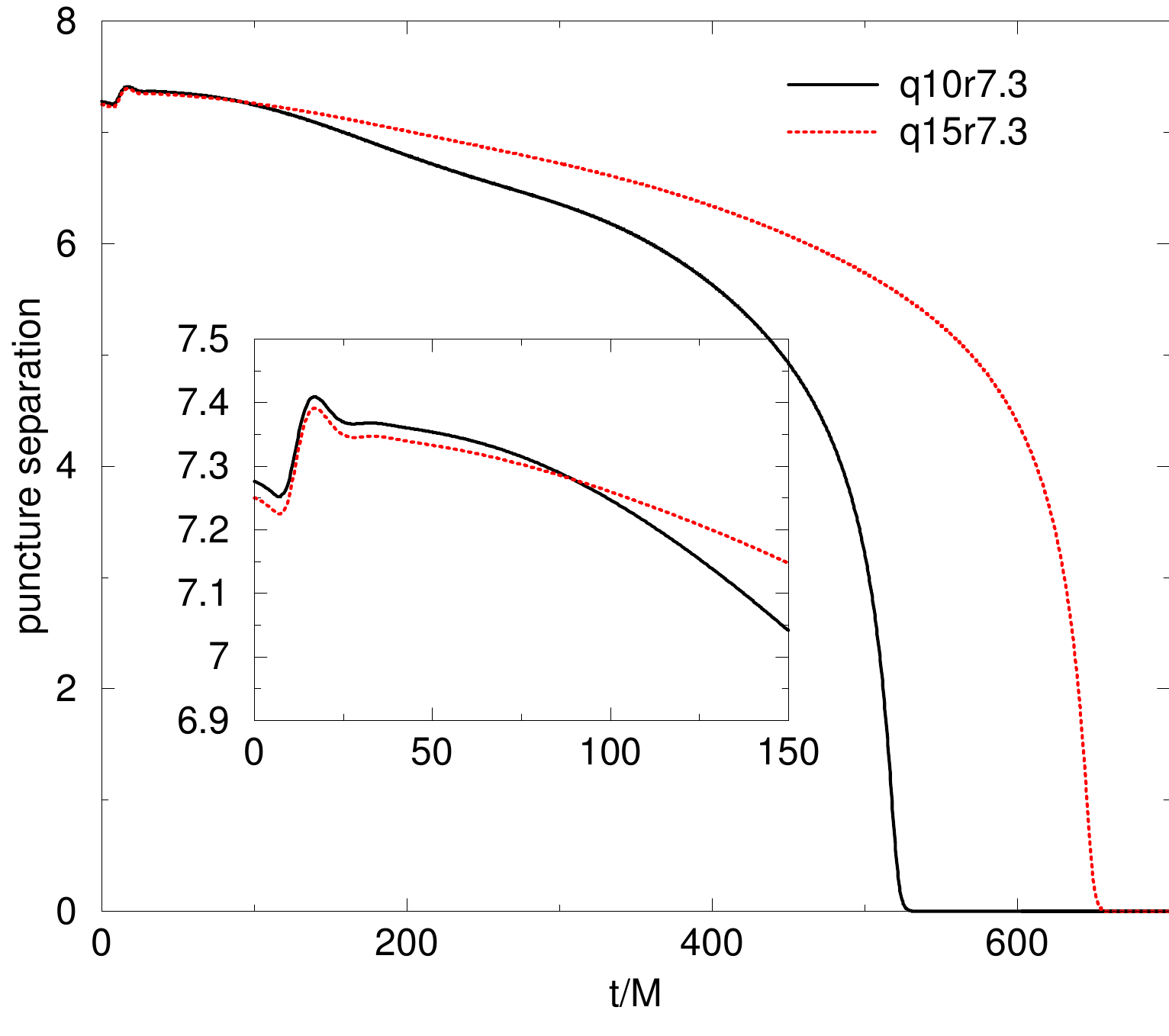}
  \label{fig:q10_q15_r}
\end{figure}

In Fig.~\ref{fig:q10_r} we show the orbital separation as a function
of time for the  $q10r8.4$ and  $q10r7.3$ configurations, as well as a
high-eccentricity configuration obtained by directly using PN parameters
in the initial data ($q10r7.3PN$) that we used for the proof-of-concept
in Ref.\ \cite{Lousto:2010tb}. Note that the highly eccentric
$q10r7.3PN$ binary merges sooner than the lower eccentricity
$q10r7.3$. From the plot we can also see that the initial jump in the
orbit is not a function of either initial separation or eccentricity. 
In Fig.~\ref{fig:q10_q15_r} we compare the orbital separation for the
$q10r7.3$ and $q15r7.3$ configurations. From the plot it is clear that
the initial jump in the orbit is not a strong function of mass ratio
either. This indicates that the initial jump will become more
problematic as the mass ratio is reduced (and hence the inspiral
becomes weaker).
We also observe that, quite independent of the initial separation and the
initial eccentricity, the track displays a universal behavior during the
final plunge. This confirms that the tracks are gravitational radiation
driven and we are numerically resolving this radiation accurately.

\begin{figure}
  \caption{An (xy) projection of the puncture separation ($\vec x_1 -
  \vec x_2$)   for 
  a $q=1/10$ and $q=1/15$ binary at similar initial separations. The
  trajectories have been rotated so that they overlap during the
  plunge and merger. Note the ``universal'' plunge trajectory.}
  \includegraphics[width=3.2in]{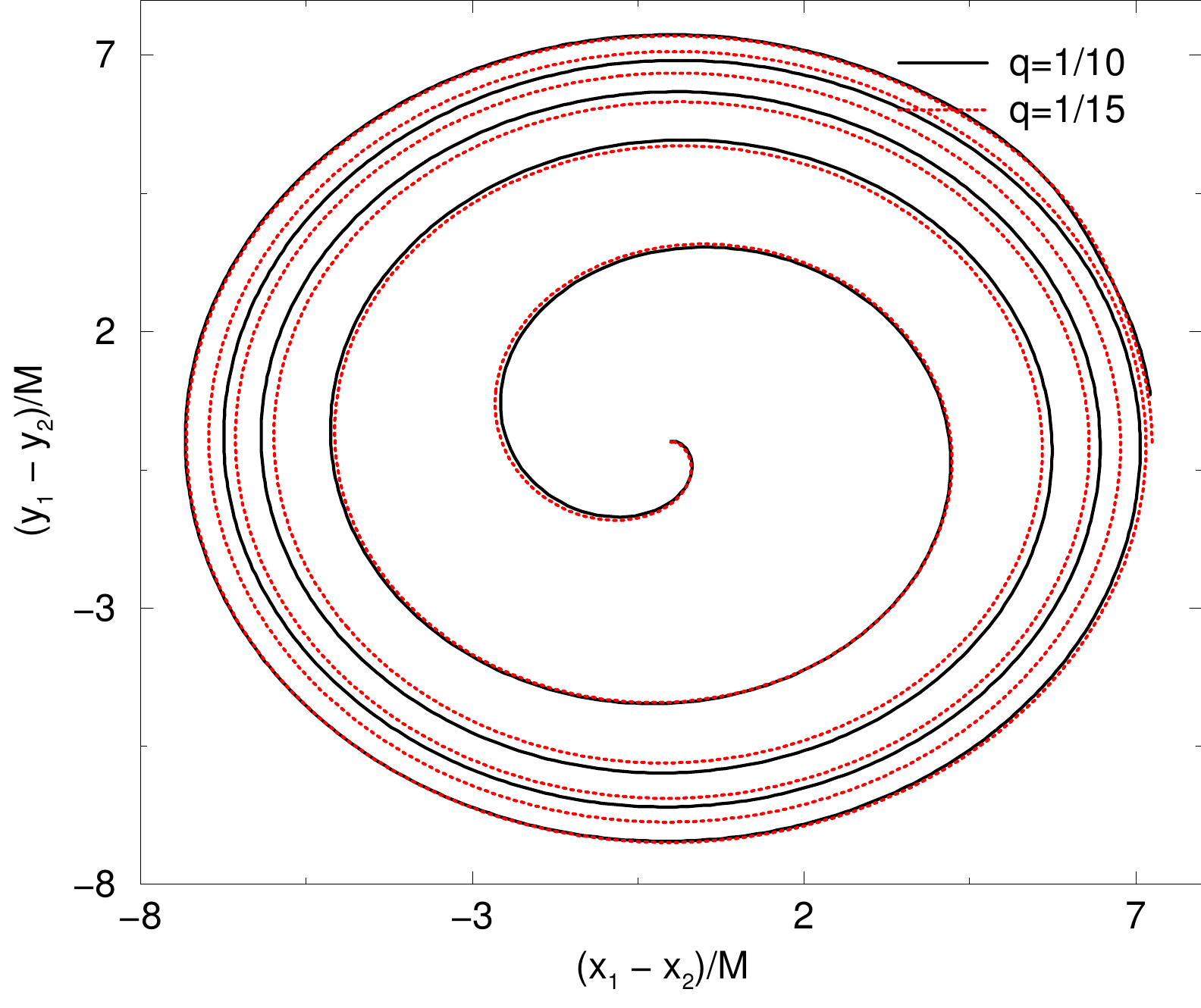}
  \label{fig:q10_q15_track}
\end{figure}

\begin{figure}
  \caption{The real part of the $(\ell=2,m=2)$ mode of $\psi_4$
   for a $q=1/10$ and $q=1/15$ binaries starting at similar
  separations. The waveform from the $q=1/15$ binary was rescaled by
  a factor of 1.5 (15/10).}
  \includegraphics[width=3.3in]{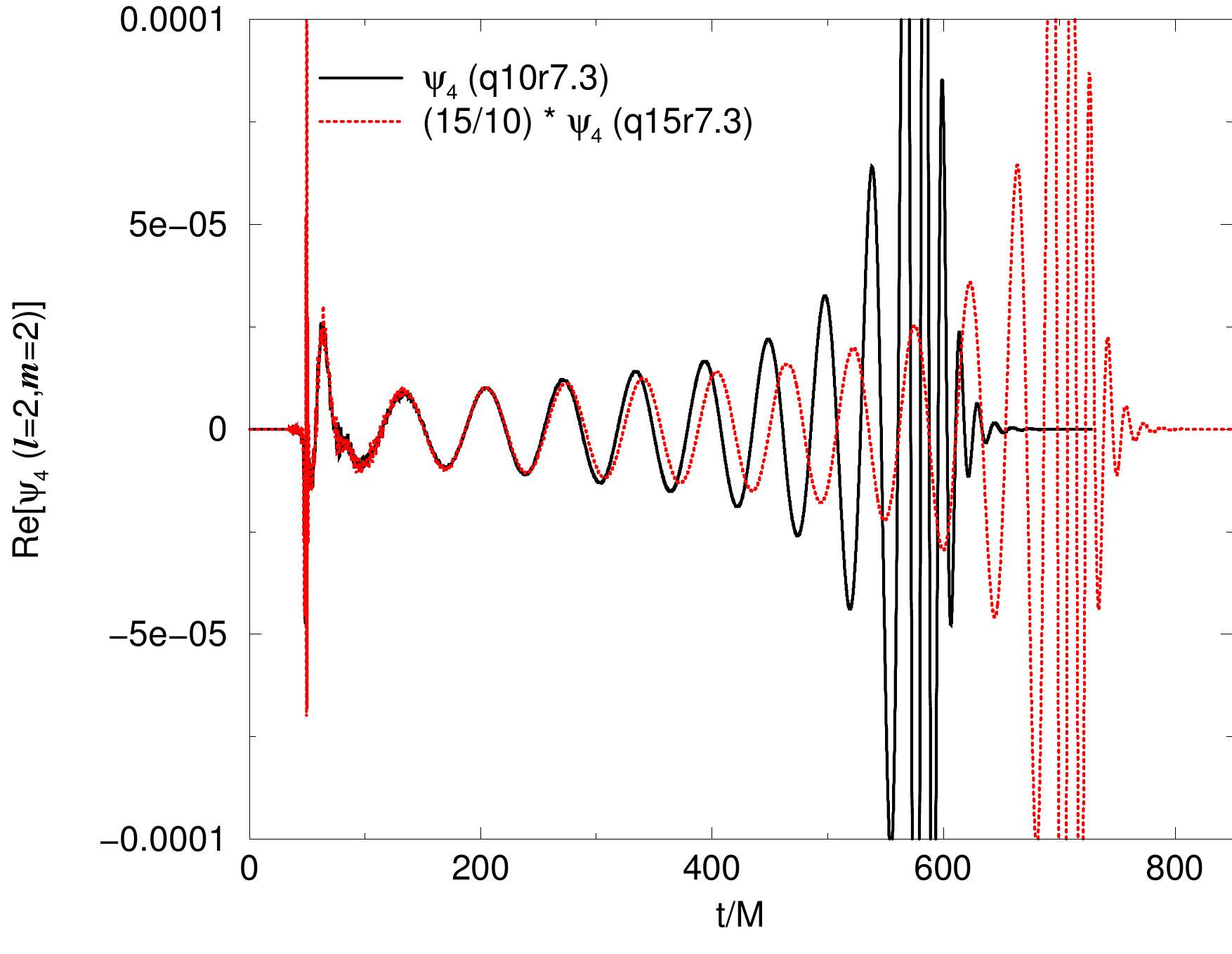}
  \label{fig:q10_q15_wave}
\end{figure}

\begin{figure}
  \caption{The convergence of the phase and amplitude of
    the $(\ell=2,m=2)$ mode
   of $\psi_4$ for the $q10r7.3PN$ configuration. Note that here the
   three resolutions consist of a low resolution with grid-spacing 1.2
times larger than the low resolution runs for $q10r7.3$, $q10r8.4$,
   $q15r7.3$ configurations.
 Eighth-order convergence implies $\psi_4(1.2 h_0) -
\psi_4(h_0) = 4.29982 (\psi_4(h_0) - \psi_4(h_0/1.2))$, while
fourth-order convergence implies
$\psi_4(1.2 h_0) - \psi_4(h_0) = 2.0736 (\psi_4(h_0) -
\psi_4(h_0/1.2))$.
 Initially, the error in $\psi_4$
is very small and dominated by grid noise. Eighth-order convergence in
the amplitude is apparent beginning at $t=320M$, while eighth-order
convergence in the phase becomes apparent at $t=420M$.
The dashed vertical line shows the time when the wave
frequency is $M\omega=0.2$. The phase error at this frequency is
$\delta \phi \leq 0.2$ rad.
}
  \includegraphics[width=3.3in]{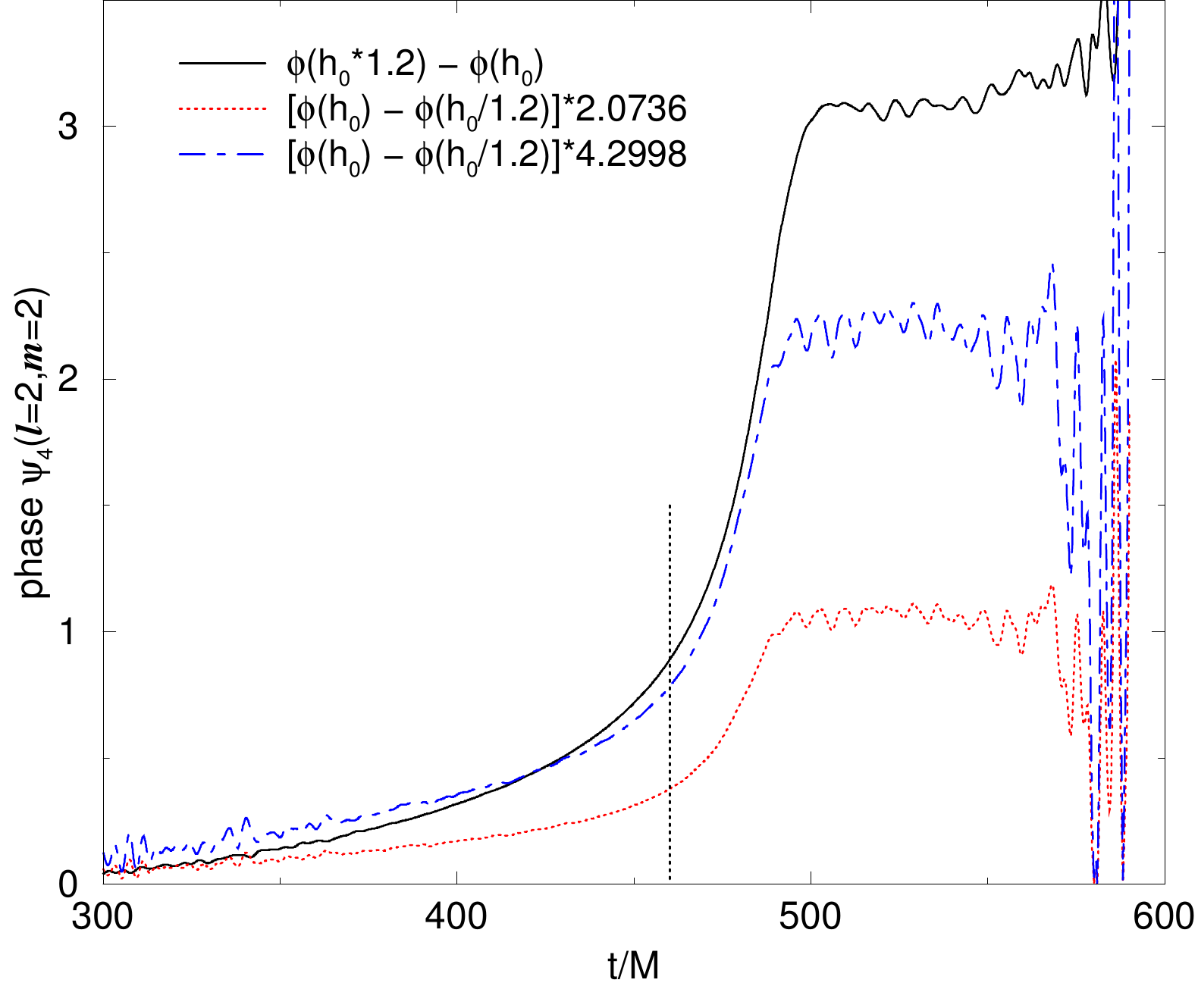}
  \includegraphics[width=3.3in]{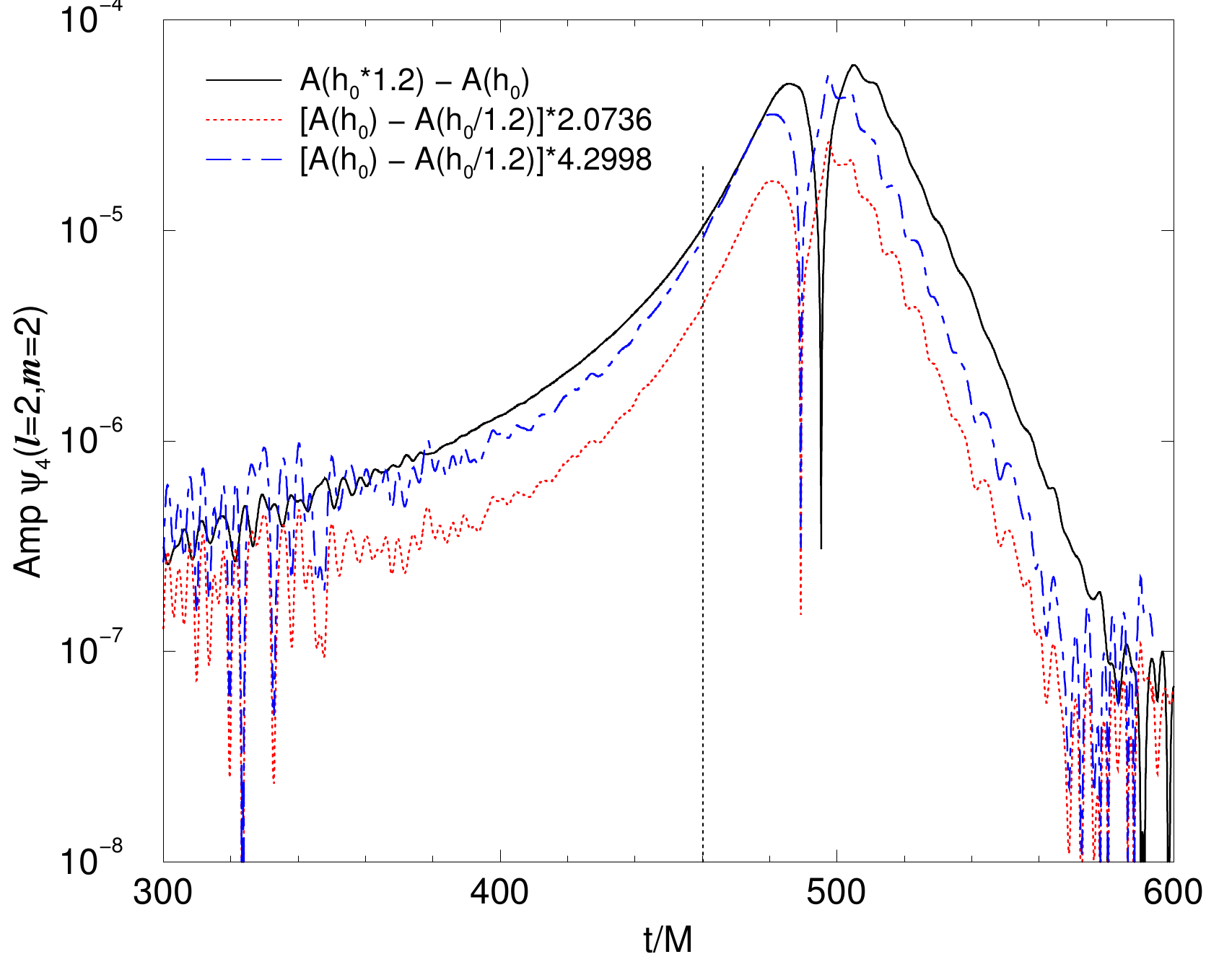}
  \label{fig:q10_wave_conv}
\end{figure}

\begin{figure}
  \caption{(Top) The phase of $(\ell=2,m=2)$ mode of $\psi_4$ 
   for a $q=1/15$ BHB for three resolutions. Note that the
  phase error only converges to fourth-order and that the highest
resolution is refined by a factor of $1.2^2$ rather than $1.2$ with
respect to the medium resolution.
  (Bottom) A convergence plot showing the initial (better than) 
    fourth-order convergence of the waveform. Note here that 
     the differences $\psi_4(1.2 h_0) - \psi_4(h_0) = 1.39895
(\psi_4(h_0) - \psi_4(h_0/1.2^2))$ if the waveform is fourth-order
convergent.
}
  \includegraphics[width=3.3in]{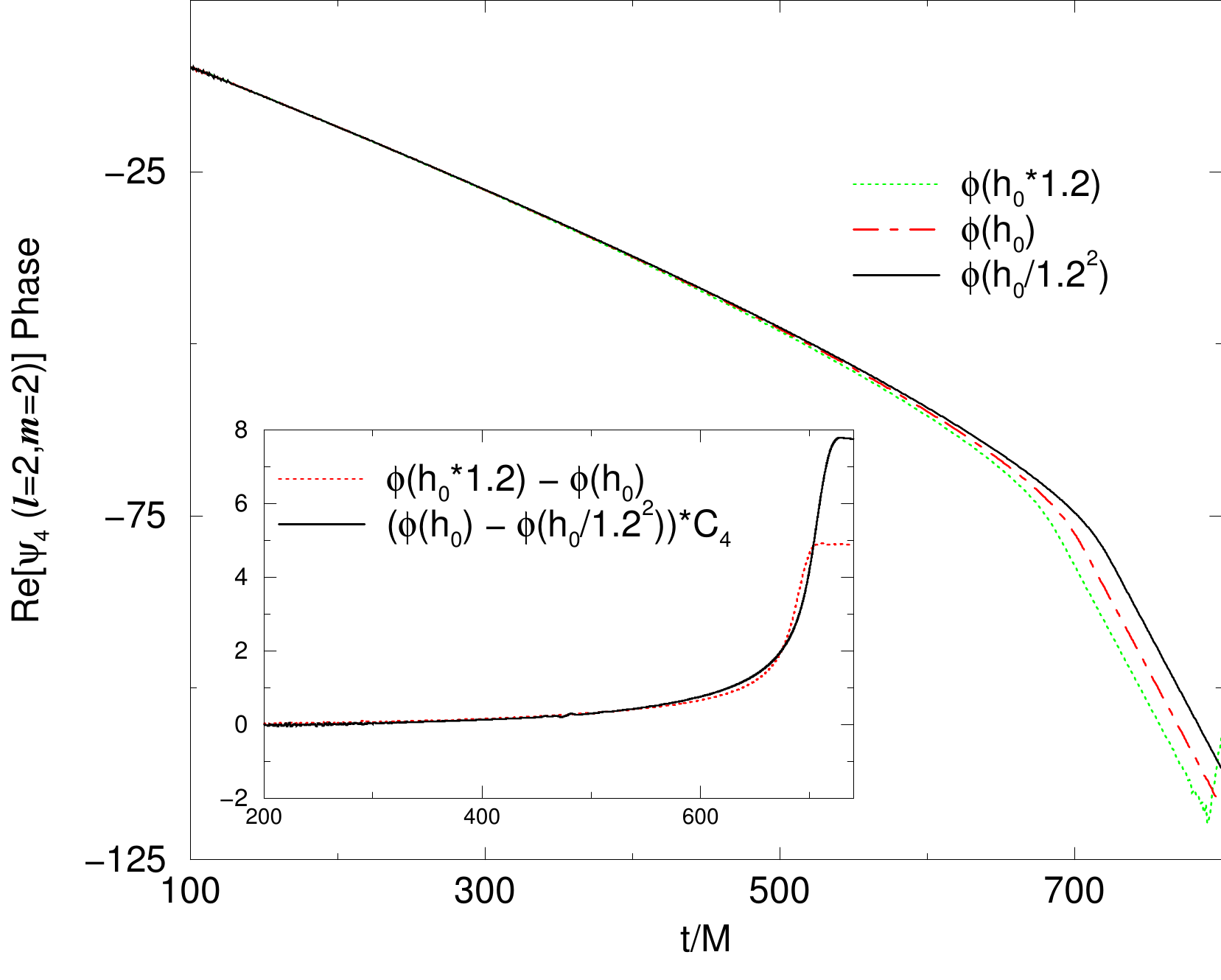}
  \includegraphics[width=3.3in]{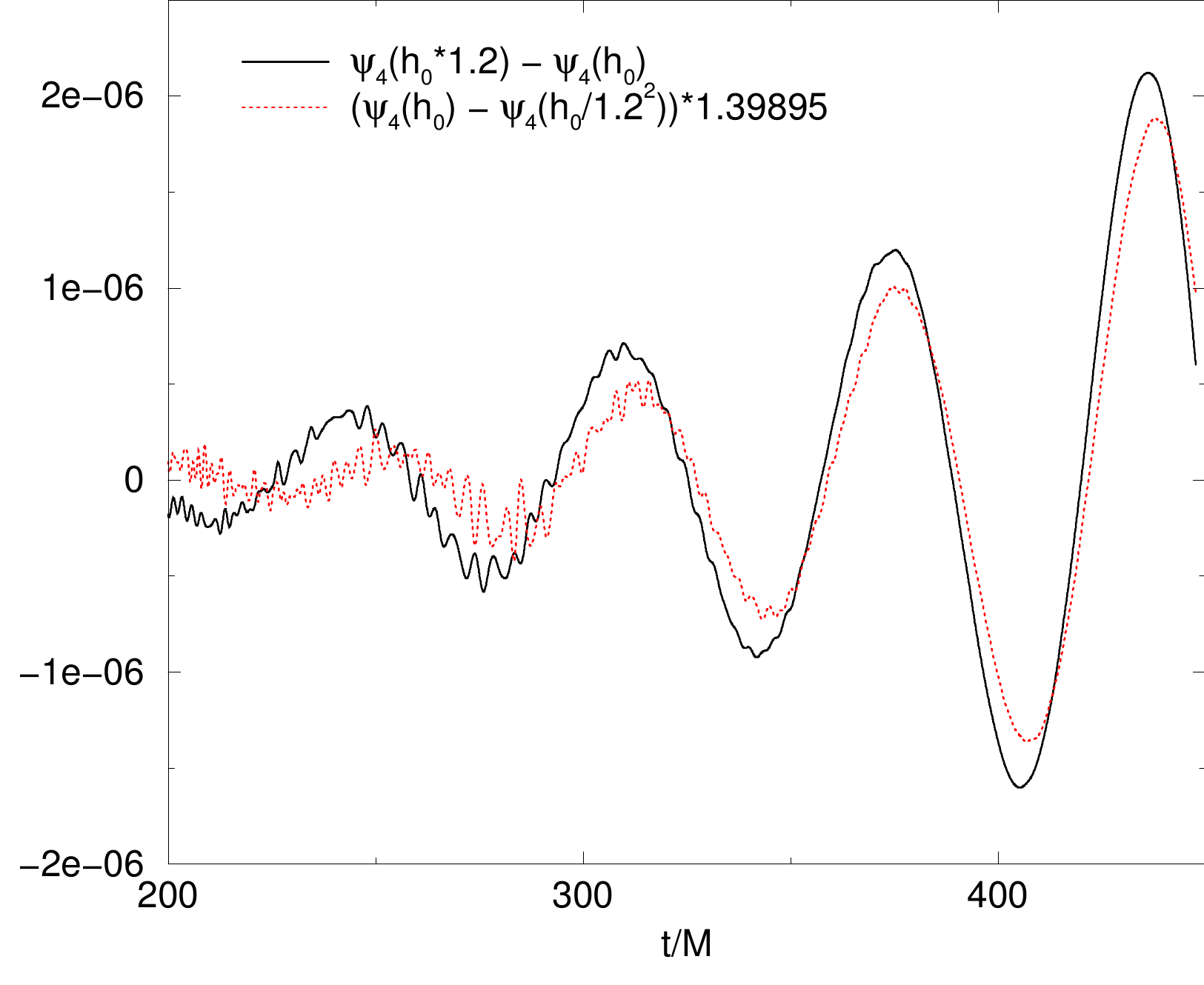}
  \label{fig:q15_wave_conv}
\end{figure}

In Fig.~\ref{fig:q10_q15_track} we show the orbital trajectories of the
$q10r7.3$ and $q15r7.3$ configuration. In the plot the curves have
been rotated to maximize the overlap during the plunge. From the plot
we see a ``universal'' plunge behavior at small separations with
distinctly different orbital dynamics at larger separations. As
expected, the small mass ratio binary merges more slowly. In
Fig.~\ref{fig:q10_q15_wave} we show the real part of the
$(\ell=2,m=2)$ mode of $\psi_4$ for the $q10r7.3$ and $q15r7.3$
configurations. Here the we rescaled $\psi_4$ for $q15r7.3$ 
by a factor of 1.5. Note that the good overlap of the rescaled
$\psi_4$ indicates that the amplitude of $\psi_4$ scales with $q$
(before the different orbital dynamics of $q=1/10$ and $q=1/15$ cause
the $q10r7.3$ to merge sooner). 
In Fig.~\ref{fig:q10_wave_conv} we show the convergence of the
$q10r7.3PN$ configuration for three resolutions. Note that in this
plot, the low resolution actually corresponds to a grid-spacing 1.2
times larger than the low resolutions for the other configurations.
From the plot we can see that at later time the convergence is eigth-order.
 The earlier time fourth-order convergence is due to
finite-difference and interpolation errors in the extraction routines.
At later times, the phase error dominates the errors in the waveform,
and this error converges to eighth-order.
Finally, in
Fig.~\ref{fig:q15_wave_conv} we
show the phase of the waveform for $q15r7.3$ for three resolutions.
The phase errors near the plunge are reported in
 Table~\ref{table:num_phase_err}.

\begin{table}[!h]
  \caption{The phase error in the $(\ell=2,m=2)$ mode of $\psi_4$
[extracted at $R=100M$ and extrapolated to $\infty$ using
Eq.~(\ref{eq:asymtpsi4ext})]
when the waveform frequency is $M\omega = 0.2$ for the medium and
high-resolution runs. The table shows the
predicted phase errors extrapolating to infinite resolution and
assuming eigth- and fourth-order convergence. 
}
  \label{table:num_phase_err}
\begin{ruledtabular}
\begin{tabular}{lll}
$Config$ & Eigth-order & Fourth-order \\
\hline
$q10r8.4$ ($h=h_0 $) & 0.205133 & 0.630496 \\
$q10r8.4$ ($h=h_0 / 1.2 $, pred) & 0.0477073 & 0.304058 \\
$q10r8.4$ ($h=h_0 / 1.2^2 $, pred) & 0.0110952 & 0.146633 \\

$q15r7.25$ ($h=h_0/1.2^2$) & 0.1406 & 0.762 \\

\end{tabular}
\end{ruledtabular}

\end{table}

\section{Perturbative Techniques}\label{Sec:Perturbations}

In this section we describe in some detail the use of perturbative
techniques to produce BHB waveforms from a small mass ratio
system. We summarize the key formulae used (for more details see, 
for instance, \cite{Lousto:2005ip}), and extend the formalism to add the spin
of the large black hole as a second-order perturbation, coupling
it to the radiative first-order perturbations. We neglect quadratic terms 
in the radiative modes of the order ${\cal O}(q^2)$. The resulting
equations are still of the Regge-Wheeler and Zerilli form (we are
still doing perturbations around a Schwarzschild background), but
they now include extended source terms with linear dependence on the
spin in addition to the local (Dirac's deltas) source terms already
present in the first-order formalism. We plug into these latter terms
the full numerical trajectories (hence indirectly also adding a
spin dependence). We denote the resulting formalism as 
Spin-Regge-Wheeler-Zerilli (SRWZ).

\subsection{Metric perturbations and particle's orbit}

\subsubsection{Spin as a perturbation}

We consider the Kerr metric up to $O(a^1)$. 
Here $a$ denotes the spin of the black hole which has the dimension of mass. 
In the usual Boyer-Lindquist coordinates, this is given by 
\begin{eqnarray}
ds^2 &=& -{\frac {r-2\,M}{r}} {dt}^{2}
-4\,{\frac {M a \sin^2 \theta  {d\phi}\,{dt}}{r}}
+ {\frac {r}{r-2\,M}} {{dr}}^{2} 
\nonumber \\ && 
+ {r}^{2} {{d\theta}}^{2}
+ {r}^{2} \sin^2 \theta {{d\phi}}^{2}
+ O(a^2) \,. 
\label{eq:Sch+spin}
\end{eqnarray}
In the above metric, the terms which depend on $a$  
are treated as the perturbation in the background Schwarzschild spacetime. 
\begin{eqnarray}
g_{\mu\nu} &=& g_{\mu\nu}^{\rm Sch} + h_{\mu\nu}^{\rm (1,spin)}  
\,. 
\label{eq:pert}
\end{eqnarray}
For the above metric perturbations, we consider 
the tensor harmonics expansion defined 
using the tensor harmonics of \cite{Nakano:2007cj}.
We find that the first-order perturbation, $O(a^1)$, is related to 
the $\ell=1,\,m=0$ odd parity mode, 
and the coefficient of the tensor harmonics is given by 
\begin{eqnarray}
h_{0\,10}^{\rm (1,spin)}(t,r) &=& \sqrt{\frac{4\pi}{3}}\frac{2 S}{r} 
\,,
\label{eq:1stS}
\end{eqnarray}
where $S=Ma$. 
The other components are zero.

\subsubsection{Second-order formulation}

In the following, we treat spin-radiation couplings 
in the second-order perturbation. 
Therefore, we consider the Einstein equation in the 
second perturbative order. 
\begin{eqnarray}
&&G_{\mu\nu}^{(1)}[h^{(1)}] 
+ G_{\mu\nu}^{(1)}[h^{(2)}]
+ G_{\mu\nu}^{(2)}[h^{(1)},h^{(1)}] 
\nonumber \\ && \quad 
= 8\,\pi\,
\left( T_{\mu\nu}^{(1)} + T_{\mu\nu}^{(2)} 
\right) 
= 8\,\pi\,T_{\mu\nu}
\,,
\end{eqnarray}
According to~\cite{Mino:2007ft}, and the fact that we use the
Numerical Relativity (NR) trajectory, 
we do not separate the first and second-order energy-momentum tensor 
of the particle. And the second-order metric perturbation, $h^{\rm (2,wave)}$
is created by the spin, $h^{\rm (1,spin)}$-radiation, $h^{\rm (1,wave)}$ couplings. 
In this case, we may solve 
\begin{eqnarray}
G_{\mu\nu}^{(1)}[h^{\rm (1,wave)}] &=& 8\,\pi\,T_{\mu\nu} \,,
\label{eq:formal1stE}
\\
G_{\mu\nu}^{(1)}[h^{\rm (2,wave)}] &=& - G_{\mu\nu}^{(2)}
[h^{\rm (1,wave)},h^{\rm (1,spin)}] \,,
\label{eq:formal2ndE}
\end{eqnarray}
up to $O(a^1)$, 
where we ignore the square of the first-order wave functions.  

As discussed below, we solve Eqs.~(\ref{eq:formal1stE}) and (\ref{eq:formal2ndE}) 
for the even parity perturbation of the Regge-Wheeler-Zerilli formalism 
in the following form.
\begin{eqnarray}
&&
G_{\mu\nu}^{(1)}[h^{\rm (1,wave)}+ h^{\rm (2,wave)}] 
+ G_{\mu\nu}^{(2)}[h^{\rm (1,wave)},h^{\rm (1,spin)}] 
\nonumber \\ && \quad 
= 
G_{\mu\nu}^{(1)}[h^{\rm (wave)}] 
+ G_{\mu\nu}^{(2)}[h^{\rm (wave)},h^{\rm (1,spin)}] 
\nonumber \\ && \quad 
= 8\,\pi\,T_{\mu\nu} 
\,,
\label{eq:formal_even}
\end{eqnarray}
where $h^{\rm (wave)}=h^{\rm (1,wave)}+ h^{\rm (2,wave)}$. 
On the other hand, for the odd parity perturbation, 
Eqs.~(\ref{eq:formal1stE}) and (\ref{eq:formal2ndE}) 
are solved for each perturbative order.

Here we consider intermediate mass ratio binaries. 
As discussed in~\cite{Lousto:2008vw}, 
we can introduce some second-order effects 
that arise purely from the particle's first-order perturbation, 
if we treat the particle as a reduced mass $\mu=m_1 m_2/(m_1+m_2)$ 
orbiting around a black hole with the total mass $M=m_1+m_2$.

\subsubsection{Orbit for inspiral}

First, we should note that the coordinates used in NR simulations 
are chosen to produce stable evolutions and  correspond, initially,  
to isotropic coordinates.  Perturbative calculations, on the other hand, 
regularly make use of the standard Schwarzschild coordinates. 
The easiest way to relate the two is to translate the numerical tracks 
into the Schwarzschild coordinates.  This can be achieved by considering 
the late-time numerical coordinates that correspond to radial isotropic 
``trumpet'' stationary $1+\log$ slices of the Schwarzschild 
spacetime~\cite{Hannam:2006vv}.  We obtain the explicit time and radial 
coordinate transformations following the procedure detailed 
in Ref.~\cite{Brugmann:2009gc}.

Thus, we consider the NR trajectory as an orbit projected 
on the Schwarzschild background. Therefore, we calculate the particle's energy, 
angular momentum etc. 
by using the Schwarzschild metric. 
Here, since we have only the three velocity $v^i(t)$ 
from the data of the NR trajectory, the time component of the four velocity $u^\mu$ 
is derived by assuming the ``instantaneous'' Schwarzschild geodesic approximation.
 
In this approximation, the energy and angular momentum are given by.  
\begin{eqnarray}
E 
&=& \left(1-\frac{2M}{R}\right)\, u^t \,,
\\
\label{eq:evalE}
L_z 
&=& R^2 \,u^{\phi} \,,
\label{eq:evalL}
\end{eqnarray}
where $u^{\mu}=dx^{\mu}/d\tau$ is the four velocity, 
$R=R(t)$ denotes the orbital radius, 
and we are considering the equatorial orbit ($\Theta_0=\pi/2$). 
To evaluate $U(t)=u^{t}$, we use 
\begin{eqnarray}
&& g_{\mu\nu} u^{\mu} u^{\nu} = -1 
\nonumber \\ && \quad 
= (U(t))^2 
\biggl[ 
- \left(1-\frac{2M}{R(t)}\right) 
+ \left(1-\frac{2M}{R(t)}\right)^{-1} (\dot {R} (t))^2
\nonumber \\ && \qquad 
+ (R(t))^2 (\dot \Phi (t))^2 
\biggr]
\label{eq:Uderi}
 \,.
\end{eqnarray}
Here, $\dot {R}=u^r/u^t=dR/dt$ 
and $\dot {\Phi}=u^\phi/u^t=d\Phi/dt$ are the three velocity of the particle. 

We note that the energy $E$ derived from the above $U(t)$
does not decrease monotonically, 
and also in the end of the orbital evolution, we 
can not calculate $U(t)$ appropriately by using Eq.~(\ref{eq:Uderi}), 
because $U(t) \to \infty$ or becomes complex.  
$U(t) \to \infty$ is, in practice, not inconsistent 
because $U(t) \sim (1-2M/R(t))^{-1}$ for Schwarzschild geodesics. 

Therefore, we fix the energy 
at some orbital radius (or time $t=t_m$) as 
\begin{eqnarray}
E_m &=& E(t_m)
\nonumber \\ 
&=& \left(1-\frac{2M}{R(t_m)}\right) U(t_m) \,,
\end{eqnarray}
and use the following expression 
to obtain $U(t)$ for smaller radii. [This may give the innermost stable circular orbit 
(ISCO) radius.] 
\begin{eqnarray}
U(t) &=& E_m\, \left(1-\frac{2M}{R(t)}\right)^{-1} 
 \,,
\end{eqnarray}
At this stage, we still use the three velocity derived from the NR trajectory. 

Here we set $R(t_m)/M=7.64$ for the $q=1/10$ case. 
This radius is obtained from the energy minimum evaluated by Eq.~(\ref{eq:evalE}). 
In the $q=1/15$ case, we do not have such an energy minimum. 
Therefore, we simply set the same radius as for the $q=1/10$ case.

\subsubsection{Orbit near merger}\label{Onm}

There are large differences between 
the coordinate system used in the NR simulation and the Schwarzschild coordinates
near the horizon. 
Although the binary merges at finite time in the NR simulation, 
the binary does not merge in the Schwarzschild coordinates.
Therefore, we need to give the orbit near the horizon. 

Here, we assume that the radiation reaction is not important near merger 
after $t=t_f$,  
and use the geodesic orbit on the Schwarzschild spacetime. 
First, we consider the conserved quantities, i.e., 
the energy and angular momentum. 
\begin{eqnarray}
&&E_m = E(t_m) = E(t_f) \,,
\nonumber \\
&&L_f = L(t_f)
\nonumber \\ && \quad 
= R(t_f)^2 \dot \Phi(t_f) E_m\, \left(1-\frac{2M}{R(t_f)}\right)^{-1} 
\,,
\end{eqnarray}
where $E_m$ is the same as the previous section. 
And then, from the above equations, we calculate  
\begin{eqnarray}
U(t) &=& E_m\, \left(1-\frac{2M}{R(t)}\right)^{-1} \,,
\nonumber \\
\dot \Phi(t) &=& \frac{L_f}{E_m}\frac{R(t)-2M}{R(t)^3} 
\,.
\end{eqnarray}

On the other hand, we use a fitting formula for the radial trajectory. 
By using $g_{\mu\nu}u^{\mu}u^{\nu}=-1$, 
we define an effective energy for the radial motion, 
\begin{eqnarray}
&& E_r^2 = 
\left(1-\frac{2M}{R(t_f)}\right)^3 
\left(1+\frac{L_f^2}{R(t_f)^2}\right)
\nonumber \\ && \quad \times 
\left( \left(1-\frac{2M}{R(t_f)}\right)^2 - \dot R(t_f)^2\right)^{-1} 
 \,,
\end{eqnarray}
and consider $E_r$ as a constant after $t=t_f$. 
The evolution of $\dot R(t)$ is derived as 
\begin{eqnarray}
&& \dot R(t) = - \left(1-\frac{2M}{R(t)}\right)
\nonumber \\ && \quad \times 
\sqrt{1- \frac{1}{E_r^2} \,\left(1-\frac{2M}{R(t)}\right)
\,\left(1+\frac{L_f^2}{R(t)^2}\right)}
 \,.
\end{eqnarray}
From this equation, we can obtain various equations if we need, 
for example, $\ddot R(t)=(\partial\dot R(t)/\partial R(t)) \dot R(t)$. 
It is noted that we may consider another treatment 
as discussed in Sec.~\ref{sec:Discussion}. 

In our perturbative code for both $q=1/10$ and $1/15$ cases, 
we set $R(t_f)/M=3.0$ which is inside the ISCO radius. 
This is because we want to use the NR trajectories as long as possible in this paper, 
and the data of the tracks become noisy inside the above orbital radius 
due to the coordinate transformation. 

\subsection{Regge-Wheeler-Zerilli equations with spin}

\subsubsection{First-order Regge-Wheeler-Zerilli equations}\label{sec:1stRWZ}

For the notation of the Regge-Wheeler-Zerilli 
formalism~\cite{Regge:1957td,Zerilli:1971wd}, 
we use \cite{Nakano:2007cj} and \cite{Lousto:2008vw}. 
In the first-order perturbation, i.e., the nonspinning case, we may solve the equations, 
\begin{eqnarray}
&&
-{\frac {\partial ^{2}}{\partial {t}^{2}}}\Psi_{\ell m}^{(1)} \left( t,r \right) 
+{\frac {\partial ^{2}}{\partial {r^*}^{2}}}
\Psi_{\ell m}^{(1)} \left( t,r \right) 
\nonumber \\ && \quad 
-V_{\ell}^{\rm (even)}(r) 
\Psi_{\ell m}^{(1)} \left( t,r \right) 
= S_{\ell m}^{\rm (even,1)} \left( t,r \right) \,,
\label{eq:1stEven}
\end{eqnarray}
for the even parity with the Zerilli function $\Psi_{\ell m}^{(1)}$, 
and
\begin{eqnarray}
&&
-{\frac {\partial ^{2}}{\partial {t}^{2}}}\Psi_{\ell m}^{\rm (o,1)} \left( t,r \right) 
+{\frac {\partial ^{2}}{\partial {r^*}^{2}}}
\Psi_{\ell m}^{\rm (o,1)} \left( t,r \right) 
\nonumber \\ && \quad 
-V_{\ell}^{\rm (odd)}(r) 
\Psi_{\ell m}^{\rm (o,1)} \left( t,r \right) 
= S_{\ell m}^{\rm (odd,1)} \left( t,r \right) \,,
\label{eq:1stOdd}
\end{eqnarray}
for the odd parity with the Regge-Wheeler function $\Psi_{\ell m}^{\rm (o,1)}$.
Here $r^*=r+2M\ln[r/(2M)-1]$ is a characteristic coordinate, 
and the first-order source terms, 
$S_{\ell m}^{\rm (even,1)}$ and $S_{\ell m}^{\rm (odd,1)}$ are given by 
\begin{widetext}
\begin{eqnarray}
S_{\ell m}^{\rm (even,1)} \left( t,r \right)  &=& 
{\frac {16\,\pi \, ( r-2\,M ) ^{2} ( r{\ell}^{2}+r\ell-4\,r+2\,M ) }
{ \ell ( \ell+1 )( r{\ell}^{2}+r\ell-2\,r+6\,M ) r }}{\cal A}_{\ell m}^{(1)} ( t,r ) 
-{\frac { 16\, \sqrt {2}\,\pi ( r-2\,M )}
{\sqrt {\ell ( \ell+1 )  ( \ell-1 )  ( \ell+2 ) }}}{\cal F}_{\ell m}^{(1)} ( t,r ) 
\nonumber \\ && 
+{\frac {32\,\pi \, ( r-2\,M ) ^{2}\sqrt {2}}
{ ( r{\ell}^{2}+r\ell-2\,r+6\,M ) \sqrt {\ell ( \ell+1 ) }}}{\cal B}_{\ell m}^{(1)} ( t,r ) 
-{\frac {32\,\pi ( r-2\,M ) ^{3}  }
{ ( r{\ell}^{2}+r\ell-2\,r+6\,M ) \ell ( \ell+1 ) }}\,{\frac {\partial }{\partial r}}
{\cal A}_{\ell m}^{(1)} ( t,r )
\nonumber \\ && 
-{\frac {16\,\pi \,r ( {\ell}^{4}{r}^{2}+2\,{r}^{2}{\ell}^{3}-5\,{r}^{2}{\ell}^{2}
+16\,r{\ell}^{2}M-6\,{r}^{2}\ell+16\,r\ell M+8\,{r}^{2}-68\,rM+108\,{M}^{2} ) }
{ ( \ell+1 ) \ell ( r{\ell}^{2}+r\ell-2\,r+6\,M ) ^{2}}}{\cal A}_{0 \ell m}^{(1)} ( t,r ) 
\nonumber \\ && 
+{\frac { 32\,\pi ( r-2\,M ) {r}^{2}}
{ ( r{\ell}^{2}+r\ell-2\,r+6\,M ) \ell ( \ell+1 ) }}\,{\frac {\partial }{\partial r}}
{\cal A}_{0 \ell m}^{(1)} ( t,r ) 
+{\frac {32\,\sqrt {2}\, \pi( r-2\,M ) ^{2} }
{ ( r{\ell}^{2}+r\ell-2\,r+6\,M ) \ell ( \ell+1 ) }}{\cal G}_{\ell m}^{(1)} ( t,r )  
\,,
\nonumber \\
S_{\ell m}^{\rm (odd,1)} \left( t,r \right)  
&=& 
{\frac {16\,\sqrt {2} \,\pi ( r-2\,M )}{\sqrt {\ell ( \ell+1 ) }
( \ell-1 )  ( \ell+2 ) }}\, {\cal Q}_{0 \ell m}^{(1)}( t,r ) 
+{\frac {16\,\sqrt {2}\,\pi \,r ( r-2\,M ) }{\sqrt {\ell ( \ell+1 ) } ( \ell-1 )  ( \ell+2 ) }}
\,{\frac {\partial }{\partial r}} {\cal Q}_{0 \ell m}^{(1)}( t,r )
\nonumber \\ && 
-{\frac {16\,\sqrt {2}\,i\pi  r ( r-2\,M ) }
{\sqrt {\ell ( \ell+1 ) } ( \ell-1 )  ( \ell+2 ) }}
\,{\frac {\partial }{\partial t}} {\cal Q}_{\ell m}^{(1)}( t,r ) 
\,,
\end{eqnarray}
\end{widetext}
where, ${\cal A}_{\ell m}^{(1)}$ etc. denote the tensor harmonics coefficient 
of the particle's energy-momentum tensor $T_{\mu\nu}$. 
It is noted that 
the even parity wave function $\Psi_{\ell m}^{(1)}$ and 
odd parity wave function $\Psi_{\ell m}^{\rm (o,1)}$ 
are related to the Moncrief's~\cite{Moncrief74} and 
the Cunningham et al.~\cite{Cunningham78} 
waveforms by a normalization factor, respectively.

\subsubsection{Even parity perturbation with spin}

When we discuss only the second-order Einstein equation 
in Eq.~(\ref{eq:formal2ndE}) for the even parity perturbation, 
the Zerilli equation with the $O(a^1)$ spin effect is 
written as  
\begin{eqnarray}\label{eq:WaveE2ndO}
&&
-{\frac {\partial ^{2}}{\partial {t}^{2}}}\Psi_{\ell m}^{(2)} \left( t,r \right) 
+{\frac {\partial ^{2}}{\partial {r^*}^{2}}}
\Psi_{\ell m}^{(2)} \left( t,r \right) 
\nonumber \\ && \quad 
-V_{\ell}^{\rm (even)}(r) 
\Psi_{\ell m}^{(2)} \left( t,r \right) = 
S_{\ell m}^{\rm (even,2)} \left( t,r \right)
\,,
\end{eqnarray}
where the second-order source term $S_{\ell m}^{\rm (even,2)}$ 
in the above equation is given by 
\begin{widetext}
\begin{eqnarray}
&&
S_{\ell m}^{\rm (even,2)} \left( t,r \right)  
= S_{\ell m}^{\rm (even,2)}(E,S)
+ S_{\ell m}^{\rm (even,2)}(O,S) \,;
\nonumber \\ 
&&
S_{\ell m}^{\rm (even,2)}(E,S) 
= 
\frac{m\,S}{{\ell}  \left( \ell+1 \right)  \left( r{\ell}^{2}+r\ell-2\,r+6\,M \right) }
 \biggl( 64\,{\frac {\sqrt {2}\pi \, \left( -r+2\,M \right) 
 \left( -2\,r+r\ell+r{\ell}^{2}+12\,M \right) }
{\sqrt {\ell \left( \ell+1 \right) }r \left( r{\ell}^{2}+r\ell-2\,r+6\,M \right) }}
{\cal B}_{0\ell m}^{(1)} \left( t,r \right) 
\nonumber \\ && 
\qquad 
+64\,{\frac {\sqrt {2}\pi \, \left( -r+2\,M \right) }{r}}
{\cal A}_{1\ell m}^{(1)} \left( t,r \right) 
+{\frac {192\,i \left( -r+2\,M \right) \pi \,\sqrt {2} }
{\sqrt {\ell \left( \ell+1 \right)  \left( \ell-1 \right)  \left( \ell+2 \right) }}}
\partial_t {\cal F}_{\ell m}^{(1)}  \left( t,r \right)
\nonumber \\ && \qquad 
+{\frac {8\,i \left( 12\,M-6\,r+{\ell}^{4}r+2\,r{\ell}^{3}+r\ell+2\,r{\ell}^{2}
 \right)  \left( -r+2\,M \right) }
{{r}^{3} \left( r{\ell}^{2}+r\ell-2\,r+6\,M \right) }}
H_{1\ell m}^{(1)} \left( t,r \right) 
+{\frac {8\,i\ell \left( \ell+1 \right) }{r}} 
\partial_t K_{\ell m}^{(1)}  \left( t,r \right) 
\biggr) 
\,,
\nonumber \\ 
&&
S_{\ell m}^{\rm (even,2)}(O,S) 
= 
\frac{4\,S}{ {\ell}  \left( \ell+1 \right)  \left( \ell-1 \right) 
 \left( \ell+2 \right) }
 \sqrt {{\frac { \left( \ell-m \right)  \left( \ell+m \right) }
 { \left( 2\,\ell-1 \right)  \left( 2\,\ell+1 \right) }}} 
\biggl( 6\, \left( r-2\,M \right)  ( 2\,{\ell}^{4}{r}^{2}-4\,{r}^{2}
{\ell}^{2}+6\,{r}^{2}+{r}^{2}{\ell}^{5}
\nonumber \\ && \qquad 
-5\,{r}^{2}\ell-28\,rM+12\,r\ell M
+4\,{\ell}^{3}rM+12\,r{\ell}^{2}M+36\,{M}^{2})  
\left( \ell-1 \right)  \left( \ell+2 \right)
/ [\left( r \ell^{2}+r\ell-2\,r+6\,M \right)^{2} {r}^{5}]
\nonumber \\ && \qquad \times  
\left( r\, \partial_t h_{1\,\ell-1 m}^{(1)}  \left( t,r \right) 
+2\,h_{0\,\ell-1 m}^{(1)} \left( t,r \right) 
-r \,\partial_r h_{0\,\ell-1 m}^{(1)}  \left( t,r \right) \right)  
\nonumber \\ && \qquad 
+32\,{\frac {\sqrt {2}\pi \, \left( \ell+2 \right) 
 \left( r{\ell}^{2}+r\ell-2\,r+3\,M \right)  \left( \ell-
1 \right) ^{2} \left( r-2\,M \right) }{r\sqrt { \left( \ell-1 \right) \ell}
 \left( r{\ell}^{2}+r\ell-2\,r+6\,M \right) ^{2}}}
{\cal Q}_{0\,\ell-1 m}^{(1)} \left( t,r \right) 
 \biggr) 
\nonumber \\ && \qquad 
+\frac{4\,S}{{\ell} \left( \ell+1 \right)  
\left( \ell-1 \right)  \left( \ell+2 \right) }
\sqrt {{\frac { \left( \ell+m+1 \right) 
\left( \ell-m+1 \right) }{ \left( 2\,\ell+1 \right)  \left( 2\,\ell+3 \right) }}}
 \biggl( 
-6\,\left( r-2\,M \right) ( 2\,{r}^{2}{\ell}^{2}+32\,rM-36
\,{M}^{2}
\nonumber \\ && \qquad 
+3\,{\ell}^{4}{r}^{2}+4\,{\ell}^{3}rM+{r}^{2}{\ell}^{5}
+2\,{r}^{2}{\ell}^{3}-8\,{r}^{2} )  
\left( \ell-1 \right)  \left( \ell+2 \right) 
/ [ \left( r{\ell}^{2}+r\ell-2\,r+6\,M \right) ^{2}{r}^{5}]
\nonumber \\ && \qquad \times  
 \left( r\,\partial_t h_{1\,\ell+1 m}^{(1)}  \left( t,r \right) r
+2\,h_{0\,\ell+1 m}^{(1)}  \left( t,r \right) 
- r\,\partial_r h_{0\,\ell+1 m}^{(1)}   \left( t,r \right) \right) 
\nonumber \\ && \qquad 
-32\,{\frac {\sqrt {2} \pi \, \left( \ell-1 \right)  
\left( r{\ell}^{2}+r\ell-2\,r+3\,M \right)  \left( \ell+2 \right) ^{
2} \left( r-2\,M \right) }{r\sqrt { \left( \ell+1 \right)  \left( \ell+2
 \right) } \left( r{\ell}^{2}+r\ell-2\,r+6\,M \right) ^{2}}} 
{\cal Q}_{0\,\ell+1 m}^{(1)} \left( t,r \right) 
\biggr) 
\,.
\end{eqnarray}
\end{widetext}
$S_{\ell m}^{\rm (even,2)}(E,S)$ and $S_{\ell m}^{\rm (even,2)}(O,S)$
mean the coupling between the black hole's spin and 
the first-order even and odd parity perturbations, respectively. 
The tensor harmonics coefficients 
of the first-order metric perturbation, $H_{1\ell m}^{(1)}$ etc. 
are written in terms of the first-order Regge-Wheeler and Zerilli functions. 

Here, we introduce the following combined function. 
\begin{eqnarray}
\Psi_{\ell m} \left( t,r \right) = \Psi_{\ell m}^{(1)} \left( t,r \right)
+ \Psi_{\ell m}^{(2)} \left( t,r \right) \,,
\label{eq:comboPsi}
\end{eqnarray}
which is the linear combination of the first- and second-order wave functions. 
This function formally satisfies
\begin{widetext}
\begin{eqnarray}
&& 
-{\frac {\partial ^{2}}{\partial {t}^{2}}}\Psi_{\ell m} \left( t,r \right) 
+{\frac {\partial ^{2}}{\partial {r^*}^{2}}}
\Psi_{\ell m} \left( t,r \right) -V_{\ell}^{\rm (even)}(r) 
\Psi_{\ell m} \left( t,r \right) 
\nonumber \\ && \qquad 
+ i \,{S}\,m\,P_{\ell}^{\rm (even,1)}(r) 
{\frac {\partial }{\partial t}}\Psi_{\ell m} \left( t,r \right)
+ i \,{S}\,m\,P_{\ell}^{\rm (even,2)}(r) 
{\frac {\partial^2 }{\partial t \partial r}}\Psi_{\ell m} \left( t,r \right)
\nonumber \\ 
&& = {S}\,\sqrt {{\frac { \left( \ell-m \right)  \left( \ell+m \right) }
{ \left( 2\,\ell-1 \right)  \left( 2\,\ell+1 \right) }}} Q_{\ell}^{\rm (even,-)}(r) 
\Psi_{\ell-1\, m}^{\rm (o,1)} \left( t,r \right) 
\nonumber \\ && \qquad 
+ {S}\,\sqrt {{\frac { \left( \ell+m+1 \right)  \left( \ell-m+1 \right) }
{ \left( 2\,\ell+1 \right)  \left( 2\,\ell+3 \right) }}} Q_{\ell}^{\rm (even,+)}(r) 
\Psi_{\ell+1\, m}^{\rm (o,1)} \left( t,r \right)  
+ S_{\ell m}^{\rm (even,L)}(t,r)
\,,
\label{eq:ComboFWE}
\end{eqnarray}
\end{widetext}
where $S_{\ell m}^{\rm (even,L)}(t,r)$ 
denotes the local source term with the Dirac's delta function and its derivative. 
The explicit expression and some detailed analysis are given in 
Appendix~\ref{app:evendetail}.

\subsubsection{Odd parity perturbation with spin}

In the first perturbative order calculation, 
we have used the Cunningham et al. waveform $\Psi_{\ell m}^{\rm (o,1)}$ 
for the odd parity as the Regge-Wheeler function. 
When we use $\Psi_{\ell m}^{\rm (o,2)}$, we have some trouble 
in the source terms of the perturbed Regge-Wheeler (odd parity) equation. 
The second-order local source term does not vanish at the horizon. 
Therefore, we use the Zerilli waveform 
$\Psi_{\ell m}^{\rm (o,Z,2)}$ 
instead of the Cunningham et al. waveform $\Psi_{\ell m}^{\rm (o,2)}$ 
in the second perturbative order
\begin{eqnarray}
&&
-{\frac {\partial ^{2}}{\partial {t}^{2}}}\Psi_{\ell m}^{\rm (o,Z,2)} \left( t,r \right) 
+{\frac {\partial ^{2}}{\partial {r^*}^{2}}}
\Psi_{\ell m}^{\rm (o,Z,2)} \left( t,r \right) 
\nonumber \\ && \quad 
-V_{\ell}^{\rm (odd)}(r) 
\Psi_{\ell m}^{\rm (o,Z,2)} \left( t,r \right) 
= S_{\ell m}^{\rm (odd,Z,2)} \left( t,r \right)
\,,
\label{eq:WaveEoZ}
\end{eqnarray}
where the second-order source term $S_{\ell m}^{\rm (odd,Z,2)}$ 
is formally given as 
\begin{eqnarray}
&&
S_{\ell m}^{\rm (odd,Z,2)} \left( t,r \right)  
= 
{\frac {8\,\sqrt {2}\,\pi  \,i \left( r-2\,M \right) ^{2} }
{{r}^{2}\sqrt {\ell \left( \ell+1 \right) }}}
{\cal Q}_{\ell m}^{(2)} \left( t,r \right) 
\nonumber \\ && \quad
-{\frac {16\,\sqrt {2}\pi \,i\,M \left( r-2\,M \right)  }
{{r}^{2}\sqrt {\ell \left( \ell+1 \right)  \left( \ell-1 \right) 
 \left( \ell+2 \right) }}}{\cal D}_{\ell m}^{(2)} \left( t,r \right)
\nonumber \\ && \quad
-{\frac {8\,\sqrt {2}\,\pi \,i  
\left( r-2\,M \right) ^{2}}{r\sqrt {\ell \left( \ell+1 \right)  
\left( \ell-1 \right)  \left( \ell+2 \right) }}}
{\frac {\partial }{\partial r}}{\cal D}_{\ell m}^{(2)} \left( t,r \right) 
\,.
\end{eqnarray}
${\cal Q}_{\ell m}^{(2)}$ and ${\cal D}_{\ell m}^{(2)}$ 
are calculated by the tensor harmonics expansion 
of $-G_{\mu\nu}^{(2)}[h^{\rm (1,wave)},h^{\rm (1,spin)}]/(8\pi)$
from the second-order Einstein tensor. 
And using ${\cal Q}_{\ell m}^{(2)}$, we have the relation 
between the two waveforms $\Psi_{\ell m}^{\rm (o,2)}$ and $\Psi_{\ell m}^{\rm (o,Z,2)}$ as
\begin{eqnarray}
&& \partial_t \Psi_{\ell m}^{\rm (o,2)} (t,r) 
= 2\,\Psi_{\ell m}^{\rm (o,Z,2)}(t,r) 
\nonumber \\ && \quad 
+ 
{\frac {16\,\sqrt {2}\,\pi  \,i \,r \left( r-2\,M \right)  }{
(\ell-1)(\ell+2) \sqrt {\ell \left( \ell+1 \right) }}}
{\cal Q}_{\ell m}^{(2)} \left( t,r \right) 
\,.
\label{eq:CtoZforOdd}
\end{eqnarray}

For the wave equation of $\Psi_{\ell m}^{\rm (o,Z,2)}$, 
we have the second-order source term as 
\begin{widetext}
\begin{eqnarray}
&& S_{\ell m}^{\rm (odd,Z,2)} \left( t,r \right)  
= S_{\ell m}^{\rm (odd,Z,2)}(E,S)
+ S_{\ell m}^{\rm (odd,Z,2)}(O,S) \,;
\nonumber \\ 
&& S_{\ell m}^{\rm (odd,Z,2)}(E,S)
= 
\frac{4\,S}{ {\ell}  \left( \ell+1 \right)  \left( \ell-1 \right) 
 \left( \ell+2 \right) }
 \sqrt {{\frac { \left( \ell-m \right)  \left( \ell+m \right) }{
 \left( 2\,\ell-1 \right)  \left( 2\,\ell+1 \right) }}} 
\biggl( 3\,{\frac {
 \left( r-2\,M \right)  \left( \ell-1 \right)  \left( \ell+2 \right) 
 \left( \ell+1 \right)  }{{r}^{4}}}
\partial_t  K_{\ell-1 m}^{(1)}  \left( t,r \right)
\nonumber \\ && \qquad 
+12\,{\frac { \sqrt {2}\, \pi \,\left( r-2\,M \right)  \left( \ell-1
 \right)  \left( \ell+2 \right)  \left( \ell+1 \right) }{{r}^{2}
\sqrt { \left( \ell-1 \right) \ell \left( \ell-2 \right)  \left( \ell+1 \right) }}}
 \partial_t {\cal F}_{\ell-1 m}^{(1)} \left( t,r \right)  
 \biggr)
\nonumber \\ && 
+\frac{4\,S}{{\ell} \left( \ell
+1 \right)  \left( \ell-1 \right)  \left( \ell+2 \right) }
\sqrt {{\frac { \left( \ell+m+1
 \right)  \left( \ell-m+1 \right) }{ \left( 2\,\ell+1 \right)  \left( 2\,\ell+3
 \right) }}}
 \biggl( -3\,{\frac { \left( r-2\,M \right) \ell \left( \ell-1
 \right)  \left( \ell+2 \right) }{{r}^{4}}}
\partial_t K_{\ell+1 m}^{(1)}  \left( t,r \right) 
\nonumber \\ && \qquad 
-12\,{\frac {\sqrt {2}\, \pi \, \left( r-2\,M \right) \ell \left( \ell-1
 \right)  \left( \ell+2 \right) }{{r}^{2}\sqrt { \left( \ell+1 \right) 
 \left( \ell+2 \right) \ell \left( \ell+3 \right) }}}
\partial_t {\cal F}_{\ell+1 m}^{(1)} \left( t,r \right)  \biggr)
\,, 
\nonumber \\ 
&&S_{\ell m}^{\rm (odd,Z,2)}(O,S)
= 
\frac{S\,m}{\ell (\ell+1)} 
\biggl( {\frac {-48\,\sqrt {2}\,i\,\pi \, \left( r-2\,M
 \right) }{\sqrt {\ell \left( \ell+1 \right) }{r}^{3}}}
 {\cal Q}_{0\ell m}^{(1)} \left( t,r \right)
 -{\frac {12\,i \left( 6\,r+r\ell+r{\ell}^{2}-14\,M \right)  \left( 
r-2\,M \right) }{{r}^{7}}}
h_{0\ell m}^{(1)} \left( t,r \right) 
\nonumber \\ && \qquad 
+{\frac {12\,i \left( r-2\,M \right)  \left( 3\,r-7\,M \right) }{{r}^{6}}}
\partial_r h_{0\ell m}^{(1)}  \left( t,r \right)  
+{\frac {4\,i \left( -9\,r+r\ell+r{\ell}^{2}+21\,M \right) 
 \left( r-2\,M \right) }{{r}^{6}}}
\partial_t h_{1\ell m}^{(1)}  \left( t,r \right) 
 \biggr) 
\,,
\label{eq:2ndSoZ}
\end{eqnarray}
\end{widetext}
without any regularization (or modification) of the wave function.  
Here, we note that $S_{\ell m}^{\rm (odd,Z,2)}(E,S)$ 
for the $\ell=3$, $m=2$ mode is the time derivative of the second-order term 
in Eq.~(76) of~\cite{Gleiser:2001in}. 
The explicit expression of Eq.~(\ref{eq:WaveEoZ}) is given in 
Appendix~\ref{app:odddetail}. 
We should note that for the $\ell=2$ mode, 
there is an ill-defined term due to the factor $(\ell-2)$ 
in the denominator. 
This is why we need a special treatment for 
the $\ell=1$ mode in the next section.

\subsubsection{For lower $\ell$ modes}

In the calculation of the second-order $\ell=2$ odd parity perturbation, 
we have the first-order $\ell=1$ mode contribution. 
In~\cite{Zerilli:1971wd}, this $\ell=1$ mode has been calculated 
under the Zerilli gauge, i.e., $K_{1m}=h_{01m}^{(e)}=h_{11m}^{(e)}=0$. 
\begin{eqnarray}
H_{0\,1m}^{(1)Z}(t,r) 
&=&  \frac{1}{3M(r-2M)^2}
\left(r^3\frac{d^2 }{dt^2}f_m(t) +M\,f_m(t) \right) 
\nonumber \\ && \times 
\theta(r-R(t)) \,,
\nonumber \\
H_{1\,1m}^{(1)Z}(t,r) 
&=&  - \frac{r}{(r-2M)^2}\frac{d}{dt}f_m(t) \,
\theta(r-R(t)) \,,
\nonumber \\
H_{2\,1m}^{(1)Z}(t,r) 
&=&  \frac{1}{(r-2M)^2} f_m(t) \,
\theta(r-R(t)) \,,
\end{eqnarray}
where 
\begin{eqnarray}
f_m(t) = 8\,\pi\,\mu\,U(t)\,\frac{(R(t)-2\,M)^2}{R(t)}
\,Y_{1 m}^*(\Theta_0,\Phi(t)) \,.
\end{eqnarray}
Here $*$ denotes the complex conjugate.
There is no contribution from the $m=0$ mode. 

Using the above first-order $\ell=1$ mode, 
we calculate the second-order source term from the coupling 
between this mode and the black hole's spin. 
Then the source term becomes finite at the horizon. 
In order to remove this finite term, we introduce a regularization function, 
\begin{eqnarray}
&& \Psi_{2 m}^{\rm (o,Z,2)} \left( t,r \right) 
= \Psi_{2 m}^{\rm (o,Z,2),R} \left( t,r \right) 
\nonumber \\ && 
- \frac{S\sqrt { 15 \left( 2-m \right)  \left( 2+m \right) }}
{30\,M\,r\left( r-2\,M \right) } \dot f_m  \left( t \right)
{\theta} \left( r-R \left( t \right)  \right) \,,
\end{eqnarray}
and we solve the wave equation for the regularized function 
$\Psi_{2 m}^{\rm (o,Z,2),R}$. 
Here, we note that the regularization function does not affect the waveform 
at infinity in our calculation. 
The regularized second-order source term is derived as 
\begin{widetext}
\begin{eqnarray}
&&
S_{2 m}^{\rm (odd,Z,2),R}(E,S,[\ell=1])
= 
{\frac {4\sqrt {15}\,\pi \,\mu\,S\,\sqrt {\left( 2-m \right) 
 \left( 2+m \right) }}{15}}
\, Y_{1 m}^*(\Theta_0,\Phi(t))
\biggl[
\biggl(
{\frac {im \left( {R} \left( t \right) -2\,M \right) ^{2}U \left( t \right)  
\left( \dot {\Phi} \left( t \right)  \right) ^{3}}
{{R} \left( t \right) M}}
\nonumber \\ && \qquad 
+{\frac {\dot {R} \left( t \right) U \left( t \right)  
\left( {R} \left( t \right) -2\,M \right)  
\left( \dot {\Phi} \left( t \right)  \right) ^{2}}
{{R} \left( t \right) M}}
-{\frac {im \left( {R} \left( t \right) 
-2\,M \right) ^{2}\dot {\Phi}  \left( t \right) }
{ \left( {R} \left( t \right)  \right) ^{3}MU \left( t \right) }}
+{\frac {\dot {R} \left( t \right)  
\left( {R} \left( t \right) -2\,M \right) }
{ \left( {R} \left( t \right)  \right) ^{3}MU \left( t \right) }}
\biggr) \, \frac{d}{dr} \,\delta(r-R(t))
\nonumber \\ && \qquad 
+ \biggl(
{\frac {im \left( {R} \left( t \right) -2\,M \right)  
\left( 13\,M-3\,{R} \left( t \right)  \right)  
\left( \dot {\Phi}  \left( t \right)  \right) ^{3}U \left( t \right) }
{ \left( {R} \left( t \right)  \right) ^{2}M}}
-{\frac {2\,im \left( 5\,M-2\,{R} \left( t \right)  \right)  
\left( {R} \left( t \right) -2\,M \right) ^{2}\dot {\Phi}  \left( t \right) 
U \left( t \right) }{ \left( {R} \left( t \right)  \right) ^{5}M}}
\nonumber \\ && \qquad 
-4\,{\frac { \left( {R} \left( t \right) -2\,M \right) ^{2}
\dot {R}  \left( t \right) U \left( t \right) }
{ \left( {R} \left( t \right)  \right) ^{5}M}}
+{\frac { \left( -12\,M+2\,M{m}^{2}-{m}^{2}{R} \left( t \right) 
+4\,{R} \left( t \right)  \right)  
\left( \dot {\Phi} \left( t \right)  \right) ^{2}
\dot {R} \left( t \right) U \left( t \right) }
{ \left( {R} \left( t \right)  \right) ^{2}M}}
\nonumber \\ && \qquad 
+{\frac {im \left( 11\,M-2\,{R} \left( t \right)  \right)  
\left( {R} \left( t \right) -2\,M \right) 
\dot {\Phi}  \left( t \right) }
{ \left( {R} \left( t \right)  \right) ^{4}MU \left( t \right) }}
-2\,{\frac { \left( 4\,M-{R} \left( t \right)  \right) 
\dot {R} \left( t \right) }
{ \left( {R} \left( t \right)  \right) ^{4}MU \left( t \right) }}
\biggr) \,\delta(r-R(t))
\biggr]
 \,. 
\end{eqnarray}
\end{widetext}
We have only the local source contributions 
as the second-order source term from this mode. 
Using the following asymptotic behavior near the horizon, 
$U ( t ) \sim (1-2M/R(t))^{-1}$, 
$\dot{R}(t) \sim (1-2M/R(t))$, and  
$\dot{\Phi}(t) \sim (1-2M/R(t))$, 
we find that the above source term vanishes at the horizon
in the integration of the wave equation.

\subsubsection{Symmetry in $\Psi_{\ell m}$ and $\Psi_{\ell m}^{\rm (o)}$} 

In this subsection, 
we use the notation $\Psi_{\ell m}^{\rm (even)}=\Psi_{\ell m}$ 
and $\Psi_{\ell m}^{\rm (odd)}=\Psi_{\ell m}^{\rm (o)}$,  
which have the following relation 
in the first perturbative order: 
\begin{eqnarray}
\Psi_{\ell -m}^{\rm (even/odd)} 
&=& (-1)^m \left(\Psi_{\ell m}^{\rm (even/odd)}\right)^*
\,.
\end{eqnarray}
This is derived from a formula for the spherical harmonics, 
\begin{eqnarray}
Y_{\ell -m}(\theta,\phi) = (-1)^m Y_{\ell m}^*(\theta,\phi) \,.
\end{eqnarray}

In the $O(a^1)$ calculation, we should have the same symmetry 
because the metric perturbations become real. 
We can check this by using the explicit form of $S_{\ell m}^{\rm (even/odd)}$.

\subsubsection{Gravitational waves}

In the above sections, we discussed the techniques to calculate
 the wave functions 
$\Psi_{\ell m}  = \Psi_{\ell m}^{(1)} + \Psi_{\ell m}^{(2)}$, 
$\Psi_{\ell m}^{\rm (o,1)}$ and $\Psi_{\ell m}^{\rm (o,Z,2)}$. 
The first-order wave functions and waveforms at infinity 
are simply related as 
\begin{eqnarray}
&& h_{+} - i\,h_{\times} 
=
\sum \frac{\sqrt{(\ell-1)\ell(\ell+1)(\ell+2)}}{2r} 
\nonumber \\ && \quad \times 
\left(
\Psi_{\ell m}^{(1)} - i\, \Psi_{\ell m}^{\rm (o,1)}
\right)
{}_{-2}Y_{\ell m}
\,,
\end{eqnarray}
where ${}_{-2}Y_{\ell m}$ denotes the spin-weighted spherical harmonics 
used in~\cite{Brown:2007jx}.  

On the other hand, in order to discuss gravitational waveforms 
in the second perturbative order, 
we need to check the asymptotic behavior of the metric perturbation 
and the contributions from the first-order gauge transformation. 
First, we evaluate the asymptotic behavior of the tensor harmonics 
coefficients of $G_{\mu\nu}^{(2)}$, because 
this information is used to construct the metric perturbation from the wave functions.  
For the odd parity-spin coupling part, we have the following behavior.
\begin{eqnarray}
{\cal A}_{0\, \ell m}^{(2)}(O,S) &\sim& O(1/r^3) \,, 
\,
{\cal A}_{1\, \ell m}^{(2)}(O,S) \sim O(1/r^3) \,, 
\nonumber \\ 
{\cal A}_{ \ell m}^{(2)}(O,S) &\sim& O(1/r^3) \,,
\,
{\cal B}_{0\, \ell m}^{(2)}(O,S) \sim O(1/r^3) \,, 
\nonumber \\ 
{\cal B}_{\ell m}^{(2)}(O,S) &\sim& O(1/r^3) \,, 
\,
{\cal G}_{\ell m}^{(2)}(O,S) \sim O(1/r^3) \,,
\nonumber \\ 
{\cal F}_{\ell m}^{(2)}(O,S) &\sim& O(1/r^4) \,, 
\,
{\cal Q}_{\ell m}^{(2)}(O,S) \sim O(1/r^3) \,, 
\nonumber \\ 
{\cal Q}_{1\,\ell m}^{(2)}(O,S) &\sim& O(1/r^3) \,,
\,
{\cal D}_{\ell m}^{(2)}(O,S) \sim O(1/r^4) \,,
\end{eqnarray}
and for the even parity-spin coupling part, 
\begin{eqnarray}
{\cal A}_{0\, \ell m}^{(2)}(E,S) &\sim& O(1/r^3) \,, 
\,
{\cal A}_{1\, \ell m}^{(2)}(E,S) \sim O(1/r^3) \,, 
\nonumber \\ 
{\cal A}_{ \ell m}^{(2)}(E,S) &\sim& O(1/r^3) \,,
\,
{\cal B}_{0\, \ell m}^{(2)}(E,S) \sim O(1/r^2) \,, 
\nonumber \\ 
{\cal B}_{\ell m}^{(2)}(E,S) &\sim& O(1/r^2) \,, 
\,
{\cal G}_{\ell m}^{(2)}(E,S) \sim O(1/r^4) \,,
\nonumber \\ 
{\cal F}_{\ell m}^{(2)}(E,S) &\sim& O(1/r^3) \,, 
\,
{\cal Q}_{\ell m}^{(2)}(E,S) \sim O(1/r^2) \,, 
\nonumber \\ 
{\cal Q}_{1\,\ell m}^{(2)}(E,S) &\sim& O(1/r^2) \,,
\,
{\cal D}_{\ell m}^{(2)}(E,S) \sim O(1/r^3) \,.
\end{eqnarray}
And the even parity-spin coupling part from the $\ell=1$ even parity has 
a different behavior.
\begin{eqnarray}
{\cal Q}_{2 m}^{(2)}(E,S,[\ell=1]) &\sim& O(1/r^3) \,, 
\nonumber \\ 
{\cal Q}_{1\,2 m}^{(2)}(E,S,[\ell=1]) &\sim& O(1/r^2) \,,
\end{eqnarray}
and ${\cal D}_{2 m}^{(2)}(E,S,[\ell=1])=0$ in the first-order Zerilli gauge.

From the above asymptotic behaviors, if we set the observer location 
to a large distance, 
we do not need to consider these tensor harmonics contributions 
because the contributions are at least $O(1/r)$ lower than the leading part. 
Note that the metric reconstruction in 
the second-order odd parity perturbation has been done 
from the Zerilli waveform $\Psi_{\ell m}^{\rm (o,Z,2)}$. 

Next, we discuss the contributions from the first-order gauge transformation.  
Formally the following gauge transformation~\cite{Bruni:1996im} 
is used in the second-order calculation. 
\begin{widetext}
\begin{eqnarray}
x^{\mu}_{RW} &\to& x^{\mu}_{AF} 
= x^{\mu}_{RW} + \xi^{(1)\mu} \left(x^{\alpha}\right)
+\frac{1}{2} \left[\xi^{(2)\mu}\left(x^{\alpha}\right)
+\xi^{(1)\nu} \xi^{(1)\mu}{}_{,\nu}\left(x^{\alpha}\right)\right] \,,
\end{eqnarray}
\end{widetext}
where comma "," in the index indicates the partial derivative 
with respect to the background Schwarzschild coordinates, 
and $\xi^{(1)\mu}$ and $\xi^{(2)\mu}$ are generators of 
the first and second-order gauge transformations, respectively. 
The subscripts $RW$ and $AF$ 
show the Regge-Wheeler gauge 
where we reconstruct the metric perturbation, 
and the asymptotic flat gauge where we obtain the gravitational waveforms, 
respectively. 
Then the metric perturbations change to
\begin{widetext} 
\begin{eqnarray}
h_{RW \mu \nu}^{(1)} \to h_{AF \mu \nu}^{(1)} &=&
h_{RW \mu \nu}^{(1)} - {\cal L}_{\xi^{(1)}} g_{\mu \nu} \,,
\label{eq:genGT1} \\
h_{RW \mu \nu}^{(2)} \to h_{AF \mu \nu}^{(2)} &=&
h_{RW \mu \nu}^{(2)} 
-\frac{1}{2} {\cal L}_{\xi^{(2)}} g_{\mu \nu}
+\frac{1}{2} {\cal L}_{\xi^{(1)}}^2 g_{\mu \nu}
-{\cal L}_{\xi^{(1)}} h_{RW \mu \nu}^{(1)} \,,
\label{eq:genGT2}
\end{eqnarray}
\end{widetext}
where ${\cal L}_{\xi^{(i)}}$ denotes the Lie derivative.

In this paper, second perturbative order means $O(\mu a)$ 
where $\mu$ and $a$ are small quantities. 
Since $\xi^{(1)}$ is $O(\mu)$, we ignore 
${\cal L}_{\xi^{(1)}}^2 g_{\mu \nu}$ 
and ${\cal L}_{\xi^{(1)}} h_{RW \mu \nu}^{(1)}$ 
with $h_{RW \mu \nu}^{(1)} \sim O(\mu)$ in Eq.~(\ref{eq:genGT2}). 
On the other hand, there is a contribution from 
${\cal L}_{\xi^{(1)}} h_{\mu \nu}^{\rm (1,spin)}$. 
The asymptotic behavior of this tensor harmonics coefficient 
becomes 
\begin{eqnarray}
&& 
\delta H_{0\ell m} \sim O(1/r) \,, \quad
\delta H_{1\ell m} \sim O(1/r) \,, 
\nonumber \\ && 
\delta H_{2\ell m} = 0 \,, \quad 
\delta h^{(e)}_{0\ell m} \sim O(r^0) \,, 
\nonumber \\ &&
\delta h^{(e)}_{1\ell m} \sim O(r^0) \,, \quad
\delta G_{\ell m} \sim O(1/r^2) \,, 
\nonumber \\ &&
\delta K_{\ell m} \sim O(1/r^2) \,, \quad
\delta h_{0\ell m} \sim O(r^0) \,, 
\nonumber \\ &&
\delta h_{1\ell m} \sim O(r^0) \,, \quad 
\delta h_{2\ell m} \sim O(r^0) 
\,.
\end{eqnarray}
For the $\ell=1$ mode in the first perturbative order
when we consider the gauge transformation to 
the center of mass coordinates, we have the same behaviors. 
These contributions to the second-order metric perturbation 
under the Regge-Wheeler gauge are also lower order 
by $O(1/r)$ at least. 

Finally, to derive the waveforms in the SRWZ formalism, 
we may consider 
\begin{eqnarray}
&& h_{+} - i\,h_{\times} 
=
\sum \frac{\sqrt{(\ell-1)\ell(\ell+1)(\ell+2)}}{2r} 
\nonumber \\ && \times 
\left(
\Psi_{\ell m} - i\, \Psi_{\ell m}^{\rm (o)}
\right)
{}_{-2}Y_{\ell m}
\,,
\label{eq:RWZwaves}
\end{eqnarray}
where again $\Psi_{\ell m}  = \Psi_{\ell m}^{(1)} + \Psi_{\ell m}^{(2)}$. 
Note that for $\Psi_{\ell m}^{\rm (o)}$ we have used 
a different wave functions for the first and second 
order odd parity calculations for the sake of simplicity of the final results. 
Using Eq.~(\ref{eq:CtoZforOdd}) 
and the above asymptotic behaviors of ${\cal Q}_{\ell m}^{(2)}$, 
we simply combine them as 
\begin{eqnarray}
\Psi_{\ell m}^{\rm (o)} 
&=& \Psi_{\ell m}^{\rm (o,1)} + 2\,\int dt \, \Psi_{\ell m}^{\rm (o,Z,2)}
\,.
\end{eqnarray}

\subsubsection{Observer location effect}\label{sss:OLE}

In~\cite{Lousto:2010tb}, we saw that the observer location effect 
was not negligible on the waveforms. 
To compare the NR and perturbative waveforms, 
we directly use Eq.~(\ref{eq:RWZwaves}) because 
we can set the extraction radius of gravitational waves 
at a sufficiently distant location, for example, $R_{\rm Obs}/M=1000$. 
On the other hand, the NR waveforms are 
obtained from the NR $\psi_4$ data
\begin{eqnarray}
\psi_4 &=& \ddot{h}_{+} - i\,\ddot{h}_{\times} 
\,,
\end{eqnarray}
We should note that these are true only at $R_{\rm Obs} \to \infty$. 

First, we discuss the asymptotic behavior of the (first-order) 
Regge-Wheeler-Zerilli functions. 
In general $\ell$ modes for both the even 
and odd parities,
which we denote by 
$\Psi_{\ell m}^{\rm (even)}$ and $\Psi_{\ell m}^{\rm (odd)}$,
are given by
\begin{eqnarray}
\Psi_{\ell m}^{\rm (even/odd)} &=& H_{\ell m} (t-r^*) 
+ \frac{\ell(\ell+1)}{2\,r} \int dt H_{\ell m} (t-r^*) 
\nonumber \\ && 
+ O(r^{-2}). 
\label{eq:asymtRWZ}
\end{eqnarray}
We note that errors due to
finite extraction radii, which arise from the integral term in
Eq.~(\ref{eq:asymtRWZ}), are larger for lower frequencies due to the
$1/\omega$ factor
obtained by integrating a function with frequency $\omega$.

Next, we discuss the relation between Regge-Wheeler-Zerilli functions and 
the mode function $\psi_4^{\ell m}$ of the Weyl scalar.
Here, although we can use the formula given in Eqs.~(C.1) and (C.2) 
of \cite{Lousto:2005xu}, we use simpler formulae
valid for the asymptotic behavior of the functions. 
If the NR Weyl scalar satisfies the Teukolsky equation in the Schwarzschild spacetime, 
we have 
\begin{eqnarray}
r\,\psi_4^{\ell m} &=& \ddot {\tilde H}_{\ell m} (t-r^*) 
+ \frac{(\ell -1)(\ell +2)}{2\,r} \dot {\tilde H}_{\ell m} (t-r^*)
\nonumber \\ && 
+ O(r^{-2}) \,,
\label{eq:asymtPsi4}
\end{eqnarray}
where the difference between ${\tilde H}_{\ell m}$ 
and $H_{\ell m}$ in Eq.~(\ref{eq:asymtRWZ}) is only the numerical factor. 

Combining the above equation with Eq.~(\ref{eq:asymtRWZ}), we have
\begin{eqnarray}
r\,\psi_4^{\ell m} &\sim& \ddot \Psi_{\ell m}^{\rm (even/odd)} 
- \frac{1}{r} \int dt \, \ddot \Psi_{\ell m}^{\rm (even/odd)} 
\nonumber \\ && 
+ O(r^{-2}) \,,
\label{eq:Psi4RWZ}
\end{eqnarray}
which is independent of $\ell$ and parity modes. 
This equation is consistent with the formula in~\cite{Lousto:2005xu}. 
Here, we have considered the correction for the RWZ functions. 
It is important, however, to calculate $H_{\ell m}$, the waveform at
infinity,
because the PN waveforms which are used to construct the hybrid waveform, 
do not have the finite observer location effects. 

Therefore, we consider 
the extrapolation of the NR $\psi_4$ from for example, $R_{\rm Obs}/M=100$ 
to infinity by using Eq.~(\ref{eq:asymtPsi4}):
\begin{eqnarray}
\ddot {\tilde H}_{\ell m} &=& 
R_{\rm Obs}\,\psi_4^{\ell m} 
- \frac{(\ell -1)(\ell +2)}{2} \int dt \, \psi_4^{\ell m}
\nonumber \\ && 
+ O(R_{\rm Obs}^{-2}) \,.
\label{eq:asymtPsi4ext}
\end{eqnarray}
Again, the above formula is derived by assuming 
the Teukolsky equation in the Schwarzschild spacetime ($a=0$). 
Since we treat only the extrapolation from $R_{\rm Obs}/M=100$ to infinity, 
we may use the wave (linear propagation) equation in the flat spacetime.
Thus, the Teukolsky equation with $M \to 0$ is sufficient 
to discuss the extrapolation. 
This calculation gives the same result as Eq.~(\ref{eq:asymtPsi4ext}).
Note that since the above formulation has been discussed 
by using the Weyl scalar in the Kinnersley tetrad, 
we need an extra factor as the explanation below Eq.~(\ref{eq:asymtpsi4ext}) 
for that in another tetrad.

Let us point out that full numerical methods using Cauchy-characteristic 
methods have been developed \cite{Reisswig:2009us}. 
Also multipatch \cite{Pollney:2009yz} and pseudospectral \cite{Szilagyi:2009qz}
techniques allow extraction radii very far from the source.

\subsubsection{Numerical integration method}

Although we have used the combination of Eq.~(\ref{eq:comboPsi}) 
for the even parity perturbation and integrate 
Eq.~(\ref{eq:ComboFWE}) in this paper, 
the basic equations are the four wave equations, (\ref{eq:1stEven}) 
and (\ref{eq:1stOdd}) for the first perturbative order, 
and (\ref{eq:WaveE2ndO}) and (\ref{eq:WaveEoZ}) for the second perturbative order. 

In order to integrate the resulting even and odd parity wave equations, 
we use the method described in \cite{Lousto:1997wf}. 
This method is second-order accurate in the grid spacing
(see \cite{Lousto:2005ip} for a fourth-order formalism), but
deals with the Dirac's delta source ``exactly'' or as accurately as needed.

Even if we considered the metric (\ref{eq:Sch+spin})
with first-order spin corrections
to the Schwarzschild metric, the method of perturbations we used still
propagates waves on the exact Schwarzschild background and lumps the spin
corrections in a source term, as if they would be second-order perturbations.
We hence apply the methods of \cite{Lousto:1997wf,Lousto:2005ip} with an added smooth source to 
integrate the first-order in spin corrected RWZ wave equations. We proved
second-order convergence of the extracted waveforms and used spatial and 
time steps that produced errors well below those acceptable for full numerical
evolutions. The runs typically take under a minute on a laptop and are very
low in memory and resources requirements. We also note that these types of
codes are amenable to  implementation on accelerated hardware such as
 GPUs or Cell processors \cite{Khanna:2010jv}.

\section{Analysis of the Numerical versus Perturbative results}\label{sec:NvsP}

Here we directly compare the waveforms generated fully numerically
with those computed by the perturbative (SRWZ) approach.
Since our full numerical evolutions
routinely extract the Weyl scalar $\psi_4$ at intermediate radii,
typically around $R=100M$ (a compromise between far enough of the
sources and high enough local resolution), and the perturbative code
evolves the Regge-Wheeler and Zerilli waveforms, we need to translate
these different measurements of the waveform into a common radiation
quantity. While analytic expressions already exists
that relate them both \cite{Lousto:2005xu}, 
such expressions involve second derivatives that lead
to some numerical noise when building up $\psi_4$, for instance. The usual
strain $h$ also involves two integration constants that are hard to fix
with accuracy \cite{Campanelli:2008nk,Reisswig:2010di}. 
Hence, as a compromise, we use the {\it news}
function, essentially $dh/dt$, which displays nicer smoothness properties
for numerical comparisons.

\begin{figure}
  \caption{The real part of the $(\ell=2,\,m=2)$ mode of $dh/dt$ for the $q=1/10$ case.
  The (black) solid, (red) dotted, and (blue) dashed curves show the NR, 
  spin-off, and spin-on calculations, respectively.} 
  \includegraphics[width=3.4in]{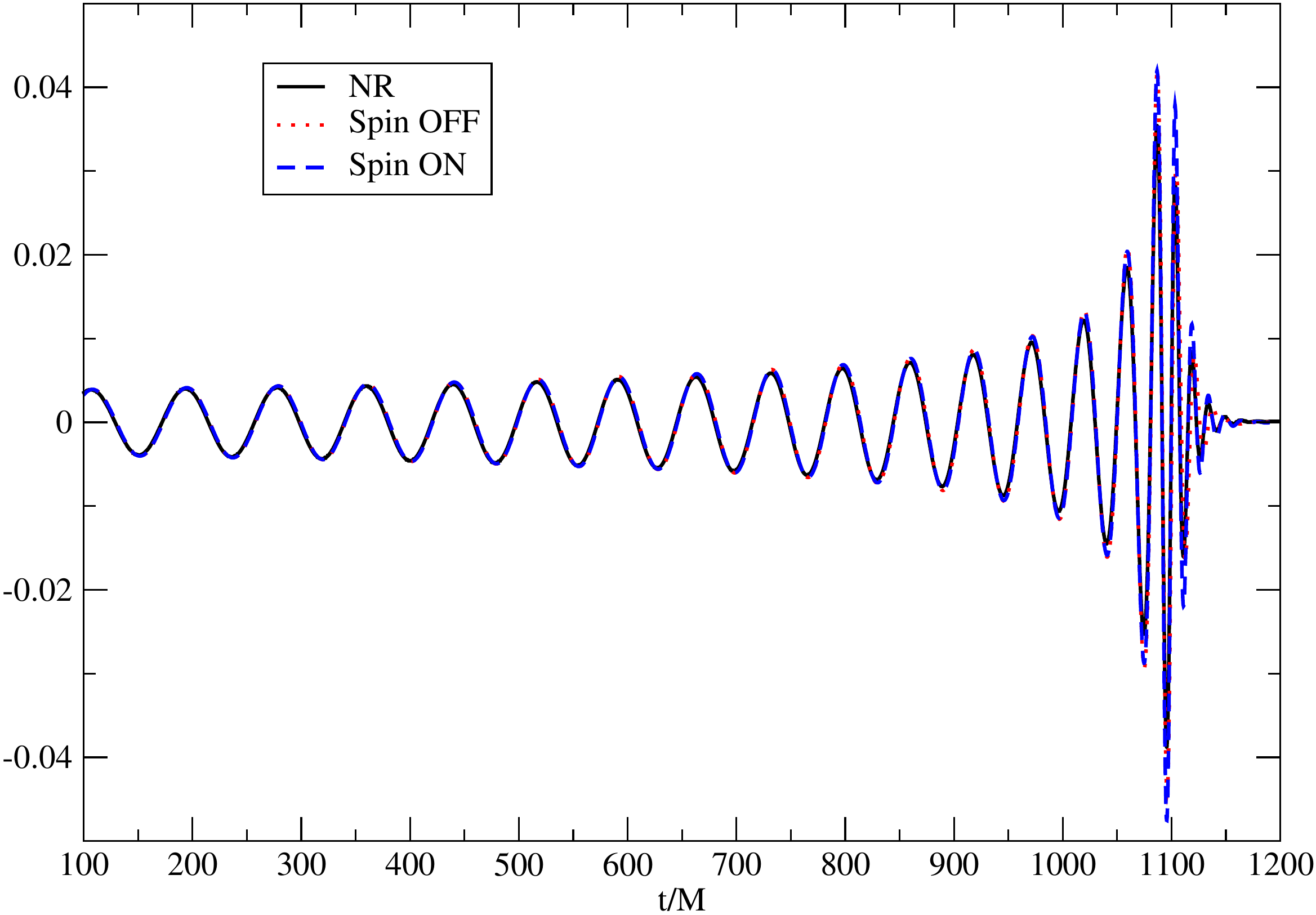}
  \label{fig:dh22_10to1}
\end{figure}

\begin{figure}
  \caption{The real part of the $(\ell=2,\,m=1)$ mode of $dh/dt$ for the $q=1/10$ case.
  The (black) solid, (red) dotted, and (blue) dashed curves show the NR, 
  spin-off, and spin-on calculations, respectively.} 
  \includegraphics[width=3.4in]{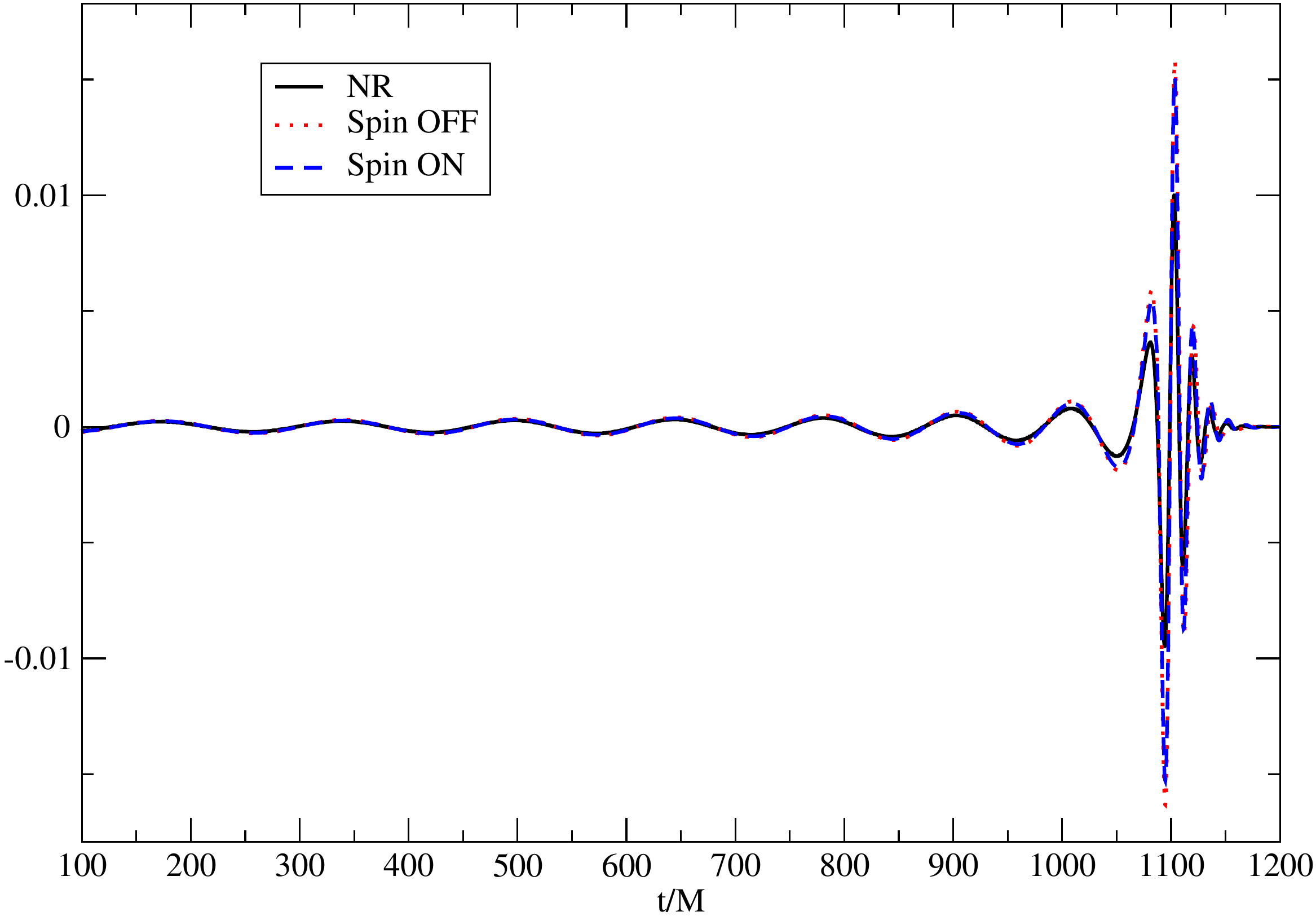}
  \label{fig:dh21_10to1}
\end{figure}

\begin{figure}
  \caption{The real part of the $(\ell=3,\,m=3)$ mode of $dh/dt$ for the $q=1/10$ case.
  The (black) solid, (red) dotted, and (blue) dashed curves show the NR, 
  spin-off, and spin-on calculations, respectively.} 
  \includegraphics[width=3.4in]{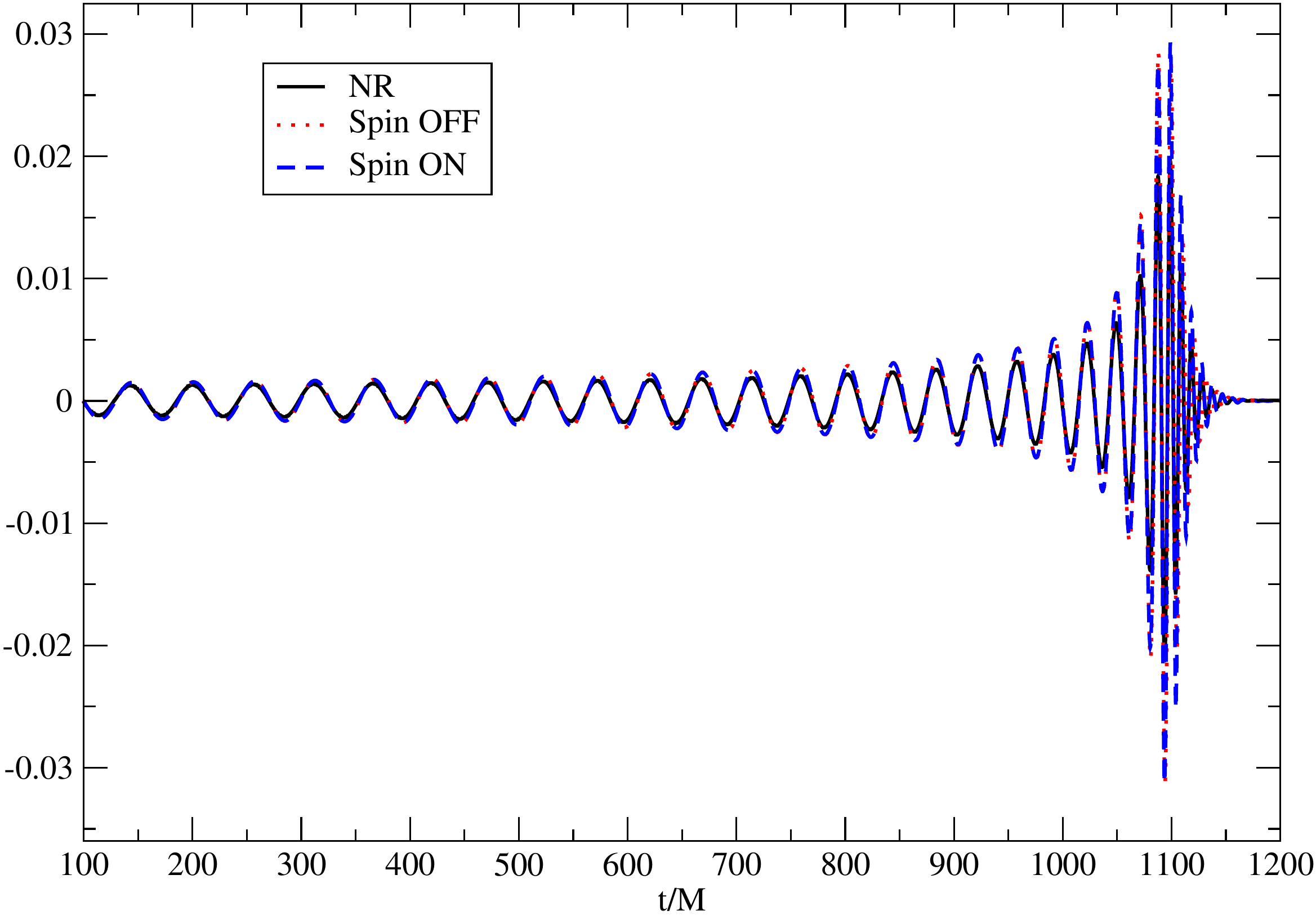}
  \label{fig:dh33_10to1}
\end{figure}

In Figs.\ \ref{fig:dh22_10to1}-\ref{fig:dh33_10to1} we superpose the
waveforms obtained for the full numerical evolution of the $q=1/10$
black-hole binary case and the perturbative waveforms as computed
by the integration of the wave equations 
(\ref{eq:WaveE2ndO}) and (\ref{eq:WaveEoZ}) both, including the spin 
corrections $(a/M=0.26)$ or simply setting it to zero. We do these
comparisons for the leading $(\ell,m)=(2,2)$ mode and the next to leading (2,1)
and (3,3) modes. Note that while (2,1) is an odd parity mode (for $a=0$)
and comes from integration of the Regge-Wheeler equation (\ref{eq:1stOdd}),
the other modes are even parity and hence obtained by integration
of the Zerilli equation (\ref{eq:1stEven}). In all cases we use the same
``full numerical'' trajectory. When spin terms are switched on, there
is a coupling of even and odd parity modes as shown in 
Eqs.~(\ref{eq:WaveE2ndO}) and (\ref{eq:WaveEoZ}).

We have computed the overlap functions, as defined in 
Ref.\ \cite{Campanelli:2008nk},
of these three sets of waveforms in order to quantify the phase agreement
between them. This provides some insight into the possibility of using
these perturbative waveforms to build up a bank of templates to support
detection and analysis of gravitational wave observatories such
as LIGO and VIRGO. Table \ref{tab:match10to1} 
shows that the agreement between numerical
and perturbative waveforms is very good in general for all three modes, 
and that including
the spin dependence improves the matching to an excellent level.
This improvement is based on the accurate description of the late
time phase, as we will discuss next, and is independent of the particle's
track. The orbital (inspiral) part of the waveforms
are not so strongly dependent on the spin terms (for our simulations)
and are correctly described by the nonspinning perturbations. It is
interesting to note here that the excellent phase agreement during the inspiral
orbit might not be so surprising since the perturbative code uses the
full numerical tracks (transformed into Schwarzschild coordinates);
however, coordinates and gauges 
in full numerical evolutions are described in quite a different way
than in (analytic) perturbative expressions and it is reassuring to find
such a good agreement in the final products of evolutions.

\begin{table*} 
\centering 
  \caption{The overlap (matching) between the NR and perturbative $dh/dt$ 
  for the $q=1/10$ case. The integration time is from $t/M=100$ to $1220$ and 
  the definition of the matching is given in Eqs.~(26) and (27) 
  of~\cite{Campanelli:2008nk}.}
\label{tab:match10to1}
\begin{ruledtabular}
\begin{tabular}{l|ccc}
   Mode & $\Re (\ell=2, m=2)$ & $\Re (\ell=2, m=1)$ & $\Re (\ell=3, m=3)$ \\
\hline
   Match (Spin OFF) & 0.980404 & 0.968137 & 0.927807 \\
\hline
   Match (Spin ON)  & 0.995055 & 0.982173 & 0.995347 \\
\hline
\hline
   Mode & $\Im (\ell=2, m=2)$ & $\Im (\ell=2, m=1)$ & $\Im (\ell=3, m=3)$ \\
\hline
   Match (Spin OFF) & 0.980379 & 0.972727 & 0.928151 \\
\hline
   Match (Spin ON)  & 0.995196 & 0.982604 & 0.995571 \\
  \end{tabular}
  \end{ruledtabular}
\end{table*}

\begin{figure}
  \caption{The real part of the $(\ell=2,\,m=2)$ mode of $dh/dt$ for the $q=1/15$ case.
  The (black) solid, (red) dotted, and (blue) dashed curves show the NR, 
  spin-off, and spin-on calculations, respectively.} 
  \includegraphics[width=3.4in]{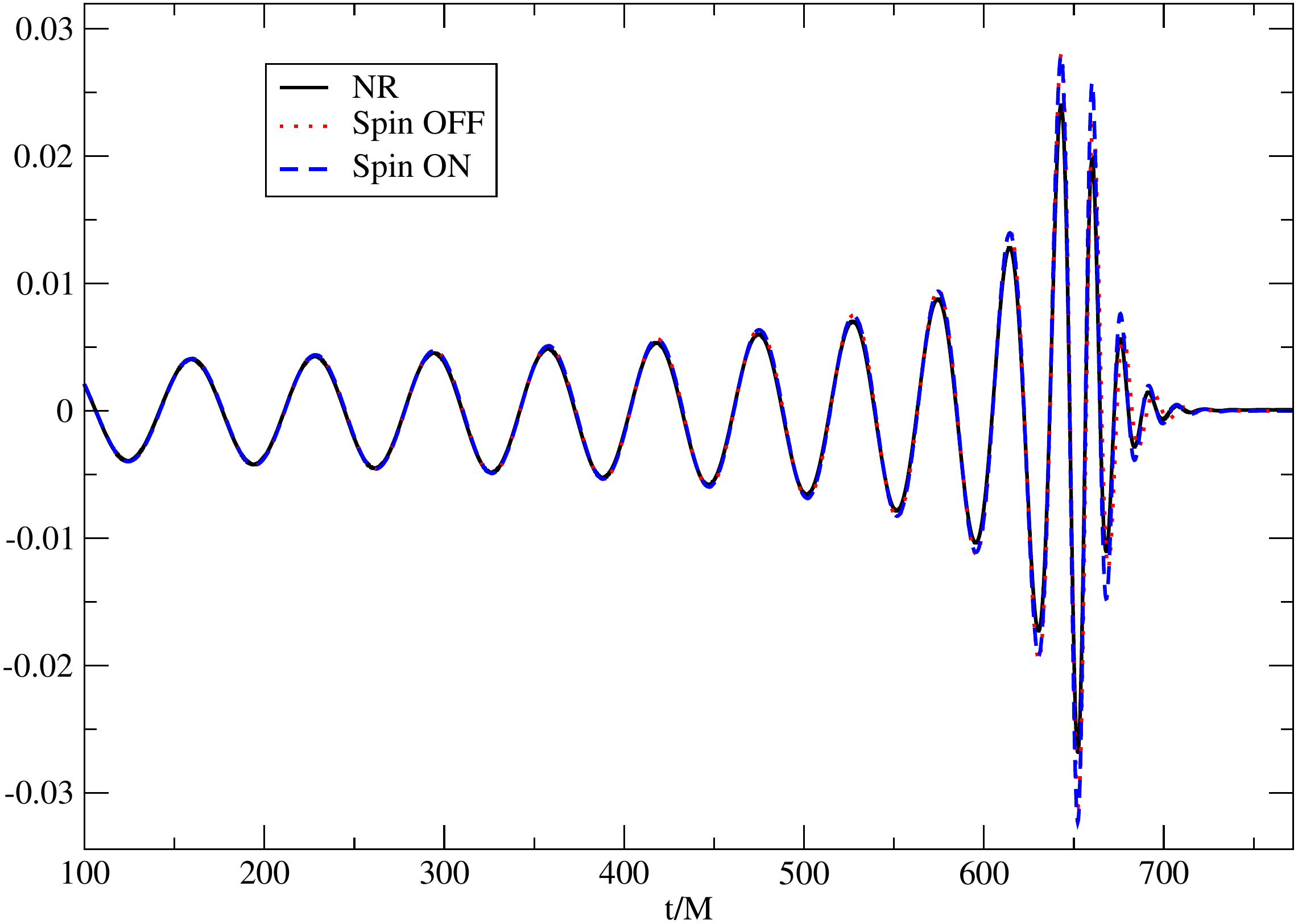}
  \label{fig:dh22_15to1}
\end{figure}

\begin{figure}
  \caption{The real part of the $(\ell=2,\,m=1)$ mode of $dh/dt$ for the $q=1/15$ case.
  The (black) solid, (red) dotted, and (blue) dashed curves show the NR, 
  spin-off, and spin-on calculations, respectively.} 
  \includegraphics[width=3.4in]{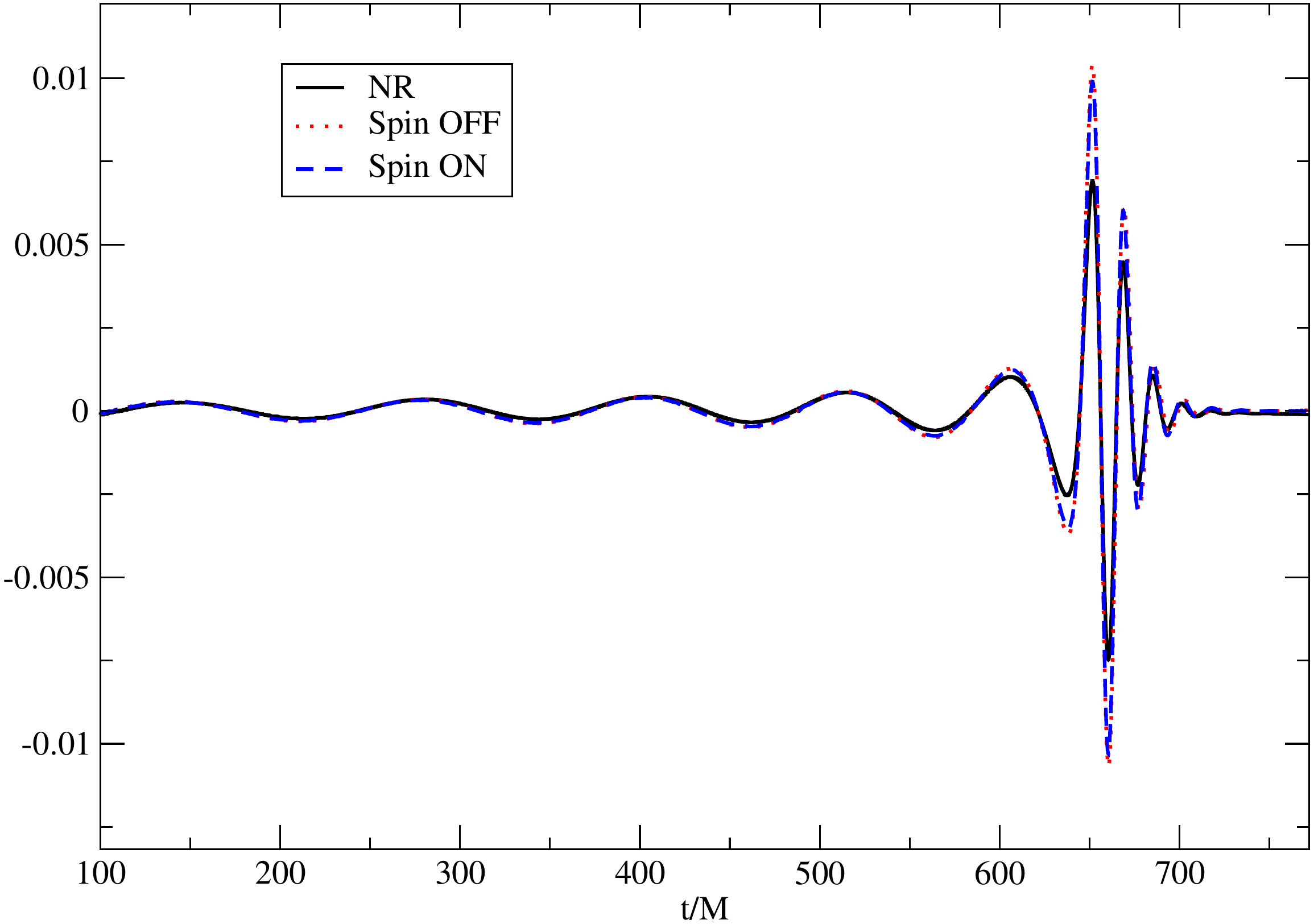}
  \label{fig:dh21_15to1}
\end{figure}

\begin{figure}
  \caption{The real part of the $(\ell=3,\,m=3)$ mode of $dh/dt$ for the $q=1/15$ case.
  The (black) solid, (red) dotted, and (blue) dashed curves show the NR, 
  spin-off, and spin-on calculations, respectively.} 
  \includegraphics[width=3.4in]{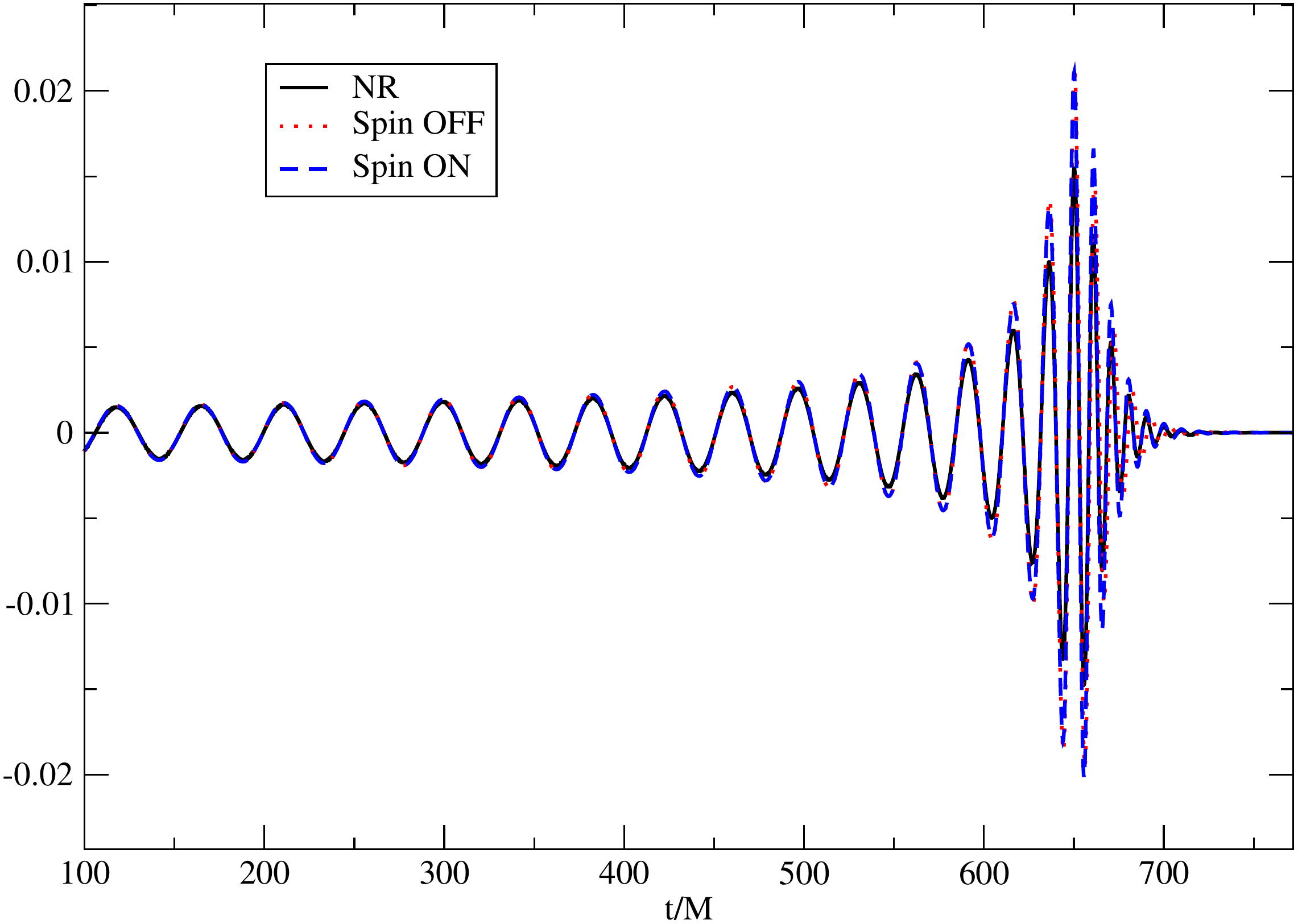}
  \label{fig:dh33_15to1}
\end{figure}

In Figs.\ \ref{fig:dh22_15to1}-\ref{fig:dh33_15to1} we superpose the
waveforms for the modes (2,2), (2,1), and (3,3)
obtained from the full numerical evolution of the $q=1/15$ case.
We included full numerical, perturbative with spin $(a/M=0.189)$
and without spin corrections $(a=0)$.
We computed the overlap functions, as defined in 
Ref.\ \cite{Campanelli:2008nk},
for these three sets of waveforms and display the results in
Table \ref{tab:match15to1}. We observe again the generally very
good agreement of the perturbative and full numerical waveforms.
The agreement is still stronger when we include the spin
dependence of the remnant black hole.

\begin{table*} 
\centering 
  \caption{The overlap (matching) between the NR and perturbative $dh/dt$ 
  for the $q=1/15$ case. The integration time is from $t/M=100$ to
$750$, and 
  the definition of the matching is given in Eqs.~(26) and (27) 
  of~\cite{Campanelli:2008nk}.}
\label{tab:match15to1}
\begin{ruledtabular}
\begin{tabular}{l|ccc}
   Mode & $\Re (\ell=2, m=2)$ & $\Re (\ell=2, m=1)$ & $\Re (\ell=3, m=3)$ \\
\hline
   Spin OFF & 0.991297 & 0.993986 & 0.969254 \\
\hline
   Spin ON  & 0.996607 & 0.997256 & 0.995974 \\
\hline
\hline
   Mode & $\Im (\ell=2, m=2)$ & $\Im (\ell=2, m=1)$ & $\Im (\ell=3, m=3)$ \\
\hline
   Spin OFF & 0.991653 & 0.996433 & 0.968889 \\
\hline
   Spin ON  & 0.996780 & 0.998178 & 0.996218 \\
  \end{tabular}
  \end{ruledtabular}
\end{table*}

In order to study in more detail the agreement of the numerical and
perturbative waveforms we will proceed to decompose them into phase
and amplitude $(\varphi,\,A)$ with the usual formula
\begin{equation}
\psi=A\exp ({i\varphi}) \,.
\end{equation}

\begin{figure}
  \caption{The phase evolution of the $(\ell=2,\,m=2)$ wave  for the $q=1/10$ case.
  The (black) solid, (red) dotted, and (blue) dashed curves show the NR, 
  spin-off, and spin-on calculations, respectively. 
  The inset shows the zoom-in for the quasinormal region.} 
  \includegraphics[width=3.4in]{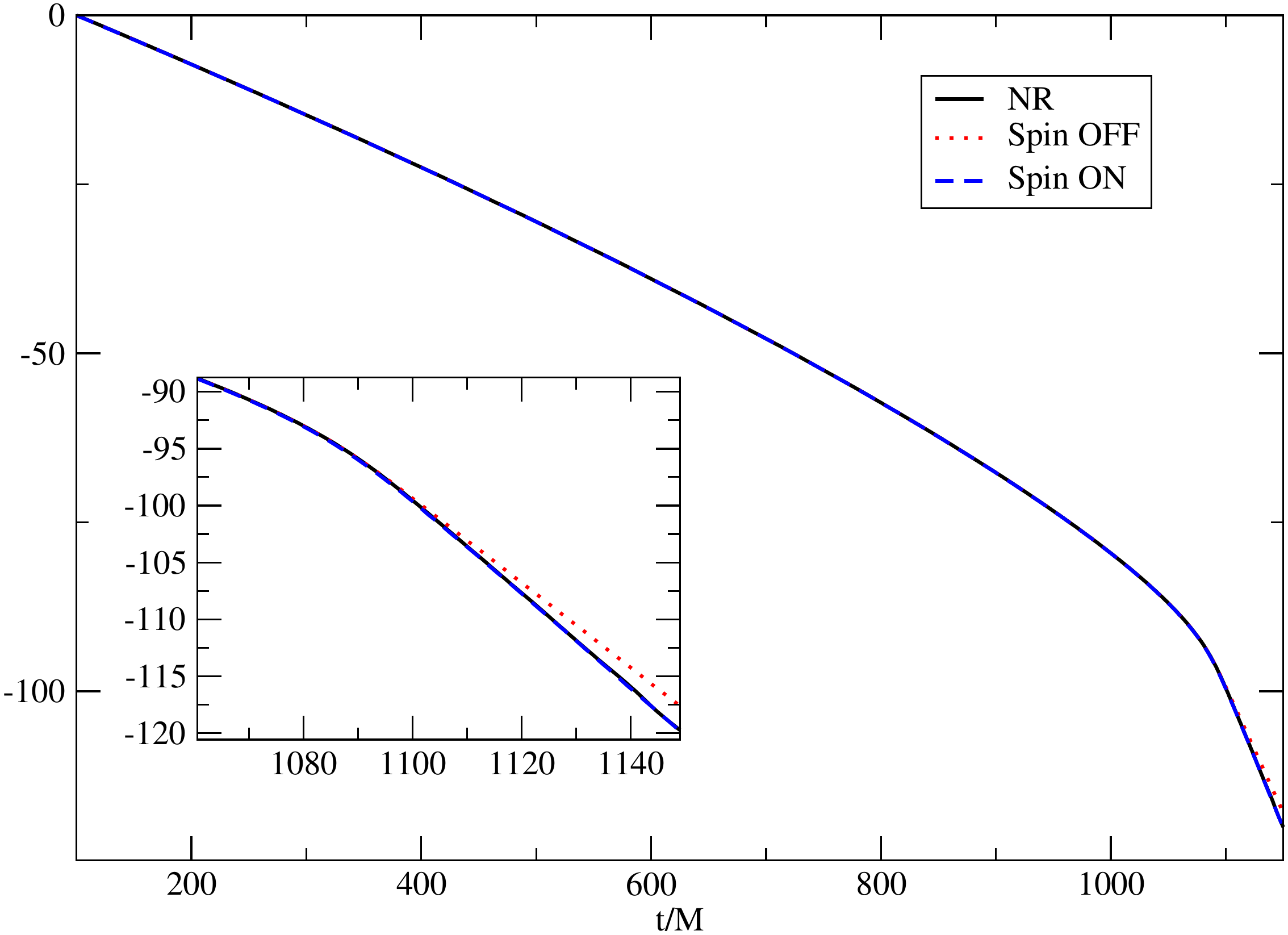}
  \label{fig:dh22_10to1_phase}
\end{figure}

\begin{figure}
  \caption{The phase evolution of the $(\ell=2,\,m=1)$ wave  for the $q=1/10$ case.
  The (black) solid, (red) dotted, and (blue) dashed curves show the NR, 
  spin-off, and spin-on calculations, respectively. 
  The inset shows the zoom-in for the quasinormal region.} 
  \includegraphics[width=3.4in]{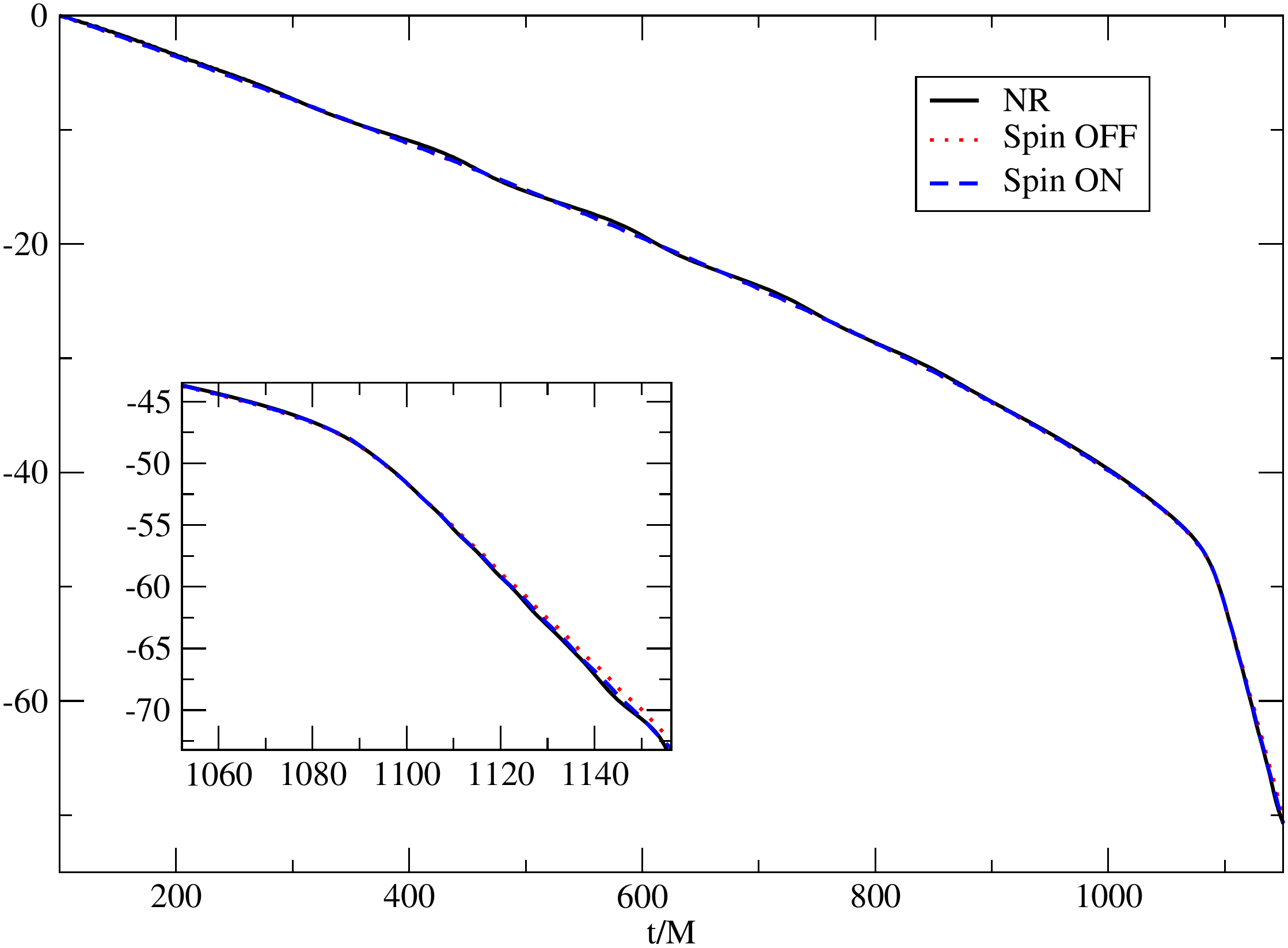}
  \label{fig:dh21_10to1_phase}
\end{figure}

\begin{figure}
  \caption{The phase evolution of the $(\ell=3,\,m=3)$ wave  for the $q=1/10$ case.
  The (black) solid, (red) dotted, and (blue) dashed curves show the NR, 
  spin-off, and spin-on calculations, respectively. 
  The inset shows the zoom-in for the quasinormal region.} 
  \includegraphics[width=3.4in]{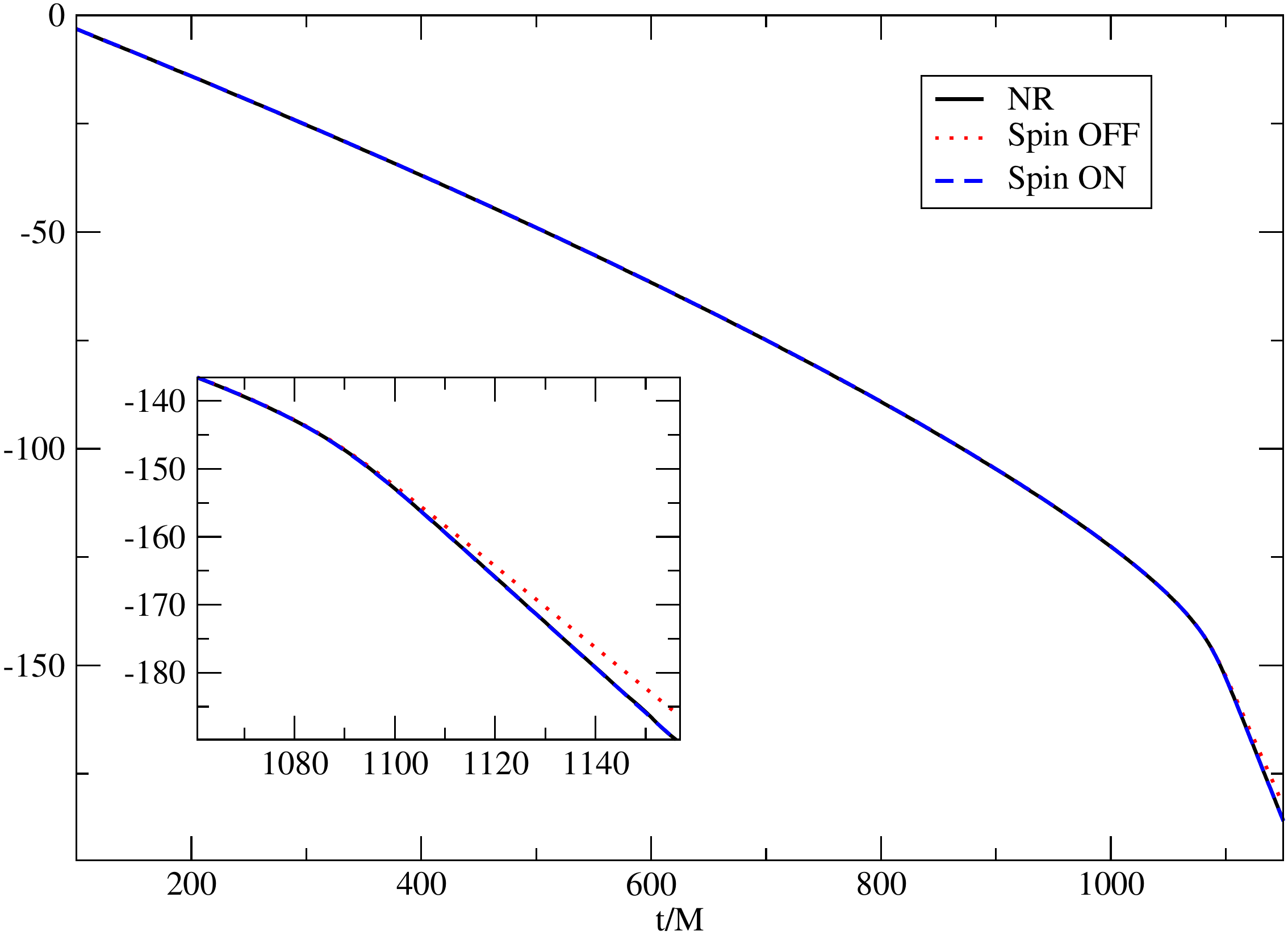}
  \label{fig:dh33_10to1_phase}
\end{figure}

We display in Figs.\ \ref{fig:dh22_10to1_phase}-\ref{fig:dh33_10to1_phase}
the phases of the (2,2), (2,1) and (3,3) modes for the $q=1/10$
case. Note the very good agreement between numerical and perturbative
waveforms for the whole range of the simulation. All the agreements 
have been found with a single
full numerical trajectory feeding the source terms of both the even
and odd parity perturbative equations. The insets in the figures zoom in
on the late time phases to display the effect of the spin correction, which in
all three modes shows improvements over the nonspinning background case.

\begin{figure}
  \caption{The phase evolution of the $(\ell=2,\,m=2)$ wave  for the $q=1/15$ case.
  The (black) solid, (red) dotted, and (blue) dashed curves show the NR, 
  spin-off, and spin-on calculations, respectively. 
  The inset shows the zoom-in for the quasinormal region.} 
  \includegraphics[width=3.4in]{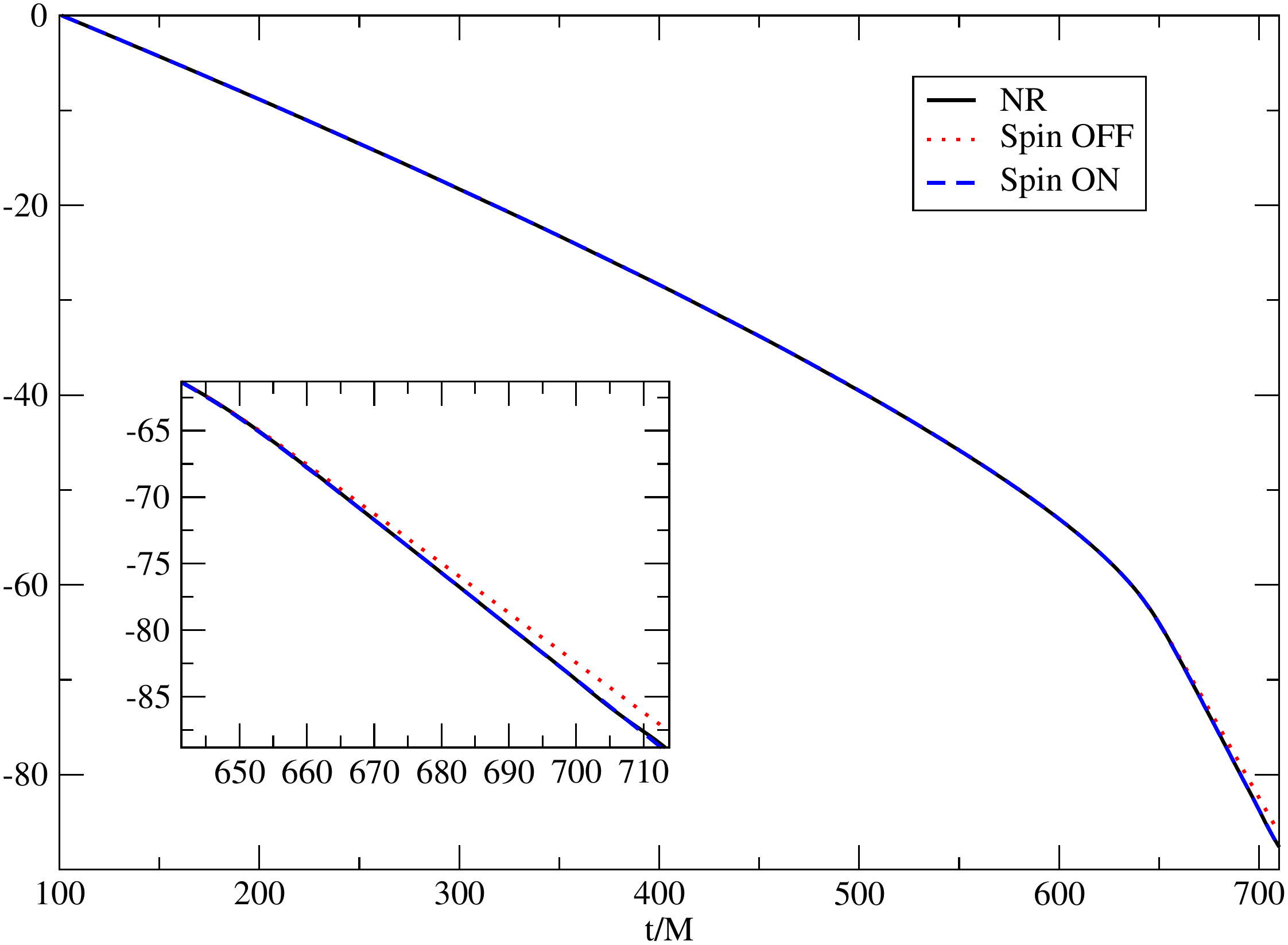}
  \label{fig:dh22_15to1_phase}
\end{figure}

\begin{figure}
  \caption{The phase evolution of the $(\ell=2,\,m=1)$ wave  for the $q=1/15$ case.
  The (black) solid, (red) dotted, and (blue) dashed curves show the NR, 
  spin-off, and spin-on calculations, respectively. 
  The inset shows the zoom-in for the quasinormal region.} 
  \includegraphics[width=3.4in]{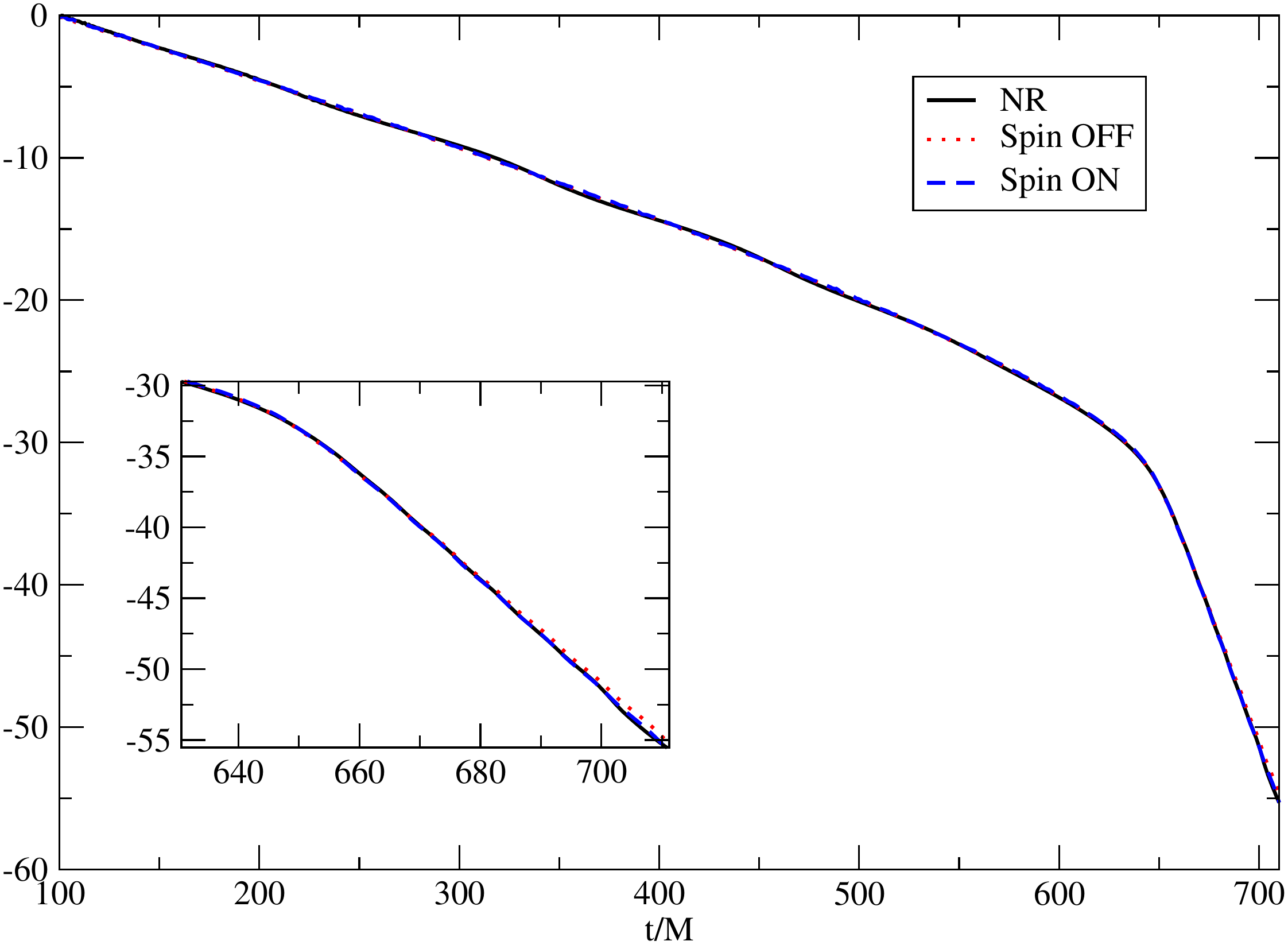}
  \label{fig:dh21_15to1_phase}
\end{figure}

\begin{figure}
  \caption{The phase evolution of the $(\ell=3,\,m=3)$ wave  for the $q=1/15$ case.
  The (black) solid, (red) dotted, and (blue) dashed curves show the NR, 
  spin-off, and spin-on calculations, respectively. 
  The inset shows the zoom-in for the quasinormal region.} 
  \includegraphics[width=3.4in]{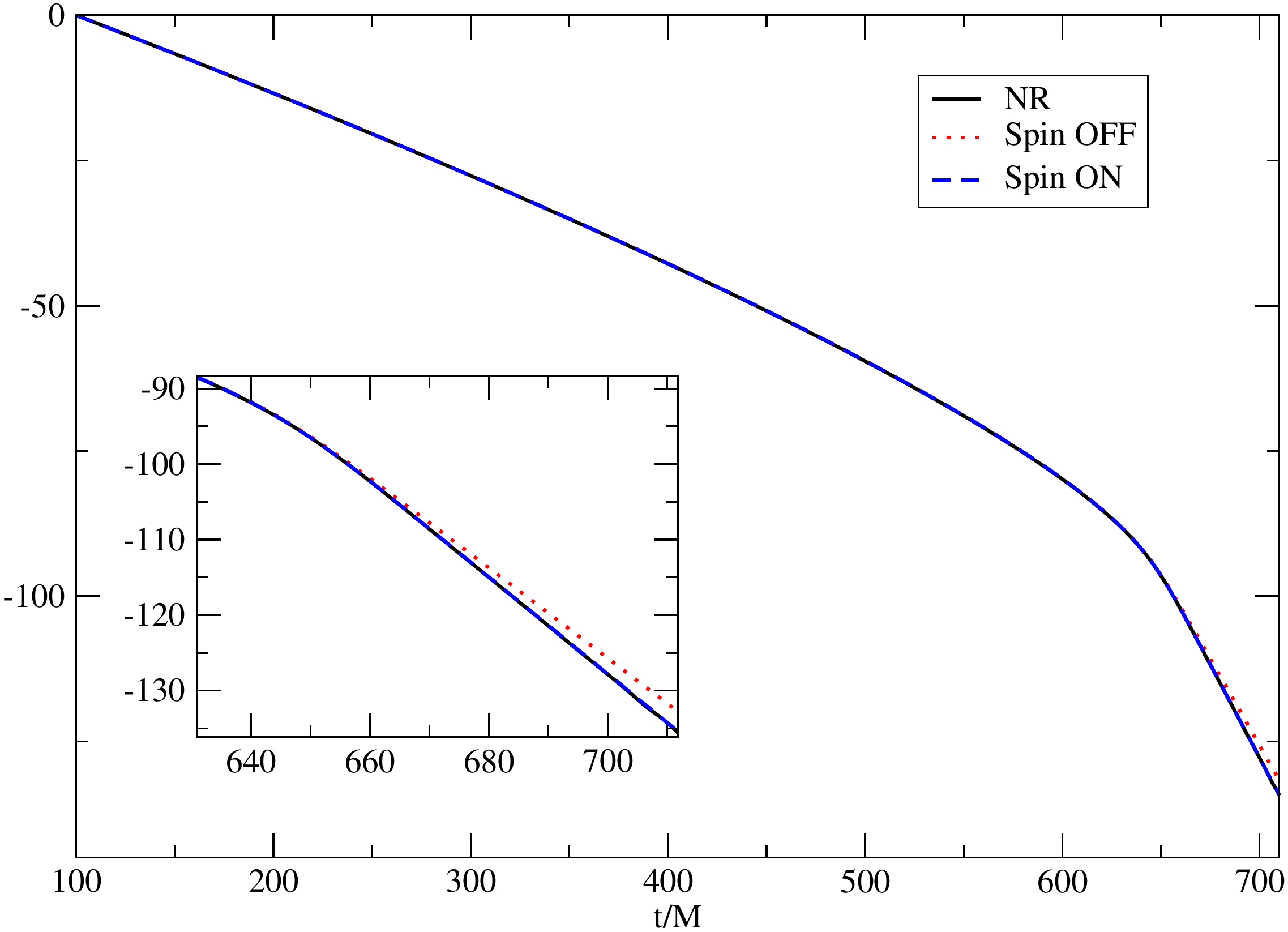}
  \label{fig:dh33_15to1_phase}
\end{figure}

Figsures\ \ref{fig:dh22_15to1_phase}-\ref{fig:dh33_15to1_phase} show
the phases of the (2,2), (2,1) and (3,3) modes for the $q=1/15$
case. Again very good agreement is seen for the whole range of the
full numerical simulation between perturbative and numerical results.
The insets show that the spin correction, even if smaller than for the
$q=1/10$ case, still improves the late time phase, correctly capturing 
the quasinormal frequencies of the slowly rotating Kerr black hole ($a/M=0.189$).

\begin{figure}
  \caption{The amplitude of the $(\ell=2,m=2)$ mode 
  of the NR $dh/dt$ for the $q=1/10$ and $1/15$ cases.
  The (black) thick solid and (red) solid curves show the $q=1/10$ and $1/15$ amplitudes, 
  respectively. The (red) dashed curve denotes $\eta(q=1/10)/\eta(q=1/15) \sim 1.41$ 
  times the $q=1/15$ amplitude.} 
  \includegraphics[width=3.4in]{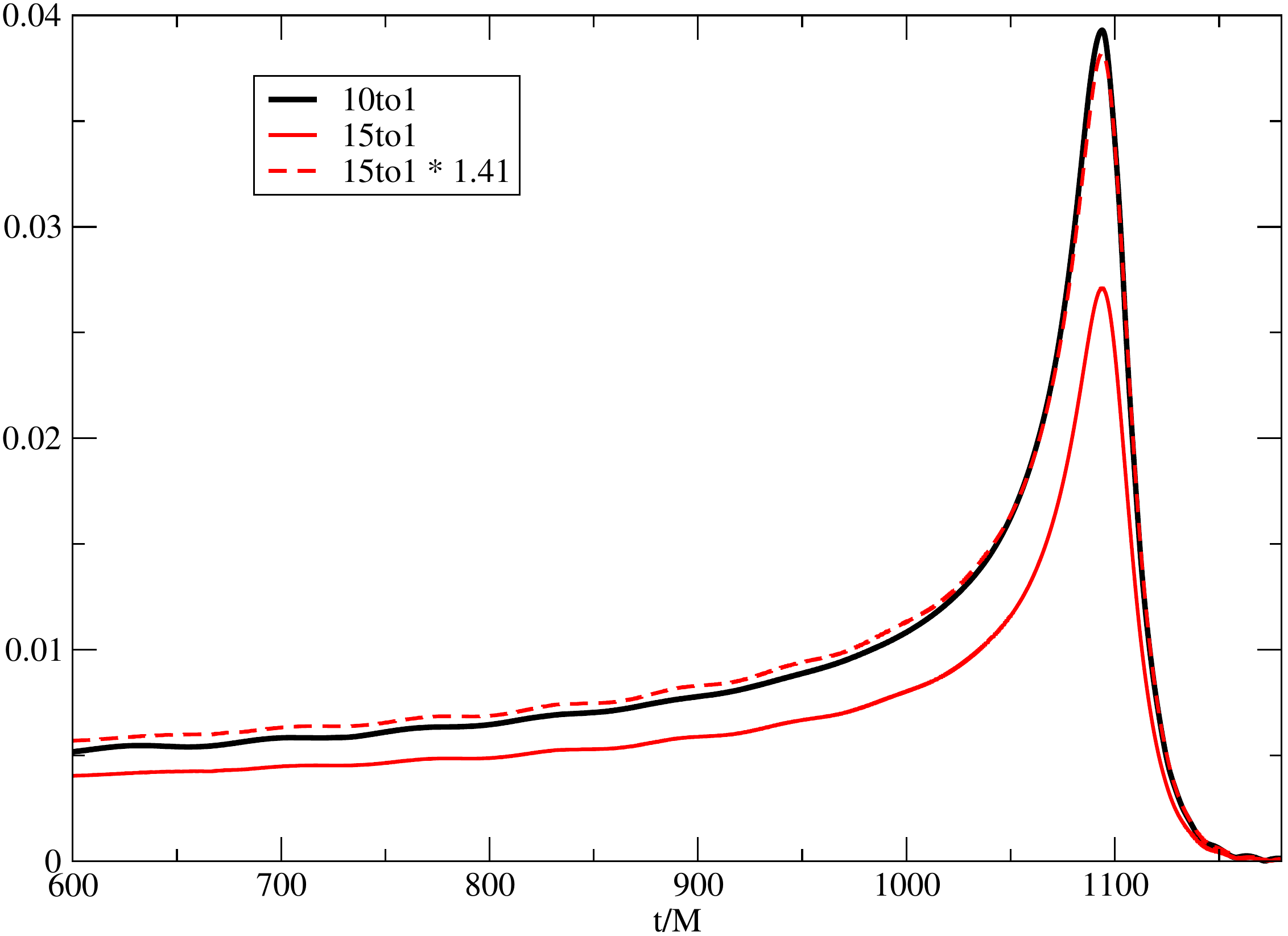}
  \label{fig:dh22_NR_amp}
\end{figure}

We now turn to compare amplitudes of waveforms. Although for gravitational
wave detection by the LIGO and VIRGO observatories the most important indicator is
the phase, the amplitude agreement is particularly important in the modeling
of the sources. Figure~\ref{fig:dh22_NR_amp} directly compares the amplitudes
of the $q=1/10$ and $q=1/15$ cases, shifted in time to agree at the
peaks of their amplitudes. We then rescale the amplitudes of the $q=1/15$
waveform by the factor $\mu(q=1/10)/\mu(q=1/15)\approx1.41$ to verify
a linear rescaling. We find that the rescaled amplitude of the $q=1/15$
wave is very close to the actual  $q=1/10$ amplitude showing that the
systems are close to behaving linearly at these mass ratios.

\begin{figure}
  \caption{The amplitude difference in the $(\ell=2,m=2)$ mode
  between the NR and perturbative $dh/dt$ 
  for the spin-off cases. The (black) thick solid curve shows
  the $q=1/10$ case. 
  The (red) solid, dotted, and dashed curves show the
amplitude differences for the $q=1/15$ case rescaled by factors of
 $1$, $1.41$, and $1.41^2$, respectively.
} 
  \includegraphics[width=3.4in]{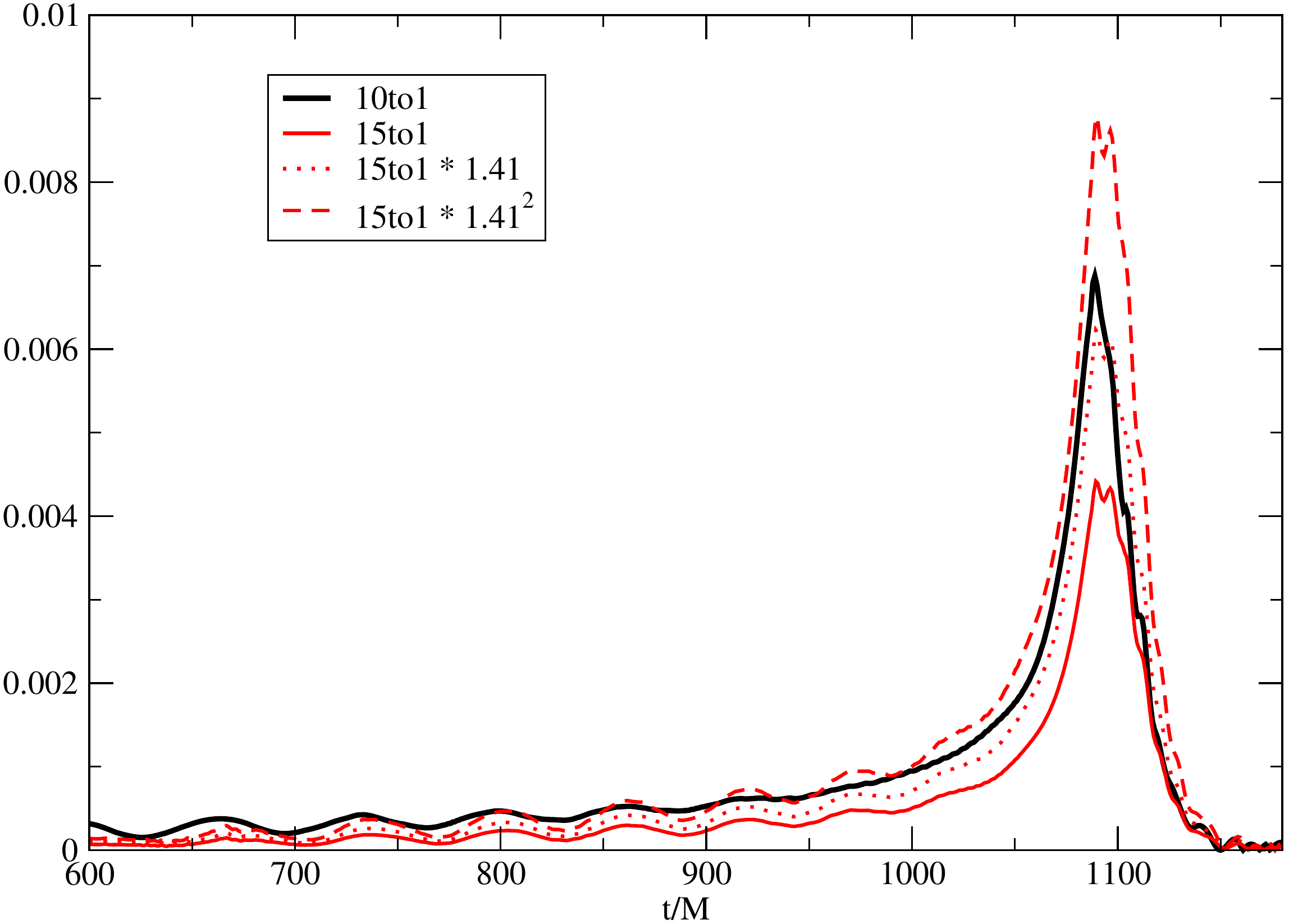}
  \label{fig:dh22_diff_Soff}
\end{figure}

\begin{figure}
  \caption{The amplitude difference in the $(\ell=2,m=2)$ mode
  between the NR and perturbative $dh/dt$ 
  for the spin-on cases. The (black) thick solid curve shows the $q=1/10$ case. 
  The (red) solid, dotted, and dashed curves show the
amplitude differences for the $q=1/15$ case rescaled by factors of
 $1$, $1.41$, and $1.41^2$, respectively.}
  \includegraphics[width=3.4in]{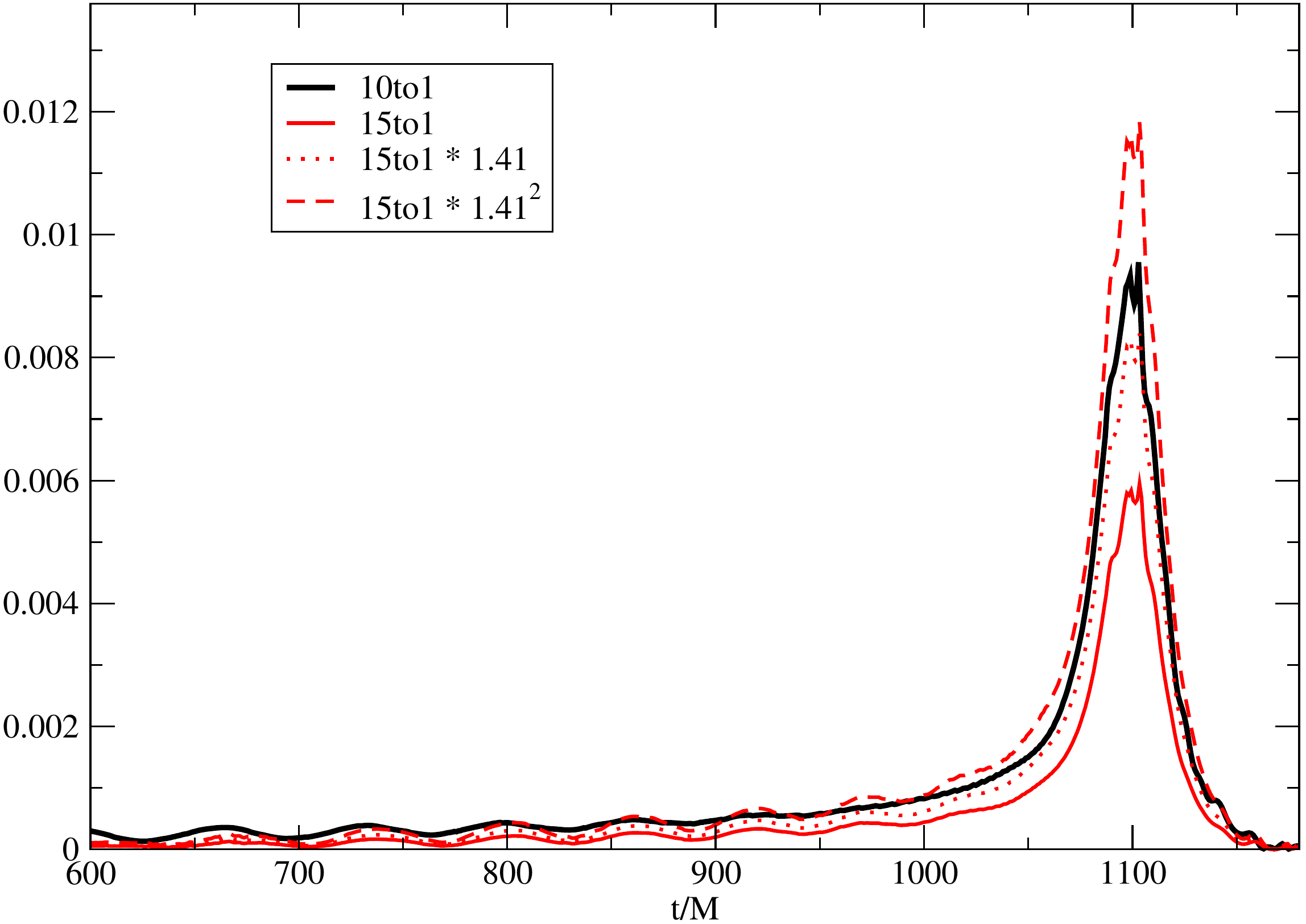}
  \label{fig:dh22_diff_Son}
\end{figure}

In order to assess this last point 
in more detail, we compute the differences of
the numerical and perturbative waveforms for each case, $q=1/10$ and $q=1/15$,
and study how this ``error'' scales with $q$ (or more precisely $\mu$).
We display the results of such computations in 
Figs.\ \ref{fig:dh22_diff_Soff} and \ref{fig:dh22_diff_Son} for the
cases of neglecting the spin of the final hole and that of taking it into
account, respectively. The plots show that the inspiral phase scales
like $\mu^2$ as one would predict if the system would be completely
linearized. While in the final merger region, near the peak of the amplitude,
the rescaled differences display a dependence in $\mu$ 
between linear and quadratic,
as if there are still nonlinearities present. One would
expect this behavior for values of $q$ that are in the intermediate
mass ratio regime, where the linear approximation is good but
small nonlinear effects can still be observed.

\section{Discussion}\label{sec:Discussion}

In this paper we have described in detail the techniques used to
compute gravitational waveforms with the perturbative approach using
full numerical trajectories in the source terms of the perturbative
wave equations. The program was successfully tested in the $q=1/10$ case
in Ref.~\cite{Lousto:2010tb}. We have taken it 
further here studying larger initial
separations for the full numerical evolutions of the $q=1/10$ case,
leading to simulations lasting for nearly eight orbits before the final plunge.
We have also studied the case $q=1/15$, the smallest mass ratio so far
in the literature, in order to assess quantitatively the $q$-dependence of
the agreement of the full numerical and perturbative evolutions in
the intermediate mass ratio regime. We have also included in our new
computations the (linear dependence) spin
of the final remnant in order to correctly reproduce
the quasinormal ringing component of the full waveform at late times
(after merger). The results are displayed in Tables 
\ref{tab:match10to1} and \ref{tab:match15to1}
and in Figs.\ \ref{fig:dh22_10to1_phase}-\ref{fig:dh33_15to1_phase}.
They show an apparent improvement in
the matching (overlap) indices when the spin correction is taken
into account compared to the vanishing spin case. 
In the Appendix we apply this linear-in-spin perturbation theory
(SRWZ) to compute the corresponding quasinormal modes and compare the
frequencies of these modes
with those obtained for a Kerr black hole background for all values of
the spin parameter.
 We observe the results in Figs.\ \ref{fig:2L} and \ref{fig:3L}.
They show that SRWZ provide reliable predictions for $a/M\leq0.3$, which
justifies their use for the cases studied here where $a/M=0.26,0.19$ for
$q=1/10,1/15$ respectively.
The generalization to arbitrary spins requires solving the 
Teukolsky equation instead of the RWZ ones \cite{Sundararajan:2010sr}.
Note that the relevant spin-effects on the waveform
are due to the spin of the large
black hole, while the effects of the spin of the small hole 
on radiation will tend to
be negligible as $q$ decreases. 
The use of numerical trajectories to describe the
motion of the small hole in the field of the larger one already
incorporates the spin dependence where the effects are stronger.

\begin{figure}[!h]
  \caption{The radial trajectory $R(t)$ obtained from the PN and NR evolutions
  for the $q=1/10$ case in the Schwarzschild coordinates. 
  The (black) solid, (red) dashed and (blue) dotted curves 
  show the NR, resummed and PN Taylor ones, respectively. From the lower-right inset, 
  we can choose the matching radius between the NR and resummed PN evolutions 
  as $R(t)/M=9.35123$. The upper-left inset is the zoom-in around the matching time.} 
  \includegraphics[width=3.4in]{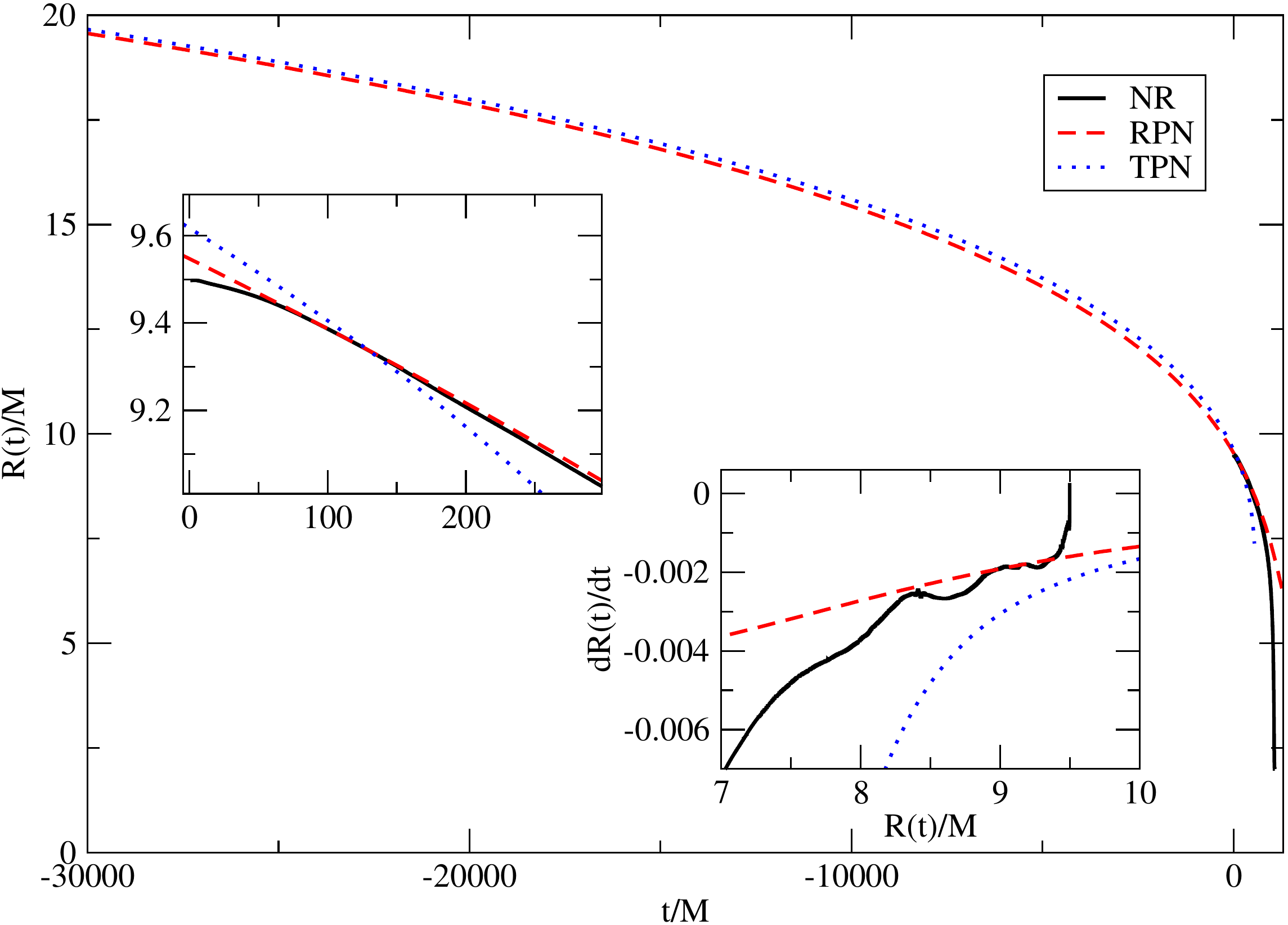}
  \label{fig:PNNR_track_10to1}
\end{figure}

\begin{figure}[!h]
  \caption{The radial trajectory $R(t)$ obtained from the PN and NR evolutions
  for the $q=1/15$ case in the Schwarzschild coordinates. 
  The (black) solid, (red) dashed and (blue) dotted curves 
  show the NR, resummed and Taylor PN ones, respectively. From the lower-right inset, 
  we can choose the matching radius between the NR and resummed PN evolutions 
  as $R(t)/M=8.28796$. The upper-left inset is the zoom-in around the matching time.} 
  \includegraphics[width=3.4in]{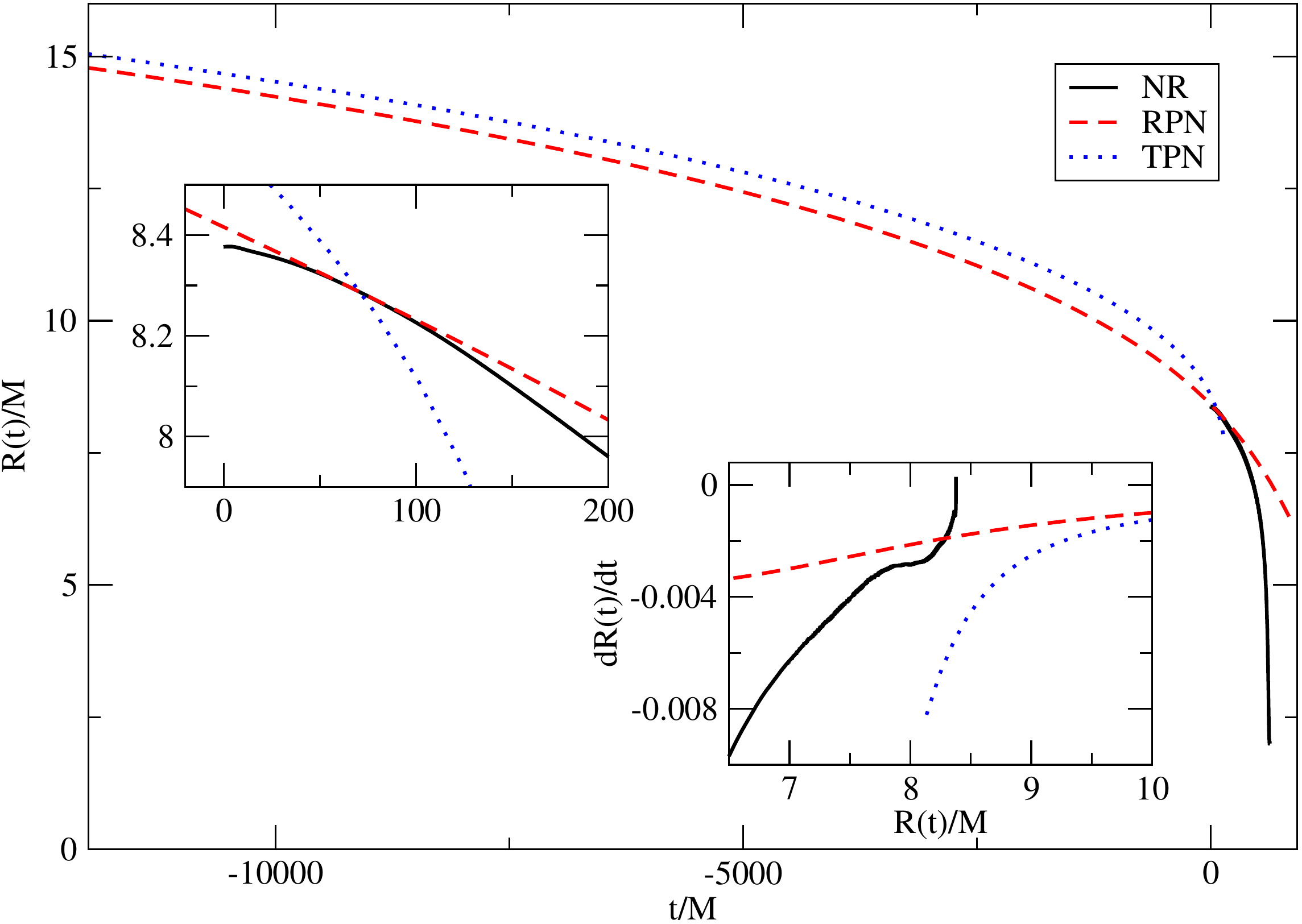}
  \label{fig:PNNR_track_15to1}
\end{figure}

After comparing the perturbative and full numerical waveforms and
verifying the accuracy of the former, there remains the question
of accurately modeling the trajectories for small $q$ BHBs.
We have stressed here an important fact, that the trajectory dependence
disappears from the perturbative formulation once the black holes merge,
reducing the need of further full numerical simulations with the
resulting saving of computational resources. This savings is
not negligible, because one not saves not only the (relatively short) time 
of evolution from merger to the end of
the ringdown, but also the evolution time required to propagate the signal
to  an observer located far away from the sources. Typically, this
should save over $500M$ of full numerical evolution. One can also
predict the parameters of the final black hole by using 
formulae for the remnant parameters, as in \cite{Lousto:2009mf,Lousto:2009ka}, 
found by empirical fitting. Still, the goal
of our project is to be able to model, empirically, the BHBs 
inspiral trajectories as a function of $q$ from a reasonably small
number of full numerical evolutions.
In particular, numerical evolutions start from a finite, relatively
close initial separation of the holes. It is hence important to provide
the large separation input from PN theory. While the full modeling of
trajectories is beyond the scope of the current paper, here we 
discuss how this interface can be achieved for the current simulations
of the  $q=1/10$ and $q=1/15$ cases.  The results are summarized 
in Figs.~\ref{fig:PNNR_track_10to1} and \ref{fig:PNNR_track_15to1}. 
We have considered the full numerical and PN trajectories in 
the Schwarzschild coordinates, i.e., correct the 
full numerical tracks for the 1+log time slice and 
the PN ones for the quasi isotropic coordinates (ADM-TT gauge). 
In the $q=1/10$ case, the full numerical
evolutions essentially start from initial separations $R_i\approx9.5M$ 
in the Schwarzschild coordinates.
We see a relatively smooth matching for the tracks and their first 
derivative in (upper-left inset) Fig.~\ref{fig:PNNR_track_10to1}.
This would lead hence to smooth waveforms in the whole range of the
evolution, i.e., from as large initial (PN) separations as needed down to
the ringdown.
Note however, that in order to achieve this smooth matching of trajectories
we had to make use of the resummed PN (RPN) evolutions 
(i.e. containing exactly
the particle limit in the Schwarzschild background). 
The RPN Hamiltonian used here is derived in the following. 
Based on the Hamiltonian formulation for the test particle 
given in~\cite{Arnowitt:1960zzc}, the resummed part $H_{\rm Sch}$ 
is calculated by using the Schwarzschild metric in the isotropic coordinates. 
Then the RPN Hamiltonian is given by 
\begin{eqnarray}
H^{\rm RPN}
&=& H_{\rm Sch}+\tilde{H}_{\rm 1PN}
+\tilde{H}_{\rm 2PN}+\tilde{H}_{\rm 3PN} \,.
\end{eqnarray}
The finite mass effects $\tilde{H}_{\rm 1PN}$, 
$\tilde{H}_{\rm 2PN}$ and $\tilde{H}_{\rm 3PN}$ in the above Hamiltonian 
are introduced by the result of the standard 3PN Taylor Hamiltonian (TPN)  
and the 3.5PN radiation reaction effects 
on the equations of motion are treated as in~\cite{Buonanno:2005xu}. 
In practice, $\tilde{H}_{\rm nPN}$ is obtained by the subtraction 
of the test particle limit from the Taylor PN Hamiltonian ${H}_{\rm nPN}$.
The PN evolutions in the figures have been obtained 
from the quasicircular initial parameter at $R(t)\sim 50M$.
A good matching, 
at this initial separation, cannot be achieved with the TPN Hamiltonian. 
Of course, at larger separations both PN expressions get 
closer to each other and a full numerical simulation started at such
large initial separations could be matched by Taylor PN expansion
as well.

This also suggest that, at even closer separations, as in the case of
the numerical evolutions for $q=1/15$ starting from $R_i\approx8.4M$,
not even the resummed PN leads to a very smooth matching of track.
This is indeed the case displayed in Fig.~\ref{fig:PNNR_track_15to1}. 
We may then conclude that, in order to simulate full inspirals of $q\sim1/10$ 
matched to resummed PN, one 
needs to start the full numerical simulations from initial separations
$R_i>9M$ in the Schwarzschild coordinates, i.e., 
$R_i^{\rm (QI)}>8M$ in the quasi isotropic coordinates. 
Alternatively, one could seek to improve the resummed PN
expansions with the effective-one-body (EOB) formalism  \cite{Buonanno:1998gg}
and its extension to incorporate full numerical results (EOBNR) 
\cite{Pan:2009wj}.
It is also relevant to cite here the works 
\cite{Nagar:2006xv,Damour:2007xr,Bernuzzi:2010ty} that make perturbative
evolutions of particle trajectories completely derived from PN expansions
and used all the way down to merger without direct input from full numerical
trajectories.

If one indeed can extend those improved post-Newtonian treatments down
to the ISCO in the particle limit, at $R=6M$ in the Schwarzschild coordinates,
i.e. $R^{\rm (ISCO)} \approx 4.95M$ in the isotropic coordinates, then one can argue that
the subsequent merger trajectory reaches a ``universal'' limit given by
the geodesic motion of quasicircular orbits. In fact this seems to
be the case for the tracks of the $q=1/10$ and $q=1/15$ simulations
as displayed in Fig.\ \ref{fig:q10_q15_track}. 
One can argue that the very low level of
radiation of those plunging orbits implies the universal form of the
track. This was also recently observed in \cite{Bernuzzi:2010ty} studying
PN orbits.
Notably, at the other extreme of the mass ratio range, i.e. for
equal (and comparable) mass BHBs the strong gravitational emission taking
place during the plunge erases any details of the preliminary evolution
and one observes a universal waveform \cite{Baker:2002qf,Campanelli:2006gf,
Baker:2006yw,Hinder:2007qu}

To see the universal behavior of geodesics inside the ISCO for quasicircular
inspirals, we use the orbits with imaginary eccentricities for timelike 
geodesics in the Schwarzschild spacetime as given on page 111 
of~\cite{Chandrasekhar83}. The initial part of these orbits can be
considered the continuation of the inspiral trajectories through the ISCO.
These geodesics have the following form near the horizon:
\begin{eqnarray}
\Phi(R) &\sim&
\frac{3}{4}\,\sqrt {6} \,\biggl[
\left( 1-\frac{1}{8}\,{e}^{2}+\frac{3}{8}\, \left( 1-{\frac {R_0}{6M}} \right) ^{2} \right)  
\nonumber \\ && \times 
\left( {\frac {R}{2M}}-1 \right)  
\nonumber \\ && 
+\left( \frac{1}{8}-{\frac {13}{64}}\,{e}^{2}+{\frac {39}{
64}}\, \left( 1-{\frac {R_0}{6M}} \right) ^{2} \right)  
\nonumber \\ && \times 
\left( {\frac {R}{2M}}-1 \right) ^{2}
\biggr] 
\,,
\end{eqnarray}
where the imaginary eccentricity ($ie$) is a small quantity, 
and $R_0 < 6M$.

The initial velocity at $R(t)=R_0$ is approximately  given by 
\begin{eqnarray}
\frac{dR(t)}{dt} &=& -\frac{\sqrt {6}}{6}
\,e\,\sqrt { \left( 1-{\frac {R_0}{6M}} \right) } 
\,,
\nonumber \\ 
\frac{d\Phi(t)}{dt} &=& 
{\frac {\sqrt {6}}{36 M}}
+\frac{\sqrt {6}}{24M} \left( 1-{\frac {R_0}{6M}} \right)
\nonumber \\ && 
+{\frac {\sqrt {6}}{288M}}\, 
\left( 15\, \left( 1-{\frac {R_0}{6M}} \right) ^{2}-{e}^{2} \right) \,,
\end{eqnarray}
which allows us to match to full numerical trajectories and 
then use the geodesic
expressions to smoothly suppress the local source terms when the
particle approaches the Schwarzschild horizon (see Sec. \ref{Onm}).

\begin{figure}[!h]
  \caption{The orbit with imaginary eccentricities discussed in~\cite{Chandrasekhar83}.
  The thick and thin curves show the $q=1/10$ and $q=1/15$ cases, respectively. 
  Here we show the orbits with various eccentricities.} 
  \includegraphics[width=3.4in]{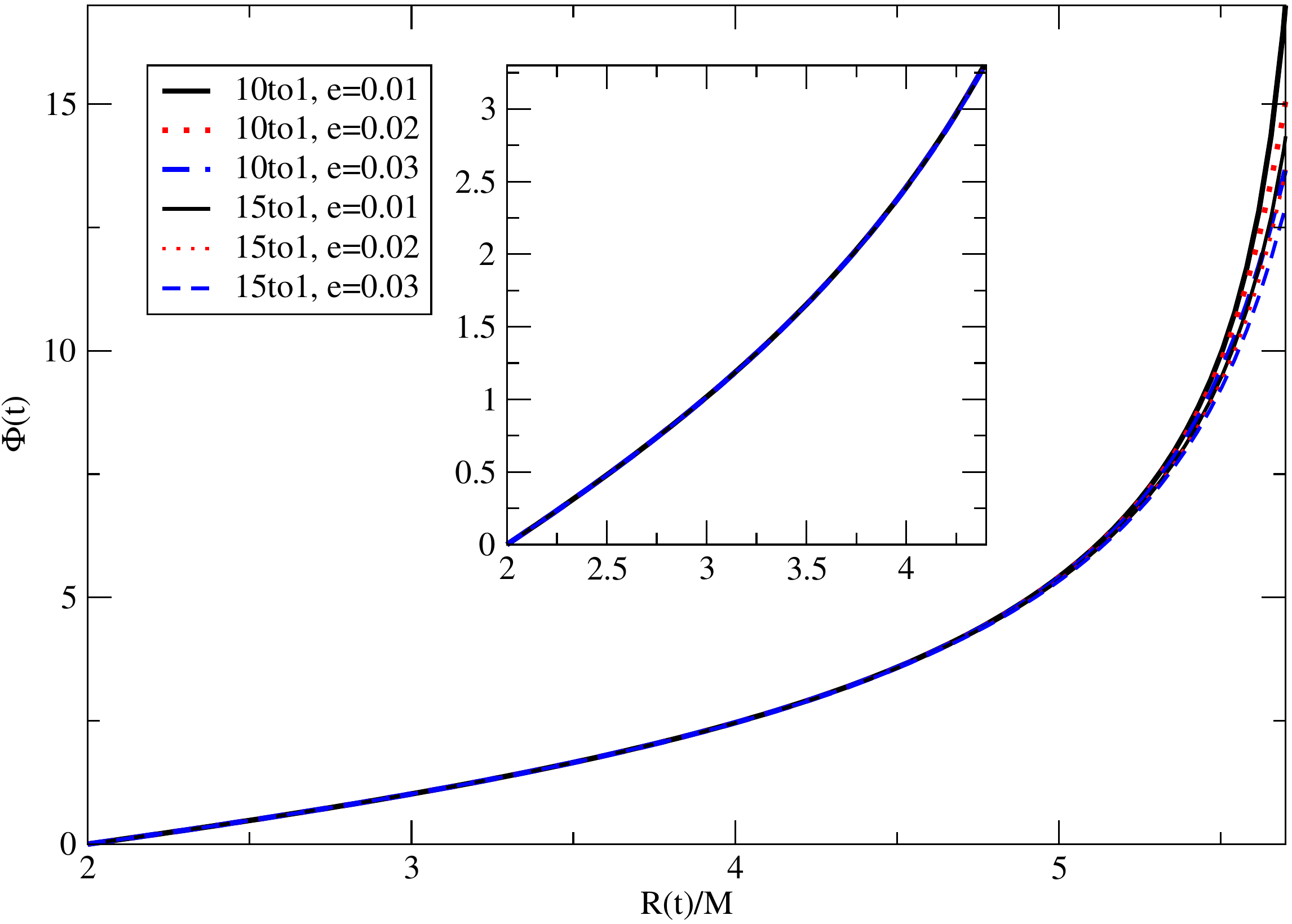}
  \label{fig:UnivTrack}
\end{figure}

In Fig.~\ref{fig:UnivTrack}, we plot the phase evolution in terms 
of the orbital radius. As a fiducial starting point, just inside the
Schwarzschild ISCO, we take the self-force corrected ISCO radius 
\begin{eqnarray}
R_0 &=& 6\,M - 3.269\,\mu \,,
\end{eqnarray}
as discussed in~\cite{Barack:2009ey}.
Although we see some differences in the initial part of the orbits,
the trajectories reach a universal limit approaching the horizon.

\acknowledgments 

We gratefully acknowledge the NSF for financial support from Grants
No. PHY-0722315, No. PHY-0653303, No. PHY-0714388, No. PHY-0722703,
No. DMS-0820923, No. PHY-0929114, PHY-0969855, PHY-0903782,
and No. CDI-1028087; and NASA for financial support from NASA Grants 
No. 07-ATFP07-0158 and No. HST-AR-11763.  Computational resources were
provided by the Ranger cluster at TACC (Teragrid allocation TG-PHY060027N) 
and by NewHorizons at RIT.

\appendix

\section{Analysis of the wave equations}\label{app:WEdetail}

The following is useful for the analytic discussion, 
especially the behavior of the source term around the horizon. 
And ths also gives a stable evolution in the numerical calculation, 
because nonvanishing contributions at the horizon in the source terms 
are canceled out analytically.

Here, we discuss the wave functions as 
\begin{eqnarray}
\Psi_{\ell m}(t,r) 
&=& \Psi_{\ell m}^{\rm (in)}(t,r) \,\theta(R(t)-r) 
\nonumber \\ && 
+ \Psi_{\ell m}^{\rm (out)}(t,r) \,\theta(r-R(t)) \,,
\nonumber \\ 
\Psi_{\ell m}^{\rm (step)}(t,r)
&=& \Psi_{\ell m}^{\rm (out)}(t,r) - \Psi_{\ell m}^{\rm (in)}(t,r) \,,
\end{eqnarray}
where $\Psi_{\ell m}$ denotes the even or odd parity wave function. 
The functions $\Psi_{\ell m}^{\rm (in)}$, $\Psi_{\ell m}^{\rm (out)}$, 
and $\Psi_{\ell m}^{\rm (step)}$ are the homogeneous solutions 
of the Regge-Wheeler-Zerilli equations. From these definition, we have 
\begin{eqnarray}
\Psi_{\ell m}(t,r) 
&=& \Psi_{\ell m}^{\rm (in)}(t,r) 
\nonumber \\ && 
+ \Psi_{\ell m}^{\rm (step)}(t,r) \,\theta(r-R(t)) \,. 
\end{eqnarray}
Therefore, for example, the time derivative of the above wave function is written as
\begin{eqnarray}
&&
\partial_t \Psi_{\ell m}(t,r) 
= \partial_t \Psi_{\ell m}^{\rm (in)}(t,r) 
\nonumber \\ && \quad 
+ \left(\partial_t \Psi_{\ell m}^{\rm (step)}(t,r)\right) \,\theta(r-R(t)) 
\nonumber \\ && \quad 
- \Psi_{\ell m}^{\rm (step)}(t,r)\,\frac{dR(t)}{dt} \,\delta(r-R(t))
\nonumber \\ && 
= \partial_t \Psi_{\ell m}^{\rm (in)}(t,r) 
\nonumber \\ && \quad 
+ \left(\partial_t \Psi_{\ell m}^{\rm (step)}(t,r)\right) \,\theta(r-R(t)) 
\nonumber \\ && \quad 
- \Psi_{\ell m}^{\rm (step)}(t,R(t))\,\frac{dR(t)}{dt} \,\delta(r-R(t))
\,. 
\label{eq:formalstepT}
\end{eqnarray}

To find the quantities of the waveforms at the particle location, 
i.e., $\Psi_{\ell m}^{\rm (step)}(t,R(t))$, 
we use 
\begin{eqnarray}
&& \partial_t \Psi_{\ell m}^{(1)} (t,r) 
= \Psi_{\ell m}^{\rm (Z,1)}(t,r) 
\nonumber \\ && \quad 
+  
{\frac {16\,\sqrt {2}\,\pi  \,i \,r^2 \left( r-2\,M \right)  }
{\ell \left( \ell+1 \right) \left( r{\ell }^{2}+ r\ell -2\,r +6\,M \right) }}
{\cal A}_{1 \ell m}^{(1)} \left( t,r \right)
\nonumber \\ 
&& \partial_t \Psi_{\ell m}^{\rm (o,1)} (t,r) 
= 2\,\Psi_{\ell m}^{\rm (o,Z,1)}(t,r) 
\nonumber \\ && \quad
+ 
{\frac {16\,\sqrt {2}\,\pi  \,i \,r \left( r-2\,M \right)  }{
(\ell-1)(\ell+2) \sqrt {\ell \left( \ell+1 \right) }}}
{\cal Q}_{\ell m}^{(1)} \left( t,r \right) 
\,,
\label{eq:Wtrans}
\end{eqnarray}
where each wave function in the left-hand and right-hand side of the above equations 
behaves as a step function at the particle's location 
because of the first-order Regge-Wheeler-Zerilli waveforms. 
Therefore, 
substituting Eq.~(\ref{eq:formalstepT}) into 
$\partial_t \Psi_{\ell m}^{(1)}$ and $\partial_t \Psi_{\ell m}^{\rm (o,1)}$, 
we obtain the analytic expression of $\Psi_{\ell m}^{\rm (step)}(t,R(t))$ 
from the coefficients of the Dirac's delta function.

\subsection{Analysis of the even parity wave equation}\label{app:evendetail}

We have introduced a new function for the even parity calculation 
to the SRWZ formalism, 
\begin{eqnarray}
\Psi_{\ell m} \left( t,r \right) = \Psi_{\ell m}^{(1)} \left( t,r \right) 
+ \Psi_{\ell m}^{(2)} \left( t,r \right) \,.
\label{eq:Psi_int}
\end{eqnarray}
The gravitational waveform with the spin effect 
is obtained directly from $\Psi_{\ell m}$. 
Therefore, we discuss the wave equation for $\Psi_{\ell m}$ in the following. 
Here, we create our numerical code for the perturbative calculation 
based on~\cite{Lousto:2005ip}. It is important to distinguish 
the cell that the particle does cross from the other cells. 

For the cell that the particle does not cross, 
we use the following homogeneous equation, 
i.e., can read the following equation from the step function part,  
which does not include the local source term:
\begin{widetext}
\begin{eqnarray}
&& 
-{\frac {\partial ^{2}}{\partial {t}^{2}}}\Psi_{\ell m} \left( t,r \right) 
+{\frac { \left( r-2\,M \right) ^{2}}{{r}^{2}}}{\frac {\partial ^{2}}{\partial {r}^{2}}}
\Psi_{\ell m} \left( t,r \right) 
+2\,{\frac { \left( r-2\,M \right) M}{{r}^{3}}}
{\frac {\partial }{\partial r}}\Psi_{\ell m} \left( t,r \right) 
\nonumber \\ && \quad 
- \left( r-2\,M \right)  
\left( 4\,{r}^{3}\ell-{\ell}^{4}{r}^{3}+3\,{\ell}^{5}{r}^{3}
-7\,{\ell}^{3}{r}^{3}+{\ell}^{6}{r}^{3}+12\,{\ell}^{3}{r}^{2}M
-24\,{r}^{2}M\ell-18\,{r}^{2}M{\ell}^{2}
+24\,{r}^{2}M
\right. \nonumber \\ 
&& \left. \qquad 
+6\,{\ell}^{4}{r}^{2}M-72\,r{M}^{2}+36\,{\ell}^{2}r{M}^{2}
+36\,\ell r{M}^{2}+72\,{M}^{3} \right) \Psi_{\ell m} \left( t,r \right) 
/
\left[ \left( r{\ell}^{2}+\ell r-2\,r+6\,M \right) ^{2}{r}^{4} \right] 
\nonumber \\ && \quad 
-4\,i {S}\,m\,\left( 4\,{r}^{3}{\ell}^{7}+144\,{M}^{3}{\ell}^{2}+16\,{r}^{3}\ell
-24\,{r}^{3}+18\,M{\ell}^{6}{r}^{2}+144\,{M}^{3}\ell+{r}^{3}{\ell}^{8}-216\,r{M}^{2}
-66\,{\ell}^{3}{r}^{2}M
\right. \nonumber \\ 
&& \left. \qquad 
-48\,{r}^{2}M\ell+144\,{M}^{3}+36\,{\ell}^{2}r{M}^{2}
+22\,{r}^{3}{\ell}^{2}+120\,{r}^{2}M+6\,{\ell}^{3}{r}^{3}-11\,{\ell}^{4}{r}^{3}
+54\,M{r}^{2}{\ell}^{5}
\right. \nonumber \\ 
&& \left. \qquad 
+72\,{M}^{2}{\ell}^{4}r+12\,{\ell}^{4}{r}^{2}M
+144\,{M}^{2}r{\ell}^{3}-90\,{r}^{2}M{\ell}^{2}-36\,\ell r{M}^{2}-14\,{\ell}^{5}{r}^{3} \right) 
{\frac {\partial }{\partial t}}\Psi_{\ell m} \left( t,r \right)
\nonumber \\ && \qquad 
/\left[{r}^{3} \left( \ell+1 \right) \ell \left( r{\ell}^{2}+\ell r-2\,r+6\,M \right) ^{3}
\right] 
+{\frac {24\,i{S}\, \left( \ell+2 \right)  \left( \ell-1 \right) m 
\left( r-2\,M \right) ^{2} }{{r}^{2} 
\left( r{\ell}^{2}+\ell r-2\,r+6\,M \right) ^{2}\ell \left( \ell+1 \right) }}
{\frac {\partial^2 }{\partial t \partial r}}\Psi_{\ell m} \left( t,r \right)
\nonumber \\ 
&& =
-12\,{S}\,\sqrt {{\frac { \left( \ell-m \right)  \left( \ell+m \right) }
{ \left( 2\,\ell-1 \right)  \left( 2\,\ell+1 \right) }}} 
\left( r-2\,M \right)  \left( \ell-2 \right)  
\left( {\ell}^{5}{r}^{2}+2\,{r}^{2}{\ell}^{4}+4\,r{\ell}^{3}M-4\,{r}^{2}{\ell}^{2}
+12\,rM{\ell}^{2}
\right. \nonumber \\ 
&& \left. \qquad 
-5\,{r}^{2}\ell+12\,rM\ell+6\,{r}^{2}-28\,rM+36\,{M}^{2} \right) 
\Psi_{\ell-1\, m}^{\rm (o)} \left( t,r \right) 
/\left[
{r}^{5}{\ell} 
\left( r{\ell}^{2}+\ell r-2\,r+6\,M \right) ^{2}
\right]
\nonumber \\ && \quad 
+12\,{S}\,\sqrt {{\frac { \left( \ell+m+1 \right)  \left( \ell-m+1 \right) }
{ \left( 2\,\ell+1 \right)  \left( 2\,\ell+3 \right) }}}  \left( r-2\,M \right) 
\left( \ell+3 \right)  
\left( {\ell}^{5}{r}^{2}+3\,{r}^{2}{\ell}^{4}+4\,r{\ell}^{3}M+2\,{r}^{2}{\ell}^{3}
+2\,{r}^{2}{\ell}^{2}
\right. \nonumber \\ 
&& \left. \qquad 
+32\,rM-8\,{r}^{2}-36\,{M}^{2} \right) 
\Psi_{\ell+1\, m}^{\rm (o)} \left( t,r \right)  
/\left[
\left( \ell+1 \right) {r}^{5} 
\left( r{\ell}^{2}+\ell r-2\,r+6\,M \right) ^{2}
\right] 
\,.
\label{eq:genLevenH}
\end{eqnarray}
\end{widetext}

And then, we need the following local source term 
which is added to the right hand side of the above equation, 
for the cell that the particle does cross:
\begin{eqnarray}
S_{\ell m}^{\rm (even,L)} &=& 
S_{\ell m}^{\rm (even,1,L)}+S_{\ell m}^{\rm (even,2,L)} \,,
\end{eqnarray}
where the first-order source term $S_{\ell m}^{\rm (even,1,L)}$ 
is the same as $S_{\ell m}^{\rm (even,1)}$ in Section~\ref{sec:1stRWZ}
and given  in Eq.~(A.5) of \cite{Lousto:2005ip} as 
\begin{widetext}
\begin{eqnarray}
S_{\ell m}^{\rm (even,1,L)} &=&
\Biggl[ 
{\frac { 32\,\pi \,\mu\, \left( {R} \left( t \right) -2\,M \right)  
\left( 2\,M-{R} \left( t \right) 
-\dot {R} \left( t \right) {R} \left( t \right)  \right)  
\left( 2\,M-{R} \left( t \right) +\dot {R} \left( t \right) 
{R} \left( t \right)  \right) U \left( t \right) }
{ \ell (\ell+1) \, \left( {R} \left( t \right)  \right) ^{2} 
\left( {R} \left( t \right) {\ell }^{2}+{R} \left( t \right) \ell 
-2\,{R} \left( t \right) +6\,M \right) }}
\frac{d}{dr} \delta(r-R(t))
\nonumber \\ &&
+\frac{\pi \,\mu}{\ell (\ell+1)} 
\biggl(
{\frac { 32\,{m}^{2}\left( {R} \left( t \right) -2\,M \right) U \left( t \right)  
\left( \dot {\Phi} \left( t \right)  \right) ^{2}}
{ \left( \ell -1 \right)  \left( \ell +2 \right) }}
-{\frac {64\,i\,m\,\dot {R} \left( t \right)  
\left( {R} \left( t \right) -2\,M \right) U \left( t \right) 
\dot {\Phi} \left( t \right) }{{R} \left( t \right) {\ell }^{2}
+{R} \left( t \right) \ell -2\,{R} \left( t \right) +6\,M}}
\nonumber \\ && 
-{\frac { 16\,\left( {R} \left( t \right) -2\,M \right)   
U \left( t \right)  \left( \dot{\Phi} \left( t \right)  \right) ^{2}}
{ \left( {R} \left( t \right) {\ell }^{2}+{R} \left( t \right) \ell 
-2\,{R} \left( t \right) +6\,M \right)  
\left( \ell -1 \right)  \left( \ell +2 \right) }}
\left( -8\,M+10\,M{\ell }^{2}+10\,M\ell -3\,{R} \left( t \right) {\ell }^{2}
\right. 
\nonumber \\ && \left. 
+2\,{R} \left( t \right) {\ell }^{3}+4\,{R} \left( t \right) 
+{R} \left( t \right) {\ell }^{4}-4\,{R} \left( t \right) \ell  \right)
+{\frac {16\,U \left( t \right)  \left( \dot {R} \left( t \right)  \right) ^{2}}
{{R} \left( t \right)  \left( {R} \left( t \right) {\ell }^{2}
+{R} \left( t \right) \ell -2\,{R} \left( t \right) +6\,M \right) ^{2}}}
\nonumber \\ && 
\times 
\left( -2\, \left( {R} \left( t \right)  \right) ^{2}\ell 
- \left( {R} \left( t \right)  \right) ^{2}{\ell }^{2}
+2\,{\ell }^{3} \left( {R} \left( t \right)  \right) ^{2}
+{\ell }^{4} \left( {R} \left( t \right)  \right) ^{2}
+12\,{R} \left( t \right) {\ell }^{2}M+12\,{R} \left( t \right) \ell M
+12\,{M}^{2} \right)  
\nonumber \\ && 
-{\frac { 16\,\left( {R} \left( t \right) -2\,M \right) ^{2} 
 U \left( t \right) }
{ \left( {R} \left( t \right)  \right) ^{3} \left( {R} \left( t \right) {\ell }^{2}
+{R} \left( t \right) \ell -2\,{R} \left( t \right) +6\,M \right) ^{2}}}
\left( 60\,{M}^{2}+12\,{R} \left( t \right) {\ell }^{2}M
-24\,{R} \left( t \right) M+12\,{R} \left( t \right) \ell M
\right. 
\nonumber \\ && \left. 
-2\, \left( {R} \left( t \right)  \right) ^{2}\ell 
- \left( {R} \left( t \right)  \right) ^{2}{\ell }^{2} 
+2\,{\ell }^{3} 
\left( {R} \left( t \right)  \right) ^{2}
+{\ell }^{4} \left( {R} \left( t \right)  \right) ^{2} \right)
\biggr) \delta \left( r-{R} \left( t \right)  \right)
\Biggr] \, Y_{\ell m}^*  \left( \Theta_0, \Phi(t) \right) 
\,.
\end{eqnarray}
The second-order local source term $S_{\ell m}^{\rm (even,2,L)}$ 
has the following expression:
\begin{eqnarray}
S_{\ell m}^{\rm (even,2,L)} &=&
{\frac {192\,i \,m\,S \,\pi \,\mu\,U \left( t \right)\, 
\left( {R} \left( t \right) -2\,M \right) 
\dot{R} \left( t \right)  \left( { \ell }^{2}+ \ell -2\,{m}
^{2} \right) \left( \dot {\Phi} \left( t \right)  \right) ^{2}}
{ \left(  \ell +2 \right)  \left(  \ell -1 \right)  \left(  \ell +1 \right) ^{2}{ \ell }^{2} 
\left( {R} \left( t \right) { \ell }^{2}+{R} \left( t \right)  \ell 
-2\,{R} \left( t \right) +6\,M \right) }}
 Y_{\ell m}^*  \left( \Theta_0, \Phi(t) \right) 
\frac{d}{dr} \delta(r-R(t))
\nonumber \\ && 
+ \biggl[
{\frac {-24\,i \,m\,S \,\left(  \ell +2 \right)  \left(  \ell -1 \right) 
 \left( {R} \left( t \right) -2\,M \right) ^{2}}
{ \left( {R} \left( t \right)  \right) ^{2} \left( {R} \left( t \right) { \ell }^{2}
+{R} \left( t \right)  \ell -2\,{R} \left( t \right) 
+6\,M \right) ^{2} \ell  \left(  \ell +1 \right) }}
\left. \frac{\partial}{\partial t} \Psi_{\ell m}^{\rm (step)} \left( t,r \right) 
\right|_{r=R(t)}
\nonumber \\ && 
+i\,m \,S\, \pi \,\mu\,Y_{\ell m}^*  \left( \Theta_0, \Phi(t) \right) 
\biggl( -384\, \left( { \ell }^{2}+ \ell -2\,{m}^{2} \right)  
\bigl( 30\,{M}^{2}+6\,{R} \left( t \right) { \ell }^{2}M
+6\,{R} \left( t \right)  \ell M
\nonumber \\ && 
-21\,{R} \left( t \right) M
-2\, \ell  \left( {R} \left( t \right)  \right) ^{2}
-2\,{ \ell }^{2} \left( {R} \left( t \right)  \right) ^{2}
+4\, \left( {R} \left( t \right)  \right) ^{2} \bigr) U \left( t \right)  
\left( \dot {\Phi} \left( t \right)  \right) ^{2} \dot{R} \left( t \right) 
\nonumber \\ && 
/ \bigl[
{R} \left( t \right)  \left(  \ell +2 \right)  \left(  \ell -1 \right)  \left(  \ell +1 \right) ^{2}
{ \ell }^{2} \left( {R} \left( t \right) { \ell }^{2}+{R} \left( t \right)  \ell 
-2\,{R} \left( t \right) +6\,M \right) ^{2} \bigr]
\nonumber \\ && 
-128\, \left( {R} \left( t \right) - 2\,M \right) 
U \left( t \right) \dot{R} \left( t \right)  
\bigl(  \left( {R} \left( t \right)  \right) ^{2}{ \ell }^{4}
+2\, \left( {R} \left( t \right)  \right) ^{2}{ \ell }^{3}-6\,{ \ell }^{2} 
\left( {R} \left( t \right)  \right) ^{2}
+18\,{R} \left( t \right) { \ell }^{2}M
\nonumber \\ && 
-7\, \ell  \left( {R} \left( t \right)  \right) ^{2}
+18\,{R} \left( t \right)  \ell M+10\, \left( {R} \left( t \right)  \right) ^{2}
-36\,{R} \left( t \right) M+36\,{M}^{2} \bigr) 
\nonumber \\ && 
/ \bigl[
\ell  \left( {R} \left( t \right)  \right) ^{3} \left( {R} \left( t \right) { \ell }^{2}
+{R} \left( t \right)  \ell -2\,{R} \left( t \right) 
+6\,M \right) ^{3} \left(  \ell +1 \right) \bigr] 
\nonumber \\ && 
+{\frac {192\,i\,m \,U \left( t \right) \left( { \ell }^{2}+ \ell -2\,{m}^{2} \right)  
\left( {R} \left( t \right) -2\,M \right)  
\left( \dot{\Phi} \left( t \right)  \right) ^{3}}
{ \left(  \ell +2 \right)  \left(  \ell -1 \right)  \left(  \ell +1 \right) ^{2}{ \ell }^{2} 
\left( {R} \left( t \right) { \ell }^{2}+{R} \left( t \right)  \ell 
-2\,{R} \left( t \right) +6\,M \right) }}
\nonumber \\ && 
+128\,i\,m\,  U \left( t \right) 
\left( {R} \left( t \right) - 2\,M \right) ^{2} 
\bigl( 72\,{M}^{2}+30\,{R} \left( t \right) { \ell }^{2}M+30\,{R} \left( t \right)  \ell M
-60\,{R} \left( t \right) M
\nonumber \\ && 
-10\, \ell  \left( {R} \left( t \right)  \right) ^{2}
+ \left( {R} \left( t \right)  \right) ^{2}{ \ell }^{4}
+2\, \left( {R} \left( t \right)  \right) ^{2}{ \ell }^{3}
+16\, \left( {R} \left( t \right)  \right) ^{2}
-9\,{ \ell }^{2} \left( {R} \left( t \right)  \right) ^{2} \bigr) 
\dot{\Phi}  \left( t \right) 
\nonumber \\ && 
/ \bigl[ \left(  \ell +1 \right) ^{2}{ \ell }^{2} 
\left( {R} \left( t \right) { \ell }^{2}+{R} \left( t \right)  \ell 
-2\,{R} \left( t \right) +6\,M \right) ^{3} 
\left( {R} \left( t \right)  \right) ^{3} \bigr] \biggr) 
\biggr] \delta(r-R(t))
\nonumber \\ && 
+
\biggl( 256\,\pi \, \mu\,S\,\sqrt {{\frac { \left(  \ell -m \right)  \left(  \ell +m \right) }
{ \left( 2\, \ell -1 \right)  \left( 2\, \ell +1 \right) }}}
\frac{U \left( t \right) \dot{\Phi}\left( t \right) 
\left( {R} \left( t \right) -2\,M \right) ^{2} \left( -2\,{R} \left( t \right) 
+{R} \left( t \right)  \ell +{R} \left( t \right) { \ell }^{2}+3\,M \right) }
{{ \ell }^{2} 
\left(  \ell +1 \right)  \left( {R} \left( t \right)  \right) ^{3} 
\left( {R} \left( t \right) { \ell }^{2}+{R} \left( t \right)  \ell 
-2\,{R} \left( t \right) +6\,M \right) ^{2}}
\nonumber \\ && 
\times 
\partial_\theta Y_{\ell-1 m}^*  \left( \Theta_0, \Phi(t) \right)  
\nonumber \\ && 
-256\,\pi \,\mu\,S\,\sqrt {{\frac { \left(  \ell +m+1 \right)  \left(  \ell -m+1 \right) }
{ \left( 2\, \ell +1 \right)  \left( 2\, \ell +3 \right) }}} 
\frac{U \left( t \right) \dot{\Phi} \left( t \right) 
\left( {R} \left( t \right) -2\,M \right) ^{2} \left( -2\,{R} \left( t \right) 
+{R} \left( t \right)  \ell +{R} \left( t \right) { \ell }^{2}+3\,M \right) }
{{ \ell } 
\left(  \ell +1 \right) ^{2} \left( {R} \left( t \right)  \right) ^{3} 
\left( {R} \left( t \right) { \ell }^{2}+{R} \left( t \right)  \ell 
-2\,{R} \left( t \right) +6\,M \right) ^{2} }
\nonumber \\ && 
\times 
\partial_\theta Y_{\ell+1 m}^*  \left( \Theta_0, \Phi(t) \right)  
\biggr) 
\delta \left( r-{R} \left( t \right)  \right) 
\,.
\end{eqnarray}
\end{widetext}
Here we have used the analytic expression of 
the wave function $\Psi_{\ell m}^{\rm (step)}(t,R(t))$ 
at the particle's location, 
and the instantaneous geodesic approximation
for the second-order source term. 

It is noted that there are remaining terms at the horizon 
in the integration of the wave equation. 
These arise from the transformation of the original wave equation 
to the above equation. 
Since the source term for the original wave equation 
does not include any remaining term at the horizon, 
these remaining terms cancel out with the derivatives of the wave function, 
i.e., $\partial\Psi_{\ell m}^{\rm (step)}/\partial t$.

\subsection{Analysis of the odd parity wave equation}\label{app:odddetail}

In the calculation for the odd parity perturbation of the SRWZ formalism, 
we have treated the first- and second-order perturbations separately. 
The first-order (local) source term, 
$S_{\ell m}^{\rm (odd,1,L)}=S_{\ell m}^{\rm (odd,1)}$,
which has been simplified with the geodesic equation, is given as 
\begin{widetext}
\begin{eqnarray}
S_{\ell m}^{\rm (odd,1,L)} &=& 
{\frac {32\,\pi \,\mu}{\ell \left( \ell+1 \right)  \left( \ell-1 \right) 
 \left( \ell+2 \right) }} 
\biggl[
\biggl( 
U \left( t \right)  \left( {R} \left( t \right) -2\,M \right) 
 \left( {R} \left( t \right)  \right) ^{2} 
\left( \dot{\Phi} \left( t \right)  \right) ^{3}
+{\frac { \left( {R} \left( t \right) -2\,M \right) 
\dot {\Phi} \left( t \right) }{U \left( t \right) }}
\biggr) \,\frac{d}{dr} \,\delta(r-R(t))
\nonumber \\ && 
+
\biggl(
 \left( 2\,{R} \left( t \right) -7\,M \right) {R} \left( t \right) 
U \left( t \right)  \left( \dot {\Phi} \left( t \right)  \right) ^{3}
-i \,m\,\dot {R}  \left( t \right) {R} \left( t \right)
U \left( t \right)  \left( \dot {\Phi} \left( t \right)  \right) ^{2}
+{\frac { \left( {R} \left( t \right) -5\,M \right) 
\dot {\Phi}  \left( t \right) }{{R} \left( t \right) 
U \left( t \right) }}
\nonumber \\ &&  
-{\frac { 2\,\left( {R} \left( t \right) -2\,M \right) ^{2}
U \left( t \right) \dot {\Phi}  \left( t \right) }
{ \left( {R} \left( t \right)  \right) ^{2}}}
\biggr) 
\,\delta(r-R(t)) 
\biggr] \, 
\partial_\theta Y_{\ell m}^*  \left( \Theta_0, \Phi(t) \right) \,. 
\end{eqnarray}
\end{widetext}

Next, we focus on the second-order wave equation. 
$\Psi_{\ell \pm 1 m}^{(1)}$ and $\Psi_{\ell m}^{\rm (o,1)}$ 
have already been derived in the first-order calculation. 
For the cell that the particle does not cross, 
we may consider only the homogeneous part of the wave equation, 
\begin{widetext}
\begin{eqnarray}
&&
-{\frac {\partial ^{2}}{\partial {t}^{2}}}\Psi_{\ell m}^{\rm (o,Z,2)} \left( t,r \right) 
+{\frac {\partial ^{2}}{\partial {r^*}^{2}}}
\Psi_{\ell m}^{\rm (o,Z,2)} \left( t,r \right) -V_{\ell}^{\rm (odd)}(r) 
\Psi_{\ell m}^{\rm (o,Z,2)} \left( t,r \right) 
\nonumber \\ && 
= 
i\,m\,S\, \biggl( -2\,{\frac { \left( 2\,r{\ell }^{3}-5\,r{\ell }^{2}+18\,r-
6\,r\ell +r{\ell }^{4}+9\,M{\ell }^{2}-42\,M+9\,\ell M \right)  \left( r-2\,M \right) }
{ \left( \ell +1 \right) {r}^{7}\ell }}
\Psi_{\ell m}^{\rm (o,1)} \left( t,r \right) 
\nonumber \\ && \quad 
-2\,{\frac { \left( 3\,r-8\,M \right)  \left( r-2\,M \right) }{{r}^{6}}}
\frac{\partial}{\partial r} \Psi_{\ell m}^{\rm (o,1)}  \left( t,r \right) 
+2\,{\frac { \left( r-2\,M \right) ^{2} }{{r}^{5}}}
\frac{\partial^2}{\partial r^2} \Psi_{\ell m}^{\rm (o,1)}  \left( t,r \right) 
\biggr) 
\nonumber \\ && 
+\frac{4\,S}{ {\ell} \left( \ell-1 \right) }
 \sqrt {{\frac { \left( \ell-m \right)  \left( \ell+m \right) }{
 \left( 2\,\ell-1 \right)  \left( 2\,\ell+1 \right) }}} 
\biggl(
3\,{\frac { \left( r-2\,M \right) ^{2} \left( \ell -1 \right) }{{r}^{5}}}
\frac{\partial^2}{\partial t \partial r} \Psi_{\ell-1 m}^{(1)}  \left( t,r \right) 
\nonumber \\ && \qquad 
+\frac{3}{2}{\frac { \left( r-2\,M
 \right)  \left( \ell -1 \right)  
 \left( {\ell }^{4}{r}^{2}-2\,{r}^{2}{\ell }^{3}-{r}^{2}{\ell }^{2}+2\,{r}^{2}\ell +6
\,r{\ell }^{2}M-6\,r\ell M-12\,rM+24\,{M}^{2} \right)}
{{r}^{6} \left( r{\ell }^{2}-r\ell -2\,r+6\,M \right) }}
\frac{\partial}{\partial t} \Psi_{\ell-1 m}^{(1)} \left( t,r \right) 
\biggr) 
\nonumber \\ && 
+\frac{4\,S}{\left( \ell+1 \right) \left( \ell +2 \right)  }
\sqrt {{\frac { \left( \ell+m+1
 \right)  \left( \ell-m+1 \right) }{ \left( 2\,\ell+1 \right)  \left( 2\,\ell+3
 \right) }}}
\biggl(
-3\,{\frac { \left( r-2\,M \right) ^{2} \left( \ell +2 \right)   }{{r}^{5}}}
\frac{\partial^2}{\partial t \partial r} \Psi_{\ell+1 m}^{(1)} \left( t,r \right) 
\nonumber \\ && \qquad 
-\frac{3}{2}\,{\frac { \left( r-2\,M \right)  \left( \ell +2 \right) 
\left( {\ell }^{4}{r}^{2}+6\,{r}^{2}{\ell }^{3}+
11\,{r}^{2}{\ell }^{2}+6\,{r}^{2}\ell +6\,r{\ell }^{2}M+18\,r\ell M+24\,{M}^{2}
 \right) }{{r}^{6} \left( r{\ell }^{2}+3\,r\ell +6\,M \right) }}
\frac{\partial}{\partial t}  \Psi_{\ell+1 m}^{(1)}   \left( t,r \right)  
\biggr)
\,.
\label{eq:genLoddH}
\end{eqnarray}
\end{widetext}

The second-order local source terms, $S_{\ell m}^{\rm (odd,Z,2,L)}$ 
which we need for the cell that the particle does cross is written as 
\begin{widetext}
\begin{eqnarray}
S_{\ell m}^{\rm (odd,Z,2,L)}
&=& 
i\,m\,S\, \biggl( 2\,{\frac { \left( {R} \left( t \right) -2\,M \right) ^{2} }
{ \left( {R} \left( t \right)  \right) ^{5}}}
\left. \frac{\partial}{\partial r} \Psi_{\ell m}^{\rm (o,1,step)}  \left( t,r \right) 
\right|_{r=R(t)}
\nonumber \\ && 
-{\frac { 32\,\pi \,\mu\,
\left( \ell+3 \right)  \left( \ell-2 \right) 
U \left( t \right) \dot \Phi  \left( t \right) 
 \left( {R} \left( t \right) -2\,M \right) ^{2}}
{ \left( {R} \left( t \right)  \right) ^{5} \left( \ell+1 \right) ^{2}
{\ell}^{2} \left( \ell-1 \right)  \left( \ell+2 \right) }}
\partial_\theta Y_{\ell m}^*  \left( \Theta_0, \Phi(t) \right) 
\biggr)\,\delta(r-R(t))
\nonumber \\ && 
+ \frac{4\,S}{ {\ell} \left( \ell-1 \right) }
 \sqrt {{\frac { \left( \ell-m \right)  \left( \ell+m \right) }{
 \left( 2\,\ell-1 \right)  \left( 2\,\ell+1 \right) }}} 
\biggl[ 
{\frac { 12\, \pi \,\mu\,\left( \ell+2 \right) \left( {\ell}^{2}-\ell-2\,{m}^{2} \right)  
\left( \dot \Phi  \left( t \right) \right) ^{2} 
\left( -{R} \left( t \right) +2\,M \right) 
U \left( t \right) \,\dot R \left( t \right)}
{ \left( {R} \left( t \right)  \right) ^{2}
 \left( \ell-2 \right) \ell}}
\nonumber \\ && \times 
Y_{\ell-1 m}^*  \left( \Theta_0, \Phi(t) \right)
\, \frac{d}{dr} \delta(r-R(t))
 + \biggl( 
{\frac {3\, \left( {R} \left( t \right) -2\,M \right) ^{2}
 \left( \ell-1 \right)  \left( \ell+2 \right)  \left( \ell+1 \right)  }
{ \left( {R} \left( t \right)  \right) ^{5}}}
\left. \frac{\partial}{\partial t} \Psi_{\ell-1 m}^{\rm (1,step)}  \left( t,r \right) 
\right|_{r=R(t)}
\nonumber \\ &&  
+\frac{\pi \,\mu\, 
\left( \ell+2 \right)}{{\ell}}  \biggl( 
- {\frac {12\,i  \,m \,\left( {R} \left( t \right) - 2\,M \right) 
U \left( t \right)\left( {\ell}^{2}-\ell-2\,{m}^{2} \right)  
\left( \dot \Phi  \left( t \right)  \right) ^{3}}
{ \left( {R} \left( t \right)  \right) ^{2} \left( \ell-2 \right) }}
\nonumber \\ &&  
-{\frac { 12\,\left( 5\,{R} \left( t \right) -14\,M \right) 
\dot R  \left( t \right) U \left( t \right)  
\left( {\ell}^{2}-\ell-2\,{m}^{2} \right)  
\left( \dot \Phi  \left( t \right)  \right) ^{2}}
{ \left( {R} \left( t \right)  \right) ^{3} \left( \ell-2 \right) }}
+ {\frac {96\,i  \,m\,\left( {R} \left( t \right)  - 2\,M \right) ^{3} 
\left( \ell+1 \right) U \left( t \right)
\dot \Phi  \left( t \right) }
{ \left( {R} \left( t \right)  \right) ^{5} \left( {R} \left( t \right) {\ell}^{2}
-{R} \left( t \right) \ell-2\,{R} \left( t \right) +6\,M \right) }}
\nonumber \\ &&  
-{\frac { 48\,\left( {R} \left( t \right) -2\,M \right) ^{2} \left( \ell+1 \right) 
U \left( t \right) \dot R  \left( t \right)  
\left( \ell-1 \right) \ell}{ \left( {R} \left( t \right)  \right) ^{5} 
\left( {R} \left( t \right) {\ell}^{2}-{R} \left( t \right) \ell-
2\,{R} \left( t \right) +6\,M \right) }} 
 \biggr)
Y_{\ell-1 m}^*  \left( \Theta_0, \Phi(t) \right)
\biggr)\, \delta(r-R(t)) 
\biggr] 
\nonumber \\ && 
+\biggl[ \ell \leftrightarrow -\ell-1 \biggr]
\,,
\label{eq:appA12}
\end{eqnarray}
\end{widetext}
where $[\ell \leftrightarrow -\ell-1]$ refers to an additional
term obtained by replacing $\ell$ with
$-\ell-1$ in all terms in Eq.~(\ref{eq:appA12}) starting from
 $\frac{4 S}{\ell (\ell-1)}\cdots$ and subsequently
replacing
$\Psi_{-\ell-2 \,m}^{\rm (1,step)}$ with $\Psi_{\ell+1 \,m}^{\rm (1,step)}$
and $Y_{-\ell-2 \,m}^*$ with $Y_{\ell+1 \,m}^*$.
The above source term is added to the right hand side 
of the homogeneous part of the wave equation. 
It is found that the right hand side of the equation vanishes 
at the horizon. Here, the instantaneous geodesic 
approximation has also been used in the above equation.

\subsection{Analysis of quasinormal modes}

In the SRWZ formalism, 
we discuss a special case 
where the $\ell=2,\,m=\pm 2$ even or odd parity mode 
is dominant and couplings with the other modes can be ignored. 
Also, the perturbed Regge-Wheeler-Zerilli equations 
with the spin effect do not have the local source terms, i.e., 
we consider the homogeneous equation. 
 
For the even parity part, we use the same equation as 
Eq.~(\ref{eq:genLevenH}) without the other mode coupling, 
\begin{widetext}
\begin{eqnarray}
&& \left[
-{\frac {\partial ^{2}}{\partial {t}^{2}}}
+\frac{\partial^2}{{\partial r^*}^2} 
-6\,{\frac {( r-2\,M) 
( 4\,{r}^{3} +4\,{r}^{2}M+6\,r{M}^{2}+3\,{M}^{3}) }{{r}^{4} ( 2\,r+3\,M ) ^{2}}}\right]
\,\Psi_{2\pm 2} (t,r)
\nonumber \\ && \quad
\pm {\frac {8\,i{S}\, \left( r-2\,M \right) ^{2} }
{{r}^{2} \left( 2\,r+3\,M \right) ^{2}}}
{\frac {\partial ^{2}}{\partial r\partial t}}\Psi_{2\pm 2} \left( t,r \right)
\mp {\frac {8\,i{S}\, \left( 6\,{r}^{3}+46\,{r}^{2}M+45\,r{M}^{2}+21\,{M}^{3} \right)  }
{{r}^{3} \left( 2\,r+3\,M \right) ^{3}}}
{\frac {\partial }{\partial t}}\Psi_{2\pm 2} \left( t,r \right) = 0 \,.
\label{eq:ESE2}
\end{eqnarray}
\end{widetext}

For the odd parity part, we use a different equation from Eq.~(\ref{eq:genLoddH}). 
This is because if we ignore the other mode coupling and the local source term, 
we can derive a simple equation by using only 
the Cunningham et al. waveform $\Psi_{2\pm 2}^{\rm (o)} $ 
(or the Zerilli waveform $\Psi_{2\pm 2}^{\rm (o,Z)} $). 
The $\ell=2,\,m=\pm 2$ odd parity wave equation with the spin effect 
becomes 
\begin{widetext}
\begin{eqnarray}
&& \left[
-{\frac {\partial ^{2}}{\partial {t}^{2}}}
+\frac{\partial^2}{{\partial r^*}^2} 
-6\,{\frac { \left( r-M \right)  \left( r-2\,M \right)}{{r}^{4}}} \right]
\,\Psi_{2\pm 2}^{\rm (o)} (t,r)
\nonumber \\ && \quad
\pm {\frac { 2\,i{S}}{{r}^{2}}}
{\frac {\partial ^{2}}{\partial r\partial t}}\Psi_{2\pm 2}^{\rm (o)} \left( t,r \right)
\mp {\frac  {2\,i{S}\, \left( 7\,{r}^{2}-17\,rM+8\,{M}^{2} \right)}
{ \left( r-2\,M \right) {r}^{4}}}
{\frac {\partial }{\partial t}}\Psi_{2\pm 2}^{\rm (o)} \left( t,r \right) = 0 \,.
\label{eq:OSO2}
\end{eqnarray}
\end{widetext}
where we have introduced 
$\Psi_{2\pm 2}^{\rm (o)} ( t,r ) = \Psi_{2\pm 2}^{\rm (o,1)} ( t,r ) + 
\Psi_{2\pm 2}^{\rm (o,2)} ( t,r )$.

We treat the above equations in the frequency domain, 
\begin{widetext}
\begin{eqnarray}
&& \left[
\omega^2
+\frac{d^2}{{d r^*}^2} 
-6\,{\frac {( r-2\,M) 
( 4\,{r}^{3} +4\,{r}^{2}M+6\,r{M}^{2}+3\,{M}^{3}) }{{r}^{4} ( 2\,r+3\,M ) ^{2}}}\right]
\,\Psi_{2\pm 2} (\omega;r)
\nonumber \\ && \quad
\pm  {\frac {8\,{S}\,\omega \left( r-2\,M \right) ^{2} }
{{r}^{2} \left( 2\,r+3\,M \right) ^{2}}}
{\frac {d}{d r}}\Psi_{2\pm 2} \left( \omega;r \right)
\mp {\frac {8\,{S}\,\omega \left( 6\,{r}^{3}+46\,{r}^
{2}M+45\,r{M}^{2}+21\,{M}^{3} \right)
  }{{r}^{3} \left( 2\,r+3\,M \right) ^{3}}}
\Psi_{2\pm 2} \left(\omega;r \right) = 0 \,,
\\
&& \left[
\omega^2
+\frac{d^2}{{d r^*}^2} 
-6\,{\frac { \left( r-M \right)  \left( r-2\,M \right)}{{r}^{4}}} \right]
\,\Psi_{2\pm 2}^{\rm (o)} (\omega;r)
\nonumber \\ && \quad
\pm  {\frac {2\,{S}\,\omega}{{r}^{2}}}
{\frac {d}{d r}}\Psi_{2\pm 2}^{\rm (o)} \left( \omega;r \right)
\mp {\frac {2\,{S}\,\omega \left( 7\,{r}^{2}-17\,rM+8\,{M}^{2} \right)}
{ \left( r-2\,M \right) {r}^{4}}}
\Psi_{2\pm 2}^{\rm (o)} \left( \omega;r \right) = 0 \,.
\end{eqnarray}
\end{widetext}

For the nonspinning ($S=0$) case of the above equations, 
we have already known the transformation between 
the Regge-Wheeler and Zerilli function. 
This is known as the Chandrasekhar
transformation~\cite{Chandrasekhar83}, given by 
\begin{eqnarray}
&& \Psi_{2\pm 2}^{\rm (o,1)} \left( t,r \right) 
= \left( 6+9\,{\frac {{M}^{2} \left( r-2\,M \right) }{{r}^{2} \left( 2\,r+3\,M \right) }} 
\right) \Psi_{2\pm 2}^{(1)} \left( t,r \right) 
\nonumber \\ && \quad 
+3\,M \left( 1-2\,{\frac {M}{r}} \right)  
{\frac {d}{dr}}\Psi_{2\pm 2}^{(1)} \left( t,r \right) \,.
\end{eqnarray}
Using these transformation, for example, we may solve only 
the Regge-Wheeler equation to obtain the quasinormal frequency. 

In order to discuss a similar treatment up to $O(a^1)$ ($a=S/M$), 
first we consider the following transformation: 
\begin{eqnarray}
\Psi_{2\pm 2} \left( \omega;r \right) 
&=& \exp\left(\pm \frac{2 \,S \,\omega}{2 r+3 M}\right) 
\tilde \Psi_{2\pm 2} \left( \omega;r \right) 
\,,
\nonumber \\
\Psi_{2\pm 2}^{\rm (o)} \left( \omega;r \right) 
&=& \exp\left(\pm \frac{S \,\omega}{r-2 M} \right) 
\tilde \Psi_{2\pm 2}^{\rm (o)} \left( \omega;r \right) 
\,,
\label{eq:ST}
\end{eqnarray}
where these transformations are consistent in the $O(a^1)$. 
Since we treat the wave functions only up to $O(a^1)$, 
we may choose another transformation here. 
From the above transformations, 
we have the simple differential equations which are similar to 
the Regge-Wheeler and Zerilli equations. 
The difference from the original Regge-Wheeler and Zerilli equations 
arises in the potential terms. 
\begin{widetext}
\begin{eqnarray}
&& \left[
\omega^2
+\frac{d^2}{{d r^*}^2} 
-6\,{\frac {( r-2\,M) 
( 4\,{r}^{3} +4\,{r}^{2}M+6\,r{M}^{2}+3\,{M}^{3}) }{{r}^{4} ( 2\,r+3\,M ) ^{2}}}\right]
\,\tilde \Psi_{2\pm 2} (\omega;r)
\nonumber \\ && \quad
\mp  {\frac {8\,{S}\,\omega \left( 4\,{r}^{3}+56\,{r}^{2}M
+36\,r{M}^{2}+15\,{M}^{3} \right)
  }{{r}^{3} \left( 2\,r+3\,M \right) ^{3}}}
\tilde \Psi_{2\pm 2} \left(\omega;r \right) = 0 \,,
\\
&& \left[
\omega^2
+\frac{d^2}{{d r^*}^2} 
-6\,{\frac { \left( r-M \right)  \left( r-2\,M \right)}{{r}^{4}}} \right]
\,\tilde \Psi_{2\pm 2}^{\rm (o)} (\omega;r)
\mp {\frac {4\,{S}\,\omega \left( 3\,{r}-2\,{M} \right)}
{ {r}^{4}}}
\tilde \Psi_{2\pm 2}^{\rm (o)} \left( \omega;r \right) = 0 \,.
\label{eq:oddF}
\end{eqnarray}

From these equations, 
we find the "Chandrasekhar" transformation as 
\begin{eqnarray}
\tilde \Psi_{2\pm 2}^{\rm (o)} \left( \omega;r \right) 
&=& 
\left( 6+9\,{\frac {{M}^{2} \left( r-2\,M \right) }{{r}^{2} \left( 2\,r+3\,M \right) }}
\mp {\frac {{S}\,M\,\omega\, \left( 45\,{M}^{2}-48\,{r}^{2} \right) }
{{r}^{2} \left( 2\,r+3\,M \right) ^{2}}} \right)  
\tilde \Psi_{2\pm 2} \left( \omega;r \right) 
\nonumber \\ && 
+3\,M \left( 1-2\,{\frac {M}{r}} \right)  
\left( 1 \pm \frac{4}{3}\,{\frac {{S}\,\omega}{M}} \right) 
{\frac {d}{dr}}\tilde \Psi_{2\pm 2} \left( \omega;r \right) \,.
\end{eqnarray}
\end{widetext}
The differential equations for the even and odd parity perturbation 
become the same form by using the above transformation. 

Next, we consider quasinormal modes derived from Eq.~(\ref{eq:oddF}). 
A recent review for quasinormal modes is given in~\cite{Berti:2009kk}. 
Here, we should note that if we use Eq.~(\ref{eq:ST}) 
to obtain the simple equation in 
Eq.~(\ref{eq:oddF}), these change the boundary behaviors 
near the horizon and at infinity. 
Therefore, although the expression is same in the $O(a^1)$ expansion, 
we should consider to do another transformation:
\begin{eqnarray}
\Psi_{2\pm 2}^{\rm (o)} \left( \omega;r \right) 
&=& 
\left[1+ \frac{r-2\,M}{r} \ln \left(1 \pm \frac{S \,\omega \,r}{(r-2\,M)^2} \right)\right] 
\nonumber \\ && \times 
\tilde \Psi_{2\pm 2}^{\rm (o)} \left( \omega;r \right) \,.
\end{eqnarray}
This does not change the boundary behaviors. 

In order to calculate the quasinormal frequencies, we use the 
Leaver's method~\cite{Leaver:1985ax}. 
As boundary conditions, the wave function $\tilde \Psi_{2\pm 2}^{\rm (o)}$ has 
the following behaviors:
\begin{eqnarray}
\tilde \Psi_{2\pm 2}^{\rm (o)} \left( \rho;r \right) &\to& 
r^{-\rho} \,e^{-\rho \,r} \quad {\rm for} \quad r \to \infty \,,
\nonumber \\ 
\tilde \Psi_{2\pm 2}^{\rm (o)} \left( \rho;r \right) &\to& 
(r-1)^{\rho+i\,\chi} \quad {\rm for} \quad r \to 1 \,,
\end{eqnarray}
where we have considered $2M=1$ and $\rho=-i\omega$ which are 
the same notation as~\cite{Leaver:1985ax}. 
Here, $\chi$ is defined 
by the nondimensional spin parameter $\chi=S/M^2$. 
Then a solution of Eq.~(\ref{eq:oddF}) can be written 
in the form of 
\begin{eqnarray}
\tilde \Psi_{2\pm 2}^{\rm (o)} \left( \rho;r \right) &=& 
r^{-\rho} \,e^{-\rho \,(r-1)} \,
(r-1)^{\rho+i\,\chi} \,r^{-(\rho+i\,\chi)}
\nonumber \\ && \times 
\sum_{n=0}^{\infty} a_n \left(\frac{r-1}{r} \right)^n
\,.
\end{eqnarray}

We obtain the recurrence relation for $a_n$ in the above equation, 
\begin{eqnarray}
\alpha_0 \,a_1 + \beta_0 \,a_0 &=& 0 \,, 
\end{eqnarray}
and for $n \geq 1$, 
\begin{eqnarray}
\alpha_n \,a_{n+1} + \beta_n \,a_n + \gamma_n \,a_{n-1} &=& 0 
\,,
\label{eq:recur}
\end{eqnarray}
where 
\begin{eqnarray}
\alpha_n &=& 
\left( 2+2\,n \right) \rho+2\,i \left( n+1 \right) \chi+ \left( n+1 \right) ^{2}
\,,
\nonumber \\ 
\beta_n &=& -8\,{\rho}^{2}+ \left( -4-8\,n-7\,i\chi \right) \rho-2\,i 
\left( 2\,n+1 \right) \chi
\nonumber \\ && 
-3-2\,{n}^{2}-2\,n
\,,
\nonumber \\
\gamma_n &=& 
4\,{\rho}^{2}+ \left( 4\,n+5\,i\chi \right) \rho+2\,i\chi\,n
\nonumber \\ && 
+ \left( n-2 \right)  \left( n+2 \right)
\,.
\end{eqnarray}
When we set $\chi=0$, the above equations reduce 
to Eq.~(8) in~\cite{Leaver:1985ax}. 

\begin{figure}[ht]
\begin{center}
\caption{The quasinormal frequencies, $\omega$ around $\chi=0$. 
We have used the same expression as \cite{Leaver:1985ax}. 
The (red) circles show our result, and 
the $+$ marks denote the values given in Table II of \cite{Glampedakis:2003dn}.}
\includegraphics[width=3.4in]{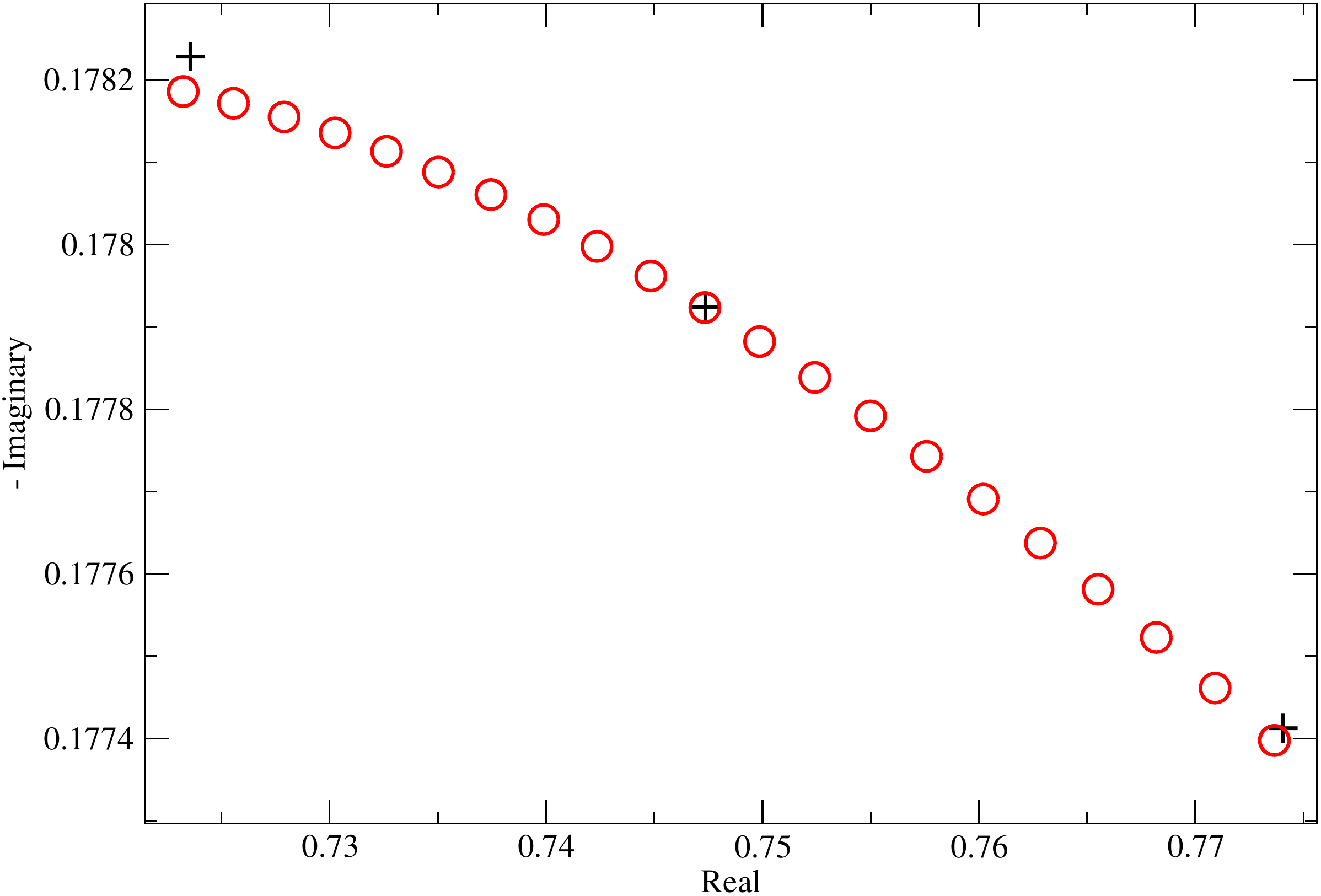}
\label{fig:1}
\end{center}
\end{figure}

\begin{figure}[ht]
\begin{center}
\caption{The real part of the quasinormal frequencies, $\omega$ around $\chi=0$. 
The horizontal axis denotes the nondimensional spin parameter, 
$\chi=S/M^2$.
The (red) circles show our result and 
the $+$ marks denote the values given in Table II of \cite{Glampedakis:2003dn}.
}
\includegraphics[width=3.4in]{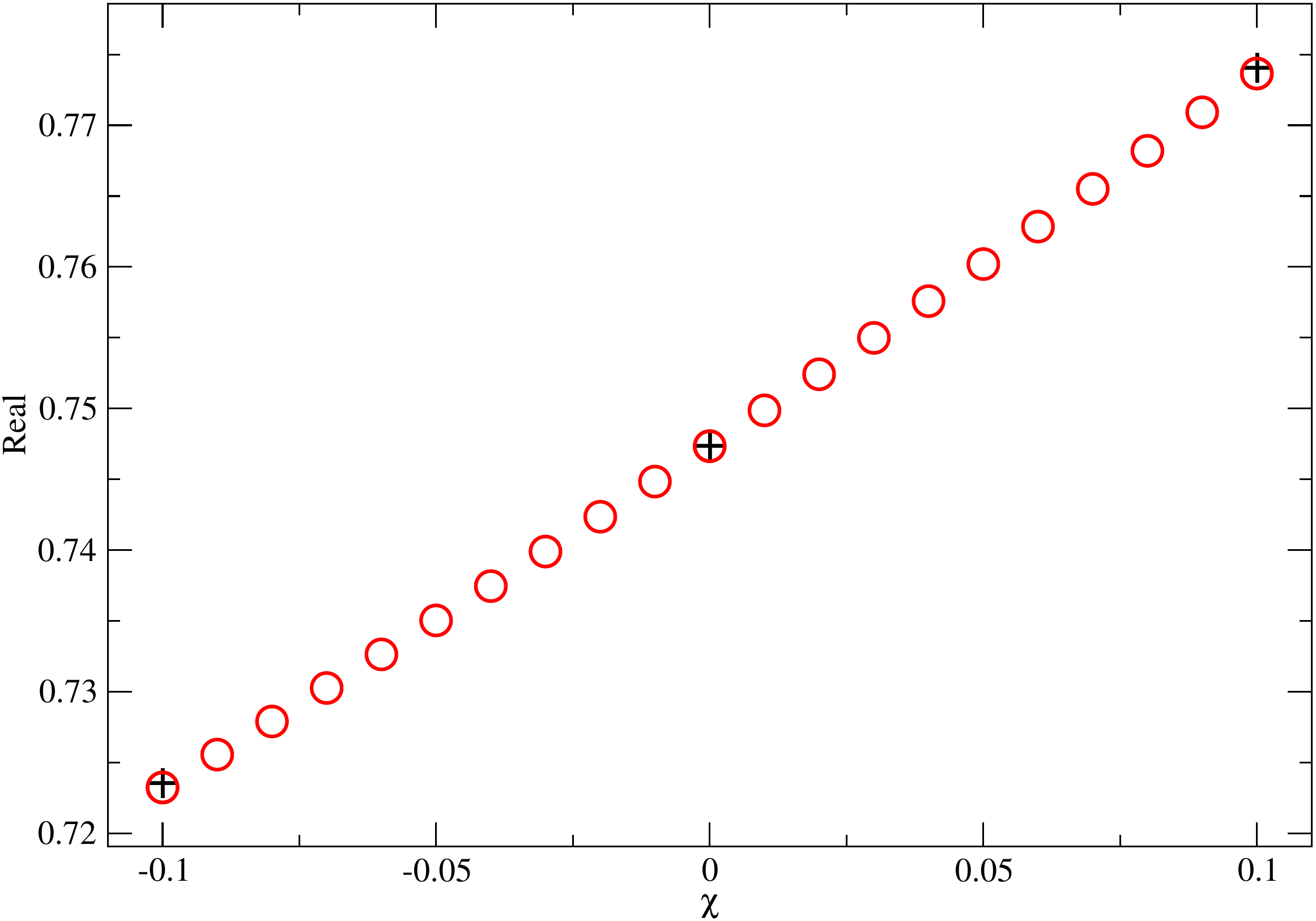}
\label{fig:2}
\end{center}
\end{figure}

\begin{figure}[ht]
\begin{center}
\caption{The (minus) imaginary part of the quasinormal frequencies, 
$\omega$ around $\chi=0$. 
The horizontal axis denotes the nondimensional spin parameter, $\chi=S/M^2$.
The (red) circles show our result and 
the $+$ marks denote the values given in Table II of \cite{Glampedakis:2003dn}.
}
\includegraphics[width=3.4in]{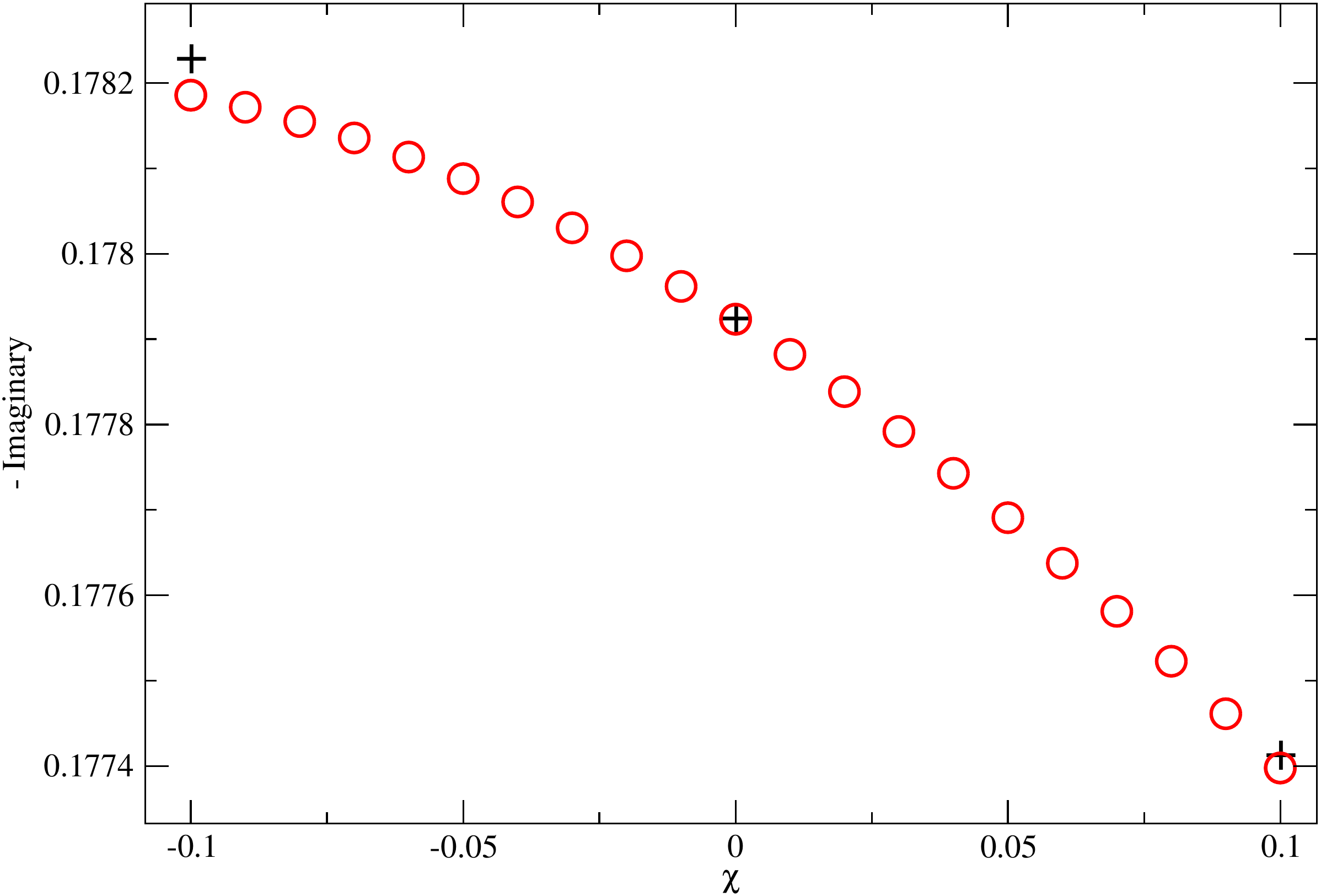}
\label{fig:3}
\end{center}
\end{figure}

\begin{figure}[ht]
\begin{center}
\caption{The quasinormal frequencies, $\omega$ for $-0.9 \leq \chi \leq 0.9$. 
We have used the same expression of \cite{Leaver:1985ax}. 
The (red) circles show our result and 
the $+$ marks denote the values given in Table II of \cite{Glampedakis:2003dn}.
}
\includegraphics[width=3.4in]{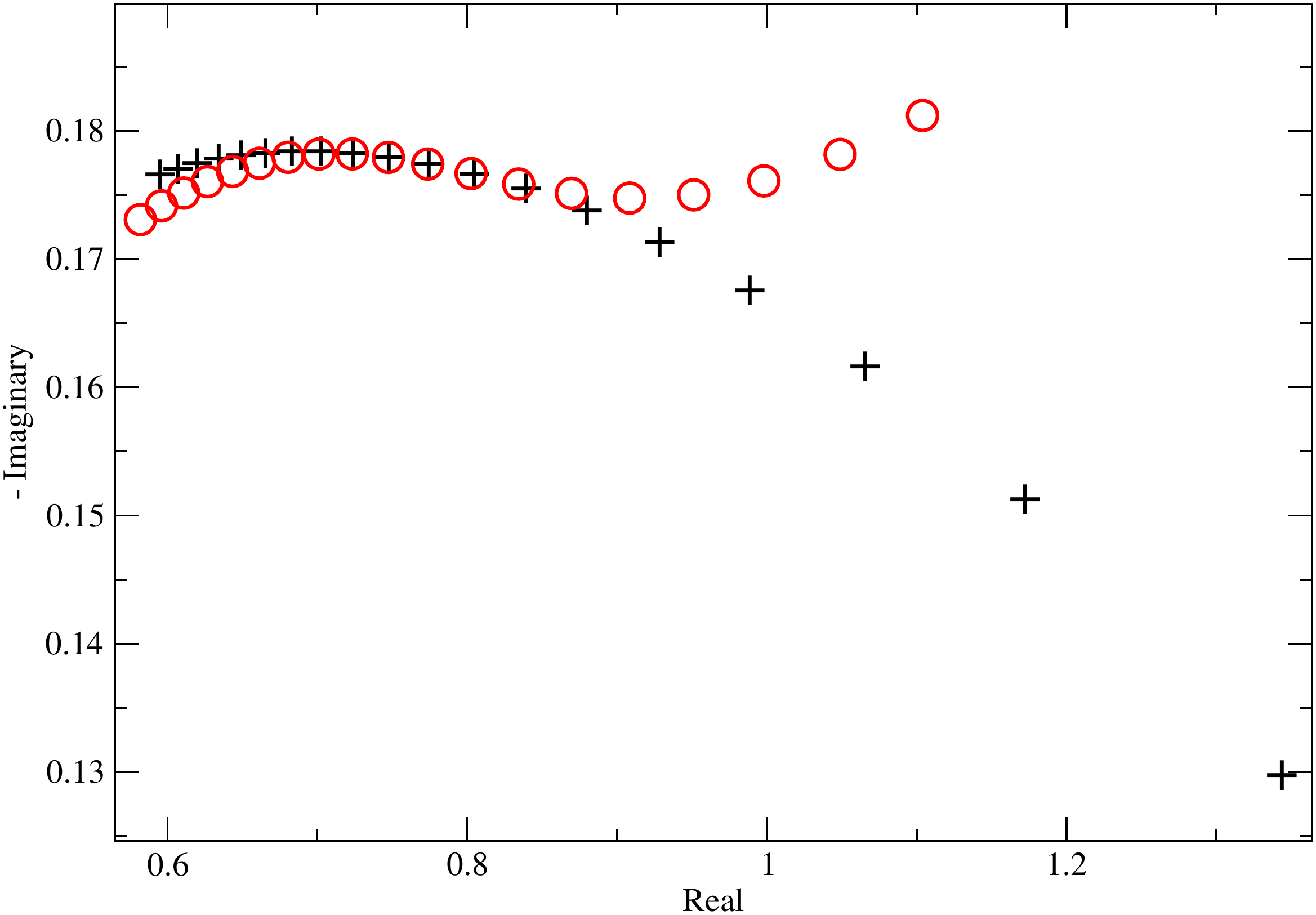}
\label{fig:1L}
\end{center}
\end{figure}

\begin{figure}[ht]
\begin{center}
\caption{The real part of the quasinormal frequencies, $\omega$ for $-0.9 \leq \chi \leq 0.9$. 
The horizontal axis denotes the nondimensional spin parameter, 
$\chi=S/M^2$.
The (red) circles show our result and 
the $+$ marks denote the values given in Table II of \cite{Glampedakis:2003dn}.
}
\includegraphics[width=3.4in]{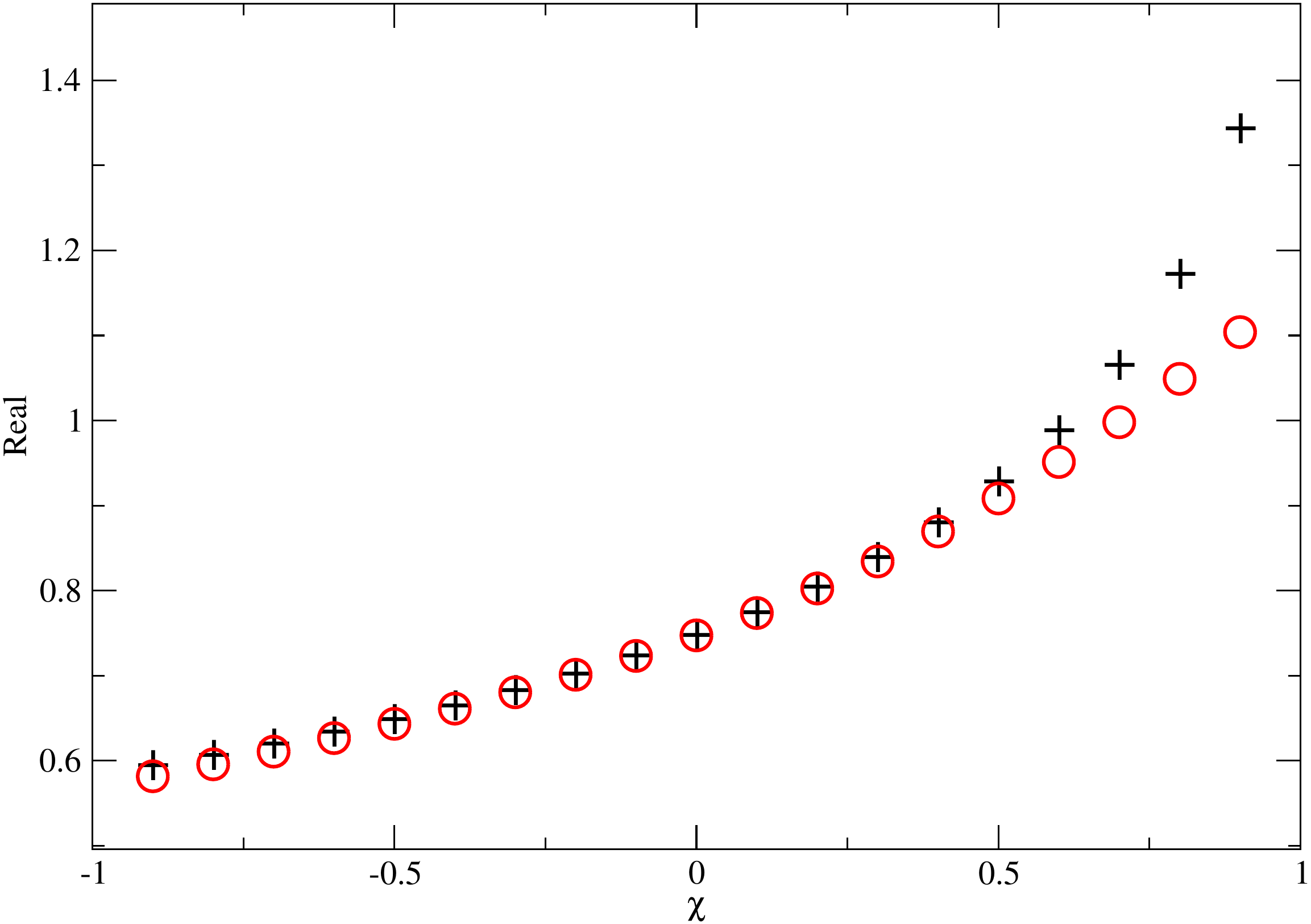}
\label{fig:2L}
\end{center}
\end{figure}

\begin{figure}[ht]
\begin{center}
\caption{The (minus) imaginary part of the quasinormal frequencies, $\omega$ 
for $-0.9 \leq \chi \leq 0.9$. 
The horizontal axis denotes the nondimensional spin parameter, 
$\chi=S/M^2$.
The (red) circles show our result and 
the $+$ marks denote the values given in Table II of \cite{Glampedakis:2003dn}.
}
\includegraphics[width=3.4in]{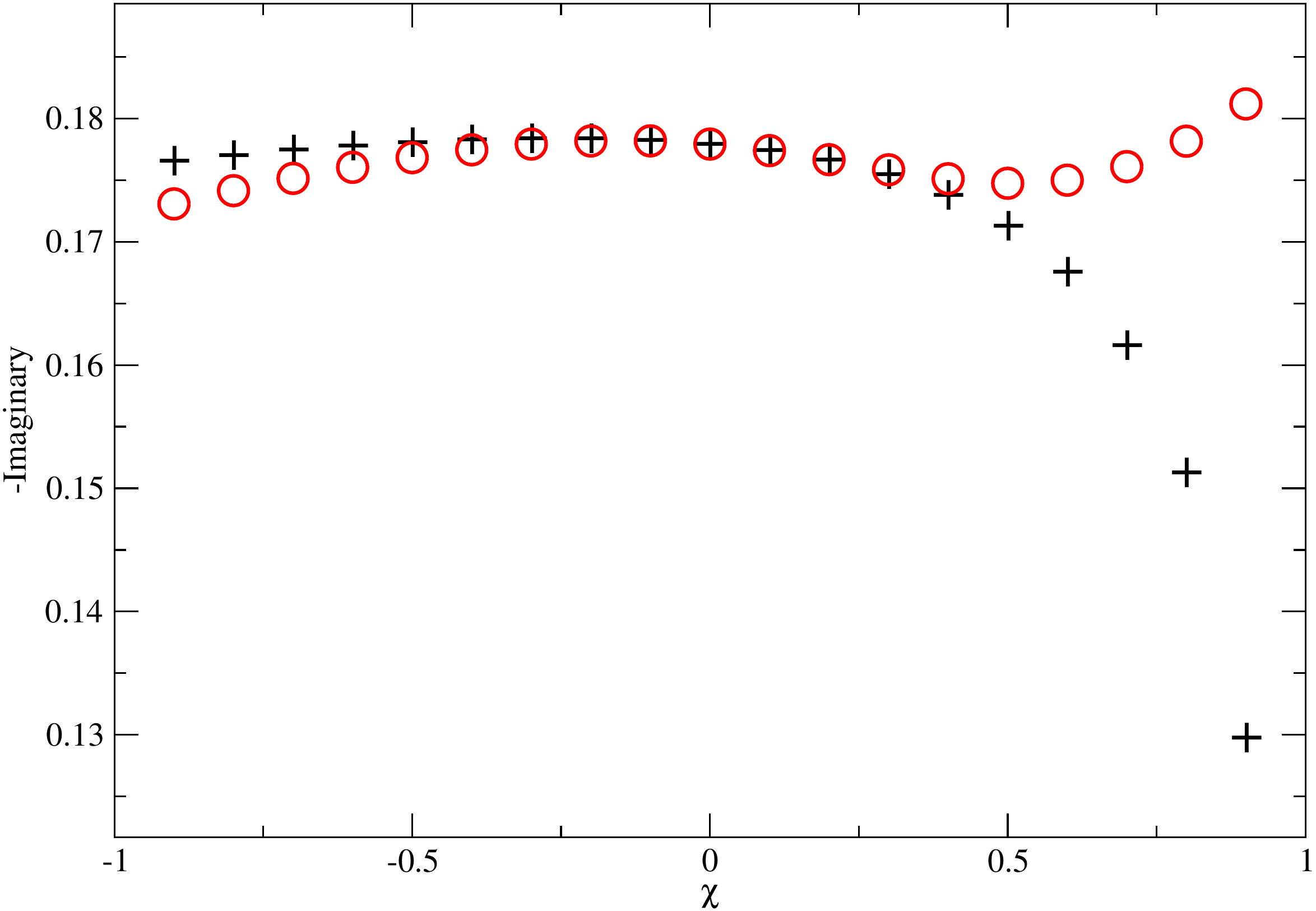}
\label{fig:3L}
\end{center}
\end{figure}

\begin{figure}[ht]
\begin{center}
\caption{The error in the real part of the quasinormal frequencies, 
i.e., $-{\rm Err}_{\Im}$. 
The horizontal axis denotes the nondimensional spin parameter, 
$\chi=S/M^2$.
}
\includegraphics[width=3.4in]{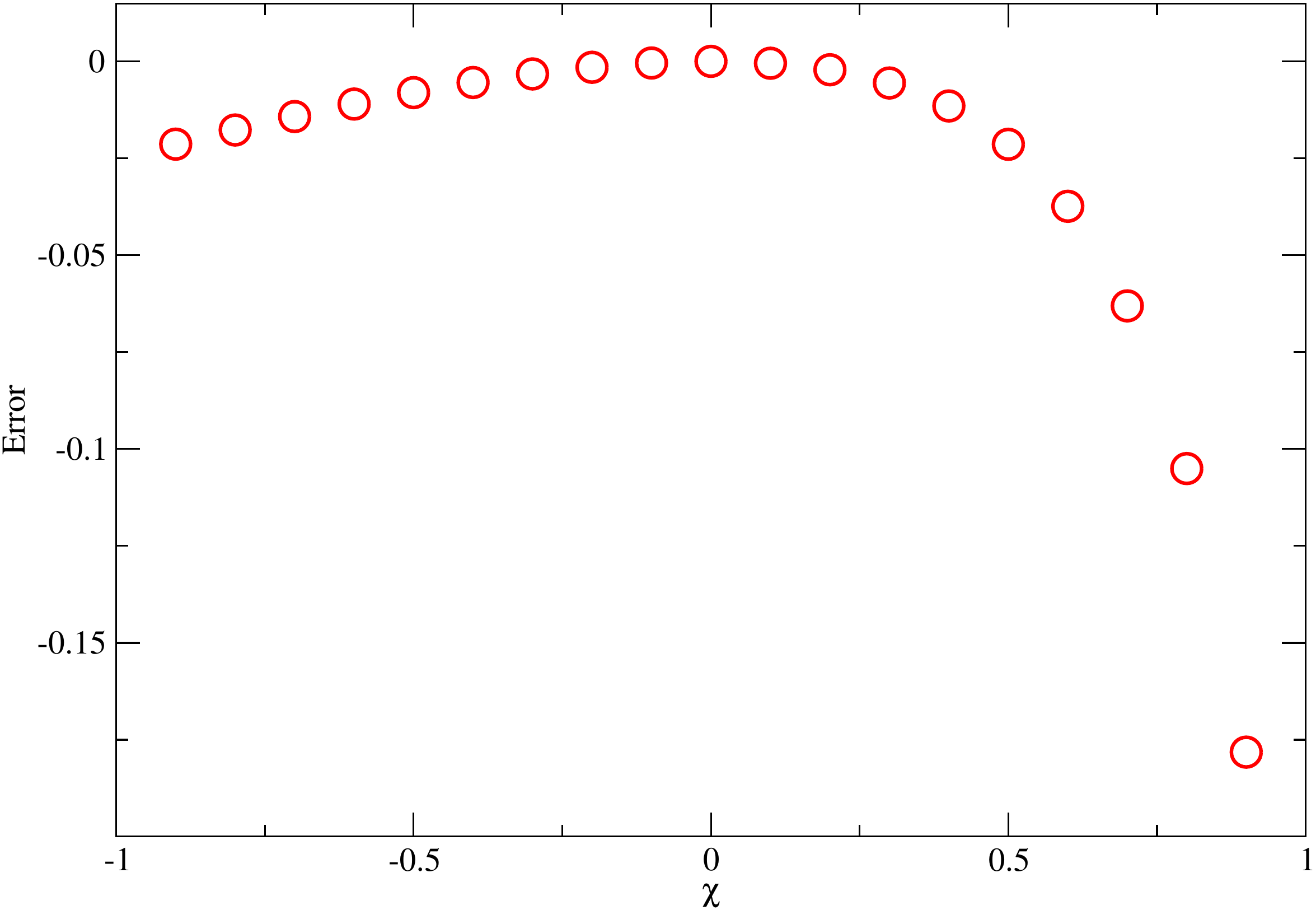}
\label{fig:ER}
\end{center}
\end{figure}

\begin{figure}[ht]
\begin{center}
\caption{The error in the (minus) imaginary part of the quasinormal frequencies, 
i.e., ${\rm Err}_{\Re}$. 
The horizontal axis denotes the nondimensional spin parameter, 
$\chi=S/M^2$.
}
\includegraphics[width=3.4in]{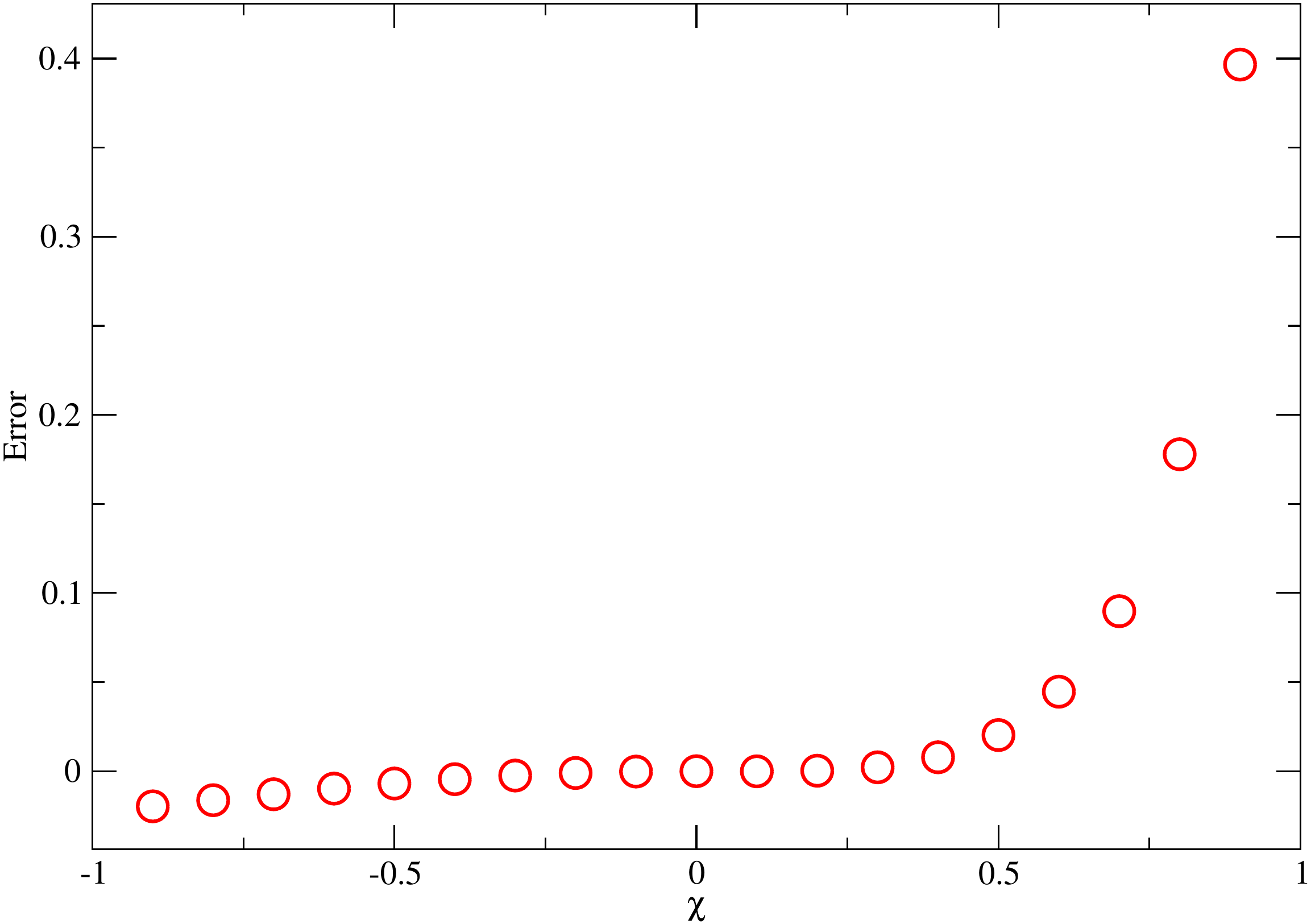}
\label{fig:EI}
\end{center}
\end{figure}

\begin{figure}[ht]
\begin{center}
\caption{The absolute value of the relative errors in quasinormal frequencies 
in the region $-0.5 \leq \chi \leq 0.5$. The (red) circles and (blue) boxes 
show those of the real and imaginary parts of the frequency, i.e., 
$|{\rm Err}_{\Im}|$ and $|{\rm Err}_{\Re}|$, respectively. 
The horizontal axis denotes the nondimensional spin parameter, 
$\chi=S/M^2$.
}
\includegraphics[width=3.4in]{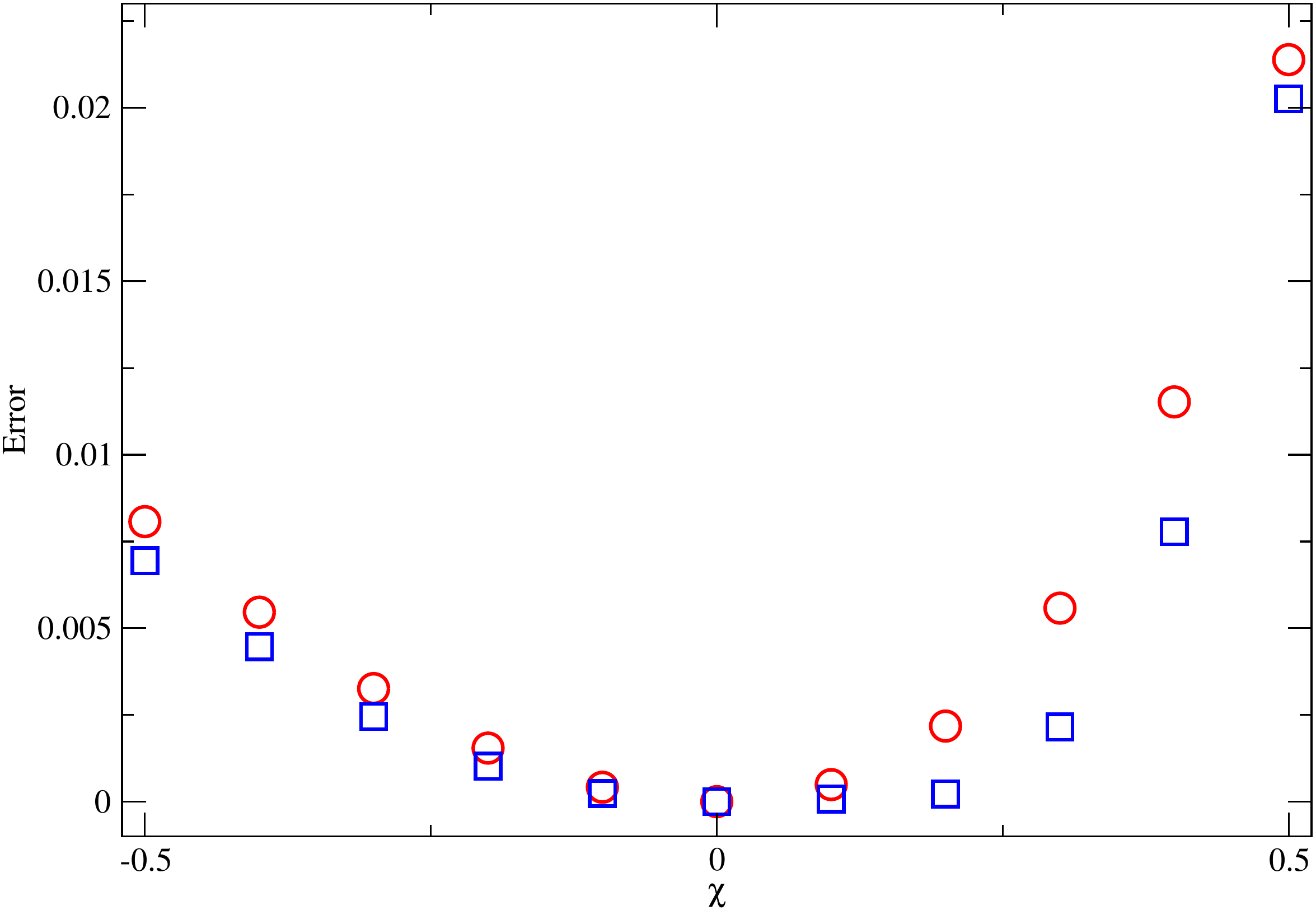}
\label{fig:absE}
\end{center}
\end{figure}

In Fig.~\ref{fig:1}, we show the result 
for the quasinormal frequencies, $\omega$ around $\chi=0$. As a reference, 
we also plot the values given in Table II of \cite{Glampedakis:2003dn}. 
Figures~\ref{fig:2} and~\ref{fig:3} show 
the real and imaginary parts of the quasinormal frequencies 
around $\chi=0$, respectively. 
The Figures \ref{fig:1L}, \ref{fig:2L}, and \ref{fig:3L} show 
the result for $-0.9 \leq \chi \leq 0.9$. 
In Table~\ref{tab:QNMF}, we show the numerical values 
and the relative errors for the real and imaginary parts of $\rho$ 
defined by 
\begin{eqnarray}
{\rm Err}_{\Re} = \frac{\Re(\rho_a)-\Re(\rho)}{\Re(\rho)} \,,
\, 
{\rm Err}_{\Im} = \frac{\Im(\rho_a)-\Im(\rho)}{\Im(\rho)} \,,
\end{eqnarray}
where $\rho_a$ and $\rho$ represent our result and that of \cite{Glampedakis:2003dn}, 
respectively. 
We plot the above errors in Figs.~\ref{fig:ER} and \ref{fig:EI}, 
and we zoom in the region $-0.5 \leq \chi \leq 0.5$ in Fig.~\ref{fig:absE} 
which shows the absolute values of the relative error. 

\begin{table*}[htbp]
\centering 
\caption{The quasinormal frequencies in terms of $\rho=-i\omega$.
The $m=-2$ mode can be considered as the $m=2$ mode 
with the inverse spin signature. 
Here we set $a_{17}=0$ in the recurrence relation of Eq.~(\ref{eq:recur}). 
This creates the numerical error in our calculation (see $\chi=0.0$).}
\label{tab:QNMF}
\begin{tabular}{c||c|c|c|c}
\hline \hline
$\chi$ & $m=2$ (This paper) & $m=2$ (\cite{Glampedakis:2003dn}) 
& ${\rm Err}_{\Re}$ & ${\rm Err}_{\Im}$ \\
\hline
$-0.9$ & $-0.173072-0.581783 \,i$ & $-0.176562-0.594488 \,i$  & $-0.019766$ & $0.021371$ \\
$-0.8$ & $-0.174141-0.595877 \,i$ & $-0.177024-0.606626 \,i$  & $-0.016285$ & $0.017719$ \\
$-0.7$ & $-0.175137-0.610783 \,i$ & $-0.177434-0.619616 \,i$  & $-0.012945$ & $0.014255$ \\
$-0.6$ & $-0.176039-0.626584 \,i$ & $-0.177784-0.633568 \,i$  & $-0.009815$ & $0.011023$ \\
$-0.5$ & $-0.176825-0.643379 \,i$ & $-0.178062-0.648614 \,i$  & $-0.006947$ & $0.008071$ \\
$-0.4$ & $-0.177466-0.661283 \,i$ & $-0.178262-0.664916 \,i$  & $-0.004465$ & $0.005463$ \\
$-0.3$ & $-0.177930-0.680440 \,i$ & $-0.178368-0.682666 \,i$  & $-0.002455$ & $0.003260$ \\
$-0.2$ & $-0.178181-0.701019 \,i$ & $-0.178364-0.702106 \,i$  & $-0.001025$ & $0.001548$ \\
$-0.1$ & $-0.178186-0.723233 \,i$ & $-0.178228-0.723536 \,i$  & $-0.000235$ & $0.000418$ \\
$0.0$ & $-0.177923-0.747340 \,i$ & $-0.177924-0.747344 \,i$   & $-0.000005$ & $0.000005$ \\
$0.1$ & $-0.177398-0.773654 \,i$ & $-0.177412-0.774036 \,i$   & $-0.000078$ & $0.000493$ \\
$0.2$ & $-0.176662-0.802534 \,i$ & $-0.176622-0.804290 \,i$   & $0.000226$ & $0.002183$ \\
$0.3$ & $-0.175836-0.834372 \,i$ & $-0.175458-0.839054 \,i$   & $0.002154$ & $0.005580$ \\
$0.4$ & $-0.175116-0.869549 \,i$ & $-0.173764-0.879684 \,i$   & $0.007780$ & $0.011521$ \\
$0.5$ & $-0.174747-0.908398 \,i$ & $-0.171278-0.928246 \,i$   & $0.020253$ & $0.021382$ \\
$0.6$ & $-0.174999-0.951162 \,i$ & $-0.167532-0.988090 \,i$   & $0.044570$ & $0.037373$ \\
$0.7$ & $-0.176094-0.997991 \,i$ & $-0.161588-1.065198 \,i$   & $0.089771$ & $0.063095$ \\
$0.8$ & $-0.178154-1.048919 \,i$ & $-0.151252-1.172030 \,i$   & $0.177862$ & $0.105040$ \\
$0.9$ & $-0.181181-1.103919 \,i$ & $-0.129726-1.343268 \,i$   & $0.396644$ & $0.178185$ \\
\hline \hline
\end{tabular}
\end{table*}

\bibliographystyle{apsrev}
\bibliography{../../../Bibtex/references}

\end{document}